\documentclass[%
 nofootinbib,
 amsmath,
 amssymb,
 twocol
]{ametsoc}

\journal{jpo}

\bibpunct{(}{)}{;}{a}{}{,}

\usepackage{graphicx}
\usepackage{dcolumn}
\usepackage{bm}
\usepackage{amsmath}
\usepackage{amssymb}
\usepackage{amsfonts}
\usepackage{mathtools}
\usepackage{lastpage}
\usepackage{ulem}
\usepackage{epsf}
\usepackage{multirow}
\usepackage{caption}
\usepackage{subcaption}
\usepackage{url}
\usepackage{natbib}
\usepackage{setspace}
\usepackage{enumerate}
\usepackage{epstopdf}
\usepackage{wrapfig}
\usepackage{morefloats}
\usepackage{float}
\usepackage{color}

\usepackage[T1]{fontenc}			
\usepackage[utf8]{inputenc}		
\usepackage{xcolor,cancel}                

\newcommand\hcancel[2][black]{\setbox0=\hbox{$#2$}%
\rlap{\raisebox{.30\ht0}{\textcolor{#1}{\rule{\wd0}{2pt}}}}#2} 

\newcommand{\rulesep}{\unskip\ \vline width 2pt}

\allowdisplaybreaks

\begin{document}

\title{ Reduced-Order Quasilinear Model of Ocean Boundary-Layer Turbulence}

\authors{Joseph Skitka and J. B. Marston}
\affiliation{\small Brown University, Department of Physics} 

\extraauthor{Baylor Fox-Kemper}
\extraaffil{\small Brown University, Department of Earth, Environmental, and Planetary Science}

\date{\today}


\abstract{The combined effectiveness of model reduction and the
quasilinear approximation for the reproduction of the low-order statistics of oceanic surface
boundary-layer turbulence is investigated. Idealized horizontally homogeneous problems 
of surface-forced thermal convection and Langmuir turbulence are
studied in detail.  Model reduction is achieved with a Galerkin projection of the governing equations onto an
subset of modes determined by proper orthogonal decomposition. 
For less than 0.2\% of the modes retained, the reduced quasilinear model is able to
reproduce vertical profiles of horizontal mean fields as well as certain energetically important
second-order turbulent transport statistics and energies to within 30\% error.  
For intermediate sizes of the basis truncation some statistics
approach those found in the fully nonlinear simulations.  Thus basis reduction 
can actually improve upon the accuracy of quasilinear dynamics.  A test
model with a small total number of modes demonstrates the non-monotonic convergence 
toward the correct statistics as the size of the basis is increased.}

\maketitle


\section{Introduction}
\label{sec:intro}

Reliable weather and climate modeling requires the accurate simulation of two turbulent components of the Earth system that have thus far only seen incremental improvements: the oceanic and atmospheric boundary layers (BLs).  Given that long duration turbulence-resolving simulation is centuries beyond the reach of present-day computers~\citep{fox-kemper14, IPCC13, Hamlington_2014},
large-scale numerical simulations of the Earth's oceans and atmospheres employ the hydrostatic primitive equations, 
in which vertical accelerations of fluid parcels are presumed negligible, 
enabling significant computational savings.  Unfortunately, predictions of \textit{vertical} transport and mixing are generally unreliable as these depend crucially on smaller-scale, non-hydrostatic 3D processes -- particularly turbulence within the planetary BL \citep{McWilliams_2000} -- that are  neglected or phenomenologically parameterized.  Errors in the simulation of low clouds, for example, are believed to be the single most important source of uncertainty in climate models and have stymied effort at improvement \citep{Bony:2015bd,2005JClimate}.  Similarly, wind- and surface-wave-driven Langmuir turbulence in the ocean surface BL should be incorporated into climate models for improved accuracy and robustness~\citep{belcher2012global}, yet a recent ocean BL simulation with horizontal grid spacing near 4 meters required millions of supercomputer hours for only weeks of simulated time to capture a region only 20km $\times$ 20km \citep{Hamlington_2014}.

Given that convective turbulence mediates vertical exchanges both within and between the atmospheric and oceanic BLs, it is perhaps not surprising that general circulation models are very sensitive to the chosen parameterizations of vertical transport \citep{Large_1997,Doney_2004,fox-kemper08,LiReichl19}.  The present unsatisfactory state-of-affairs is summarized in a review of numerical weather prediction, which asserts \textit{``the uncertainties inherent to physical parameterization, either from incomplete process understanding or the dilemma of representing the impact of unresolved processes on the resolved scales, may require a fundamentally different approach.  Elements of parameterizations or entire schemes are likely to require components that appear statistical to the large scales.
Examples are stochastic sampling of parameter probability distribution functions, stochastically driven sub-cell models, or super-parameterizations through embedding entire convection-resolving simulations at sub-grid scale $\ldots$
How radical this approach needs to be is currently not clear''} \citep{Bauer_2015}.  

The systematic development of Direct Statistical Simulation (DSS) approaches to the oceanic boundary layer is our central objective here.  As a first step in that direction we employ the quasi-linear (QL) approximation \citep{MALKUS:1954dh,Vedenov:1961us,SPIEGEL:1962vf,herring63,marston14} for which a non-local second-order closure (described below) is exact.  \cite{Chaalal:2016jx} investigated the QL approximation in a related model of the dry convective atmospheric boundary layer, demonstrating that QL can capture BL deepening and, with less accuracy, the turbulent transport of kinetic energy.   Two ocean-surface boundary layer models will be used for case studies: Surface-forced thermal convection and Langmuir turbulence \citep{Leibovich_1983,Thorpe_2004}.    Both problems are statistically homogeneous in the horizontal, while Langmuir and, if subjected to wind shear, convective turbulence additionally are anisotropic in the horizontal \citep{van2012}.   We emphasize that, owing to horizontal isotropy, this configuration constitutes a particularly challenging application for DSS, yet one of fundamental geoscientific importance.

Eschewing the traditional route of accumulating statistics from numerical simulation time series, DSS is a set of approaches for directly obtaining the statistics of dynamical systems rather than by their accumulation from numerical simulation 
\citep{Allawala:2016hx}.   One form of DSS focuses on the equations of motion (EOM) for the equal-time cumulants, which are truncated at low order.  Consider modeling the evolution of a state vector ${\bf q}(\vec{{r}}, t)$ specified by a system of master partial differential equations (PDEs). For simplicity, further consider the case where the nonlinearities are quadratic so that the system can be written as
\begin{equation}
{\bf q}_t = {\cal L}[{\bf q}]+{\cal Q}[{\bf q},{\bf q}],
\label{EOM}
\end{equation}
where ${\cal L}$ is a linear vector differential operator and ${\cal Q}$ is the operator that includes the nonlinear (quadratic) interactions such as those arising in the material derivative.  Specializing to models that are translationally invariant in one direction, henceforth denoted the zonal (or streamwise) direction, a standard Reynolds decomposition of the state vector into its zonal mean and fluctuation (denoted with an overbar and prime, respectively) is made by setting ${\bf q} = \overline{{\bf q}}+{{\bf q}^\prime}$.  
The average here is chosen to satisfy the Reynolds rules of averaging so that $\overline{\overline{\bf q}}=\overline{\bf q}\ \ {\rm and}\ \ \overline{{\bf q}^{\prime}}=0$.  
Averaging Eqn.~(\ref{EOM}), using the linearity of ${\cal L}$ and assuming the averaging operator commutes with ${\cal L}$, yields
\begin{eqnarray}
\overline{{\bf q}}_t &=& \overline{{\cal L}[{{\bf q}}]}+\overline{{\cal Q}[{\bf q},~ {\bf q}]}\;=\; {\cal L}[\overline{{\bf q}}]+\overline{{\cal Q}[{\bf q},~ {\bf q}]}.
\label{avEOM}
\end{eqnarray}
Of course, the difficulty now is to describe the average of the nonlinear term in the equation for the mean (the first cumulant). At this point local closure schemes are often adopted \citep{krauseraedler1980,ruediger1988}.  For example, in large-eddy simulations (LES) of turbulent flows,                                                                                                                                                                                                                                                                                                                                                                                                                                                                                                                                                                                                                                                                                                                                                                                                                                                                                                                                                                                                                                                                                                                                                                                                                                                                                                                                                                                                                                                                                                                                                                                                                                                                                                                                                                                                                                                                                                                                                                                                                                                                                                                                                                                                                                                                                                                                                                                                                                                                                                                                                                                                                                                                                                                                                                                                      the average of this nonlinear term is generally parameterized using Fickian diffusive schemes in which fluctuation correlations are taken to be proportional to the (local) gradient of mean variables \citep{canutominotti2001}.  Not only are such closures inaccurate and not mathematically justifiable for inhomogeneous and anisotropic flows, but in LES the ad hoc proportionality coefficients must be trained for particular flow regimes.  Standard LES schemes are particularly ineffective near walls (when additional ad hoc, e.g. Van Driest, wall models are required) and for strongly anisotropic flows, including strongly stratified turbulence. 

A more sophisticated approach is to derive an equation for the evolution of the mean of the non-local fluctuation--fluctuation (or eddy--eddy) interactions \citep{bartelloholloway1991,maltrudvallis1991}.  Unwarranted assumptions about homogeneity or isotropy are avoided, and non-local correlations of the form $\overline{{\bf q}'(\vec{r}_1){\bf q}'(\vec{r}_2)}$ are tracked, where $\vec{r}_1$ and $\vec{r}_2$ are two position vectors within the domain of interest \citep{marstonconoveretal2008}. The equation governing the evolution of this quantity (the second cumulant) is obtained by multiplying the EOM  for the system (defined at the point $\vec{r}_1$) by ${\bf q}(\vec{r}_2)$, averaging and symmetrizing.  In addition to introducing non-local quadratic terms from the $\overline{{\bf q}\,{\cal L}[{\bf q]}}$ terms, this procedure requires the evaluation of non-local cubic terms (the third cumulant) that arise from the $\overline{{\bf q}\,{\cal Q}[{\bf q}, {\bf q}]}$ terms in Eqn.~(\ref{EOM}).  The simplest and most computationally efficient approach is to truncate the hierarchy at second order by dropping the contribution of the third cumulant, $\overline{{\bf q}'(\vec{r}_1){\bf q}'(\vec{r}_2){\bf q}'(\vec{r}_3)}$, to the EOM for the second.  Termed CE2 in \cite{marstonconoveretal2008,marston10} and S3T in \cite{farrell07}, this truncation of the cumulant expansion (CE) at second order is both a conservative and realizable approximation \citep{salmon98}.  Furthermore, CE2 is quasilinear in the sense that it includes interactions of mean quantities with eddies to give eddies and interactions of eddies with eddies to give mean flows, but neglects eddy-eddy scattering.  Equivalently, it is readily shown that by invoking a QL approximation of the instantaneous dynamics, 
the same closed equations for the first two cumulants are obtained \citep{farrell07,marstonconoveretal2008,srinivasanyoung2012}. 

While ineffective for homogeneous isotropic turbulence, the quasilinear approximation has been shown to qualitatively (and in some cases, quantitatively) reproduce mean flows and two-point correlation functions for model problems describing a wide range of anisotropic and heterogeneous flows.  Specifically, QL has been successfully utilized to describe the driving of mean flows in plasmas and on giant planets \citep{tobias11}, the sustenance of wall-bounded shear-flow turbulence \citep{thomasetal2015}, the growth of a dry atmospheric convective boundary layer \citep{Chaalal:2016jx}, and even the development of the magnetorotational instability in accretion discs \citep{Squire:2015fk}.  The QL approximation is thought to be suitable to planetary flows because of the relative importance of interactions between small-scale
eddies and mean-field flow features \citep{ogorman07}.  Recently, the suitability of this approximation for planetary boundary-layer
applications has been supported by demonstrations of the emergence of zonal jets on a randomly or stochastically forced $\beta$-plane \citep{srinivasan12, tobias13} and the statistics of thermal instabilities in the atmospheric boundary layer \citep{Chaalal:2016jx} 
under QL dynamics or statistically equivalent frameworks.  Furthermore, the QL approximation has substantial capacity for improvement through
variations on the exact statistical closure of the cumulant expansion of QL systems (such as CE2.5 and CE3 \citep{marston14}) and extensions such as the generalized quasilinear approximation (GQL) \citep{marston16}.  

The quasilinear approximation has a number of mathematically attractive attributes.
\begin{itemize}
\item QL can be shown to be exact in appropriate asymptotic limits (e.g. for strong mean flows or for an extreme separation of time scales)~\citep{bouchetnardinietal2013} and, in regimes away from the strict asymptotic limit, errors can be controlled and reduced by methodically retaining higher-order cumulants.
\item QL retains the quadratic conservation properties of the underlying master PDEs and is realizable, implying a non-negative probability distribution.
\item The low-order statistics found with QL vary more smoothly than do instantaneous flow-field realizations, suggesting the possibility of dimension reduction in numerical simulations \citep{allawala17}.
\item The attractor for the low-order statistical dynamical evolution may be much simpler (e.g. an equilibrium point or limit cycle) than that characterizing the instantaneous (generally turbulent) dynamics, and the temporal evolution of the low-order statistics is comparably stiff, implying that these simple attractors should be mored readily accessible. 
\end{itemize}
For all its virtues, the quasilinear approximation nevertheless suffers from two primary deficiencies.
\begin{itemize}
\item Insufficient accuracy:  QL is deficient as the system is driven harder, reducing time-scale separation between the strict mean flow and the faster dynamics. This limitation was demonstrated in \cite{tobias13}, which compared the statistics derived from DNS for the problem of driving $\beta$-plane jets.  QL reproduced both the number and strength of jets for large time-scale separation but failed as the system was driven further away from equilibrium.  In practice, attempts to correct this deficiency by retaining higher-order cumulants are fraught with difficulty (e.g. the non-realizability of CE3).
More generally, higher-order cumulant expansions rapidly become analytically and computationally unwieldy.
\item Computational expense:  Although infrequently discussed in the research literature, naive implementations of QL/CE2 schemes actually can be as or even more expensive than full DNS because the second cumulant may have higher dimensionality than the underlying fields (depending upon the symmetries of the problem and the choice of averaging operation).  Extending the level of truncation by one order requires the calculation of correlations with an extra point in space, increasing the degrees of freedom of the system by $D$ where $D$ is the dimensionality of the system.  Consequently, in practice, attempts to increase the accuracy of CE2 by retaining higher-order cumulants are computationally prohibitive.
\end{itemize}

We show below that both of these deficiencies can be addressed simultaneously by  dimensional reduction of the boundary-layer models using Proper Orthogonal Decomposition (POD). POD, a method closely related to Empirical Orthogonal Functions (EOF) and Principal Component Analysis (PCA), is a common means to generate a reduced basis for applications in computational fluid dynamics in which the representation of a flow is decomposed into a set of modes based on the amount of variation in the flow that they describe in terms of energy \citep{holmes12}. Often, POD reduced models will retain only the most energetic of these modes. When the QL approximation is defined with respect to a spatial average over statistically homogeneous dimensions, it eliminates the direct flow of energy between scales in those dimensions; as a result, energy tends to populate fewer modes than with fully nonlinear (NL) dynamics. The use of energy optimization is a particularly suitable pairing for the QL approximation then because fewer modes must be retained to capture a similar portion of the variation in the flow than with NL dynamics.  

Another motivation for combining POD with QL when defined with a spatial average over a homogeneous dimension is that the statistics are invariant under a phase shift of individual Fourier modes in that dimension, implying that the phases of different Fourier modes are uncorrelated and localized structures cannot be captured in the flow solutions (see section \ref{sec:theory} of this paper, also \cite{skitka19_thesis}).  Because POD modes must be horizontal Fourier modes due to translational symmetry in the ensemble of flow realizations (see appendix \ref{app:pod_fourier}), POD washes away localized coherent structures that could otherwise be represented with very few coherent modes.  (\citet{sirovich89} notes a $\mathcal{O}\left(10^3\right)$ - $\mathcal{O}\left(10^4\right)$ reduction in the POD dimensionality of Rayleigh-B\'{e}nard convection when rigid walls break the symmetry of an otherwise periodic case.) The resulting bases have no encoded phase correlations, making the application of POD to homogeneous dimensions suited to the application of quasilinear approximation to spatial averages. 

Recently, \cite{allawala17} have demonstrated that accurate statistics for 2D rotating geostrophic flow on a sphere (a zonally-homogeneous, anisotropic, stationary problem) can be retained at massive dimensional reduction with a POD reduced model of CE2, an exact statistical closure of QL. Additionally, NL POD reduced models often suffer from a buildup of energy because energy pathways to the myriad small scales must be omitted, similar to an LES-simulation near the cutoff length \citep{wang12}. In QL POD reduced models, small scales only exchange energy with with the mean flow along statistically homogeneous dimensions, thus there can only be an indirect buildup of energy at cutoff scales through the modification of the mean flow rather than a direct local buildup between scales. In light of this, it might be expected that QL POD reduced models would have better stability properties than NL POD reduced models. 

The particular problems used as test cases here, surface-driven
convection into stratification and Langmuir turbulence, are classic
problems in planetary boundary layers
\citep{foken200650,large94,mcwilliams97}.  They are of immediate
importance to global climate modeling in both the ocean
\citep{belcher2012global} and atmosphere
\citep{kang2009tropical}. From the oceanographic perspective, other
problems such as submesoscale mixed layer eddy restratification
\citep{fox-kemper08} or mixing by symmetric instability
\citep{bachman2017parameterization} are seemingly similar: inhabiting
the boundary layer and affecting the balance between mixing and
restratification. However, from the generalized symmetry perspective
submesoscale turbulence differs fundamentally, as it only exists in
the presence of horizontal density gradients, thus these problems have
horizontal homogeneity in only one direction.  The general approach
here applies to boundary layer turbulence that oceanographers call
``vertical mixing'' and atmospheric modelers call ``pencil models'', both of which are horizontally homogeneous in both directions and
heterogeneous in the vertical.

In one important way, the approach taken here resembles that of
\cite{mellor74}: it relies on a structured sequence of approximations
of increasingly higher moments.  However, Mellor-Yamada SGS
models \citep[and relatives for Langmuir turbulence,
  e.g.,][]{harcourt2013second} tend to emphasize moments that are
co-located in vertical position.  Doing so reduces the dimensionality
of the problem, but it prevents non-local effects from playing a
leading role in the scheme dynamics.  Non-local effects and relative
motions are fundamental to the statistics of turbulence, as frequently
demonstrated using structure functions and two-point correlations
\citep{kolmogorov1941dissipation,batchelor1952diffusion}.  However,
allowing a SGS scheme to depend on non-local statistics means 
that the dimensionality of the analysis increases.  For example, for
a horizontally-homogeneous, stationary problem, one-point statistics 
will depend only on vertical position, while two-point, equal-time 
statistics depend on two independent vertical positions and their horizontal separation vector. Thus, a non-local approach is 
potentially more powerful, but not obviously efficient.

This article first describes the theory underpinning the proposed
model framework in section \ref{sec:theory}, including that of POD
Reduced Modeling, the QL approximation, and how basic assumptions
appropriate for ocean surface boundary layers, specifically horizontal
homogeneity, can be used to make simplifications.  Section
\ref{sec:numerics} covers the numerical implementation of the this
framework, including an introduction to a custom-built reduced-model solver.
Section \ref{sec:systems} defines the two examples of ocean surface
boundary-layer turbulence that are used to test this technique.  The
results of the QL reduced-model runs are presented in section
\ref{sec:results} while discussion of the implications of these
results as well as promising areas of future research are reserved for
section \ref{sec:discussion}.


\section{Theoretical Framework}
\label{sec:theory}

In POD, the flow is decomposed using the eigenvectors of the
uncentered, equal-time covariance matrix, $\mathsf{R}$.  Let the flow
variables be written $\mathbf{q} = \left(q_1\left(t\right) \; , \; \;
\; \dots \; \; \; q_N\left(t\right)\right)^\top$, where the index runs
over discretized spatial coordinates as well as fields (e.g. velocity,
temperature), $t$ is the undiscretized time, and $N$ is the total
number of flow variables.  
Returning to the Reynolds decomposition introduced in section \ref{sec:intro}),
$\mathbf{q} = \overline{\mathbf{q}} + \mathbf{q}'$ where
$\overline{\mathbf{q}}$ indicates the mean fields under a generic
averaging procedure and $\mathbf{q}'$ indicates the fluctuating
component of the fields, then
\begin{equation}
\mathsf{R}_{ij} = \overline{q_i q_j} \; \textrm{;} \; \mathsf{R} = \frac{1}{M}\sum_{1\le s\le M} \boldsymbol{q_s} \boldsymbol{q_s}^{\top}
\end{equation}
where the indices $i$ and $j$ indicate both field and position.  $s$
indicates the field snapshot, of which there are $M$; these are
sampled from different times and/or ensemble members.  The
eigendecomposition of $\mathsf{R}$ is:
\begin{equation}
\mathsf{R} \boldsymbol{\Phi}^\alpha = \lambda^\alpha \boldsymbol{\Phi}^\alpha \label{eq:pod_eig}
\end{equation}
A truncated basis retains the most energetic modes,
$\boldsymbol{\Phi}^\alpha$, as indicated by the associated
eigenvalues, $\lambda^\alpha$.  The naked $\phi^\alpha$ will be reserved for the
amplitudes of a flow field in the POD representation, whereas
$\boldsymbol{\Phi}^\alpha$ can be thought of as the unit vector of a
basis for flow states,
\begin{equation*}
q_i = \sum_{1\le \alpha \le N} \phi^{\alpha}\boldsymbol{\Phi}_i^\alpha,
\end{equation*}  
for some coefficients $\phi^\alpha$, where $N \le M$ is the number of
independent modes spanned by the $M$ snapshots. 

Because the cases of SGS turbulence are intended to be small in
horizontal extent, ignoring horizontal inhomogeneities in the flow
forcing is a reasonable approximation that will be important for
allowing the POD reduced model to be performed efficiently on a large,
three-dimensional problem.  For horizontally homogeneous cases, which
will be used exclusively in this work, covariances may be constrained
to depend on only the separation vector between two flow variables, 
 and it follows that POD
modes are constrained to be horizontal Fourier modes (see appendix
\ref{app:pod_fourier} for a derivation of this.)  This allows for the
POD eigendecomposition (equation \ref{eq:pod_eig}) to be performed
individually for each horizontal Fourier mode. To apply this to
discrete flow fields as they are represented in a solver, the fields
are written in terms of their horizontal Fourier wavenumber vector
index, $\mathbf{m}_i = \left\{m_i,l_i\right\}$ and their vertical
position and field index, $z_i$.  Here $m_i$ and $l_i$ are the
horizontal, $x$ and $y$ directed wavenumber indices.  Individual POD
modes from equation \ref{eq:pod_eig} now need only be labeled with
their horizontal wavenumber and an additional index, $k_\alpha$,
corresponding to the degrees of freedom associated with the vertical
direction and the different fields: $\boldsymbol{\Phi}^\alpha
= \boldsymbol{\Phi}_{\mathbf{m}_\alpha,k_\alpha}$ and may be defined by a
discrete vertical profile:
\begin{equation}
\boldsymbol{\Phi^\alpha} = \left(\Phi_{\mathbf{m}_\alpha,k_\alpha,z_1} \; , \; \; \; \dots \; \; \; , \; \; \; \Phi_{\mathbf{m}_\alpha,k_\alpha,z_{N_z}} \right)^\top \label{eq:pod_fourier_mode}
\end{equation}

Let the evolution of field variables, $q$, be written in terms of
generic constant, $\mathsf{C}$, linear, $\mathsf{L}$, and quadratic,
$\mathsf{Q}$, operators in the horizontal Fourier basis and vertical
position basis.
\begin{align}
\partial_t q_{\mathbf{m}_i,z_i}\left(t\right) &= \mathsf{Q}_{\mathbf{m}_i,z_i;\mathbf{m}_j,z_j;\mathbf{m}_l,z_l} q_{\mathbf{m}_j,z_j}\left(t\right) q_{\mathbf{m}_l,z_l}\left(t\right) \nonumber \\ 
& \quad + \mathsf{L}_{\mathbf{m}_i,z_i;\mathbf{m}_j,z_j} q_{\mathbf{m}_j,z_j}\left(t\right) + \mathsf{C}_{\mathbf{m}_i,z_i}  \label{eq:full_eoms}
\end{align}
where $t$ is the undiscretized time and repeated indices imply a sum.
Hereafter, $t$ will be dropped and all field variables will be assumed
to have an implicit time coordinate and for the remainder of this
section, $z$ will run over vertical positions and fields.  Note that
many elements in the implied summations will be exactly zero, allowing
for the equations to be constrained in terms of the allowable
horizontal wavenumbers in subsequent equations. 

As an example of the generic operator decomposition in equation
\ref{eq:full_eoms}, consider a basic temperature advection equation,
$\partial_t T + \left(\mathbf{u} \cdot \boldsymbol{\nabla}\right) T =
F_T + \kappa \nabla^2 T$, where $T$ is the temperature, $\mathbf{u}$
is the 3-dimensional velocity field, $F_T$ accounts for external
forcing, and $\kappa$ is the thermal diffusivity of a fluid. Here,
$\mathsf{Q}$ would be a tensor strictly defined by the discretized
advective term, $\mathsf{L}$ would be defined by the diffusive term,
and $\mathsf{C}$ would be defined by the external forcing.
In fluids, the continuum equations that govern
fully resolved flows do not have terms that are higher than second-order
in the field variables, so this generalized notation is an accurate
starting point for developing a SGS model.

The quasilinear approximation 
neglects the scattering of eddy fields, $q' =
q_{\mathbf{m}\ne\mathbf{0}}$, into other eddy fields,
thereby eliminating the scale-by-scale transfer of energy between horizontal length
scales; rather, all energy transfer to or from a given length-scale of
turbulence must be with the mean field, $\overline{q} =
q_{\mathbf{m}=\mathbf{0}}$.  This omits the scale-by-scale cascade of turbulent
energy and enstrophy.  Assuming the
evolution coefficients $\mathsf{Q}$, $\mathsf{L}$, and $\mathsf{C}$
are also horizontally homogeneous, the quadratic and linear operators
are constrained to just the interacting ``triads'' of the quadratic term interactions within vertical water columns:
\begin{align}
\mathsf{Q}_{\mathbf{m}_i,z_i;\mathbf{m}_j,z_j;\mathbf{m}_l,z_l} &= \tilde{\delta}_{\mathbf{m}_i - \mathbf{m}_j - \mathbf{m}_l}\mathsf{Q}_{\mathbf{m}_i,z_i;\mathbf{m}_j,z_j;\mathbf{m}_l,z_l} \\
\mathsf{L}_{\mathbf{m}_i,z_i;\mathbf{m}_j,z_j} &= \tilde{\delta}_{\mathbf{m}_i - \mathbf{m}_j} \mathsf{L}_{\mathbf{m}_i,z_i;\mathbf{m}_j,z_j} 
\end{align}
where $\tilde{\delta}_\mathbf{m}$ is a modified Kronecker delta:
\begin{equation}
\tilde{\delta}_{\mathbf{m}} = \delta_{\textrm{MOD}\left[\mathbf{m}_i\right],\mathbf{0}} = \delta_{ m \; mod \; N_x,0} \delta_{ l \; mod \; N_y,0}
\end{equation}
Then the subset of the coefficients not constrained to be zero within
this context may be written 
\begin{align}
\mathsf{Q}_{\mathbf{m}_i,\mathbf{m}_j,z_i,z_j,z_l} &= \mathsf{Q}_{\mathbf{m}_i,z_i;\mathbf{m}_j,z_j;\mathbf{m}_i-\mathbf{m}_j,z_l} \\
\mathsf{L}_{\mathbf{m}_i,z_i,z_j} & = \mathsf{L}_{\mathbf{m}_i,z_i;\mathbf{m}_i,z_j} 
\end{align}
Note that the wavenumber indices of the coefficients
$\mathsf{Q}_{\mathbf{m}_i,\mathbf{m}_j,\dots}$ and
$\mathsf{L}_{\mathbf{m}_i,\dots}$ implicitly have a
$\textrm{MOD}\left[\mathbf{m}_i\right]$ acting on them. Then, using the convention that sums over a basis index are implied when the associated vertical index is repeated (e.g. sum over $j$ for repeated $z_j$) and that all mean-field Fourier modes
(including differences such as $\mathbf{m}_i-\mathbf{m}_j$) are
explicitly written as $\mathbf{0}$, the full-basis quasilinear
equations are:
\begin{align}
& \partial_t q \left(\mathbf{0},z_i\right) = \nonumber \\
& \quad \mathsf{Q}_{\mathbf{0},\mathbf{m}_j,z_i,z_j,z_l} q\left(\mathbf{m}_j,z_j\right)q\left(-\mathbf{m}_j,z_l\right) \nonumber \\
& \quad + \mathsf{Q}_{\mathbf{0},\mathbf{0},z_i,z_j,z_l} q\left(\mathbf{0},z_j\right) q\left(\mathbf{0},z_l\right) \nonumber \\
& \quad + \mathsf{L}_{\mathbf{0},z_i,z_j}q\left(\mathbf{0},z_j\right) + \mathsf{C}_{\mathbf{0},z_i} \label{eq:fourier_ql_eom_1} \\
& \partial_t q \left(\mathbf{m}_i,z_i\right) = \nonumber \\
& \quad \hcancel[gray]{\mathsf{Q}_{\mathbf{m}_i,\mathbf{m}_j,z_i,z_j,z_l} q\left(\mathbf{m}_j,z_j\right)q\left(\mathbf{m}_i-\mathbf{m}_j,z_l\right)} \nonumber \\
& \quad + \mathsf{Q}_{\mathbf{m}_i,\mathbf{m}_i,z_i,z_j,z_l} q\left(\mathbf{m}_i,z_j\right) q\left(\mathbf{0},z_l\right) \nonumber \\
& \quad + \mathsf{Q}_{\mathbf{m}_i,\mathbf{0},z_i,z_j,z_l} q\left(\mathbf{0},z_j\right)q\left(\mathbf{m}_i,z_l\right) \nonumber \\
& \quad + \mathsf{L}_{\mathbf{m}_i,z_i,z_j}q\left(\mathbf{m}_i,z_j\right) + \mathsf{C}_{\mathbf{m}_i,z_i} \label{eq:fourier_ql_eom_2}
\end{align}
where the omitted nonlinear term is indicated by the strikethrough.  This expression of the quasilinear approximation in terms of Fourier modes more readily reveals a key feature and, depending on the circumstance, shortcoming of using spatial averages to define QL.  Assume that $\mathsf{C}_{\mathbf{m}_i,z_i} = 0$, as is true for all cases studied in this paper.  Noting that all terms of the HQL eddy-field evolution (equation \ref{eq:fourier_ql_eom_2}) are linear in $q'\left(\mathbf{m}_n\right)$ and all terms in the HQL mean-field evolution (equation \ref{eq:fourier_ql_eom_1}) are constant or bilinear in $q'\left(\mathbf{m}_n\right)$, the HQL equations are seen to be invariant under an independent phase shift \citep{pausch2019} for each set of all eddy modes of a specific horizontal wavenumber, $\mathbf{m}_n$.  It follows that the relative phases of horizontal Fourier modes in the eddy field must only depend on the initial conditions; if these initial conditions are randomly sampled, the phases of different Fourier modes, being blind to one another, will be uncorrelated, and it is unlikely that localized structures will appear; rather, flow features in the horizontal direction (or, by extension, any homogeneous dimension of averaging used for QL) will be strictly wavelike and nonlocal, something that will manifest visually distinctly in HQL flow fields observed in the results (section \ref{sec:results}).

A reduced model may be derived by projecting the dynamical equations
of motion (equation \ref{eq:full_eoms}) onto a truncated reduced
basis.  Let the POD field amplitude be written:
 \begin{equation}
 \phi_\alpha = \phi_{\mathbf{m}_\alpha,k_\alpha} = \sum_{j} q_{\mathbf{m}_\alpha,z_j} \Phi_{\mathbf{m}_\alpha,k_\alpha,z_j}^*
 \end{equation}
then the Galerkin projection of the equations of
motion onto this basis is
\begin{align}
\partial_t \phi_{\mathbf{m}_\alpha,k_\alpha} &= \nonumber \\ 
& \boldsymbol{\Phi}_{\mathbf{m}_\alpha,k_\alpha,z_i}^*\mathsf{Q}_{\mathbf{m}_\alpha,\mathbf{m}_\sigma,z_i,z_j,z_l} \boldsymbol{\Phi}_{\mathbf{m}_\sigma,k_\sigma,z_j} \nonumber \\ 
& \qquad \qquad \qquad \boldsymbol{\Phi}_{\mathbf{m}_\alpha - \mathbf{m}_\sigma,k_\beta,z_l} \phi_{\mathbf{m}_\sigma,k_\sigma} \phi_{\mathbf{m}_\alpha - \mathbf{m}_\sigma,k_\beta} \nonumber \\
& + \boldsymbol{\Phi}_{\mathbf{m}_\alpha,k_\alpha,z_i}^* \mathsf{L}_{\mathbf{m}_\alpha,z_i,z_j} \boldsymbol{\Phi}_{\mathbf{m}_\alpha,k_\sigma,z_j} \phi_{\mathbf{m}_\alpha,k_\sigma} \nonumber \\
& + \boldsymbol{\Phi}_{\mathbf{m}_\alpha,k_\alpha,z_i}^* \mathsf{C}_{\mathbf{m}_\alpha,z_i} \label{eq:coeff_deriv_1}
\end{align}
where summations are now implied when either vertical
index is repeated within a term (e.g., sum over $j$ for repeated
$k_j$ or $z_j$).  By defining the truncated evolution coefficients as
\begin{align}
& \tilde{\mathsf{Q}}_{\mathbf{m}_\alpha,\mathbf{m}_\sigma,k_\alpha,k_\sigma,k_\beta}  = \nonumber \\ 
& \qquad \boldsymbol{\Phi}_{\mathbf{m}_\alpha,k_\alpha,z_i}^* \mathsf{Q}_{\mathbf{m}_\alpha,\mathbf{m}_\sigma,z_i,z_j,z_l} \boldsymbol{\Phi}_{\mathbf{m}_\sigma,k_\sigma,z_j} \boldsymbol{\Phi}_{\mathbf{m}_\alpha-\mathbf{m}_\sigma,k_\beta,z_l} \label{eq:gen_quad_coeff}\\
&\tilde{\mathsf{L}}_{\mathbf{m}_\alpha,k_\alpha,k_\sigma}  = \boldsymbol{\Phi}_{\mathbf{m}_\alpha,k_\alpha,z_i}^* \mathsf{L}_{\mathbf{m}_\alpha,z_i,z_j} \boldsymbol{\Phi}_{\mathbf{m}_\alpha,k_\sigma,z_j} \label{eq:gen_lin_coeff}\\
&\tilde{\mathsf{C}}_{\mathbf{m}_\alpha,k_\alpha} = \boldsymbol{\Phi}_{\mathbf{m}_\alpha,k_\alpha,z_i}^* \mathsf{C}_{\mathbf{m}_\alpha,z_i} \label{eq:gen_cst_coeff}
\end{align}
the truncated evolution equations may be succinctly expressed as
\begin{align}
& \partial_t \phi_{\mathbf{0},k_\alpha} = \nonumber \\
& \quad \tilde{\mathsf{Q}}_{\mathbf{0},\mathbf{m}_\beta,k_\alpha,k_\beta,k_\gamma} \phi_{\mathbf{m}_\beta,k_\beta}\phi_{-\mathbf{m}_\beta,k_\gamma} \nonumber \\
& \quad + \tilde{\mathsf{Q}}_{\mathbf{0},\mathbf{0},k_\alpha,k_\beta,k_\gamma} \phi_{\mathbf{0},k_\beta} \phi_{\mathbf{0},k_\gamma} \nonumber \\
& \quad + \tilde{\mathsf{L}}_{\mathbf{0},k_\alpha,k_\beta}\phi_{\mathbf{0},k_\beta} + \tilde{\mathsf{C}}_{\mathbf{0},k_\alpha} \label{eq:fourier_red_eom_1} \\
& \partial_t \phi_{\mathbf{m}_\alpha,k_\alpha} = \nonumber \\
& \quad \hcancel[gray]{\tilde{\mathsf{Q}}_{\mathbf{m}_\alpha,\mathbf{m}_\beta,k_\alpha,k_\beta,k_\gamma} \phi_{\mathbf{m}_\beta,k_\beta}\phi_{\mathbf{m}_\alpha-\mathbf{m}_\beta,k_\gamma}} \nonumber \\
& \quad + \tilde{\mathsf{Q}}_{\mathbf{m}_\alpha,\mathbf{m}_\alpha,k_\alpha,k_\beta,k_\gamma} \phi_{\mathbf{m}_\alpha,k_\beta} \phi_{\mathbf{0},k_\gamma} \nonumber \\
& \quad + \tilde{\mathsf{Q}}_{\mathbf{m}_\alpha,\mathbf{0},k_\alpha,k_\beta,k_\gamma} \phi_{\mathbf{0},k_\beta}\phi_{\mathbf{m}_\alpha,k_\gamma} \nonumber \\
& \quad + \tilde{\mathsf{L}}_{\mathbf{m}_\alpha,k_\alpha,k_\beta}\phi_{\mathbf{m}_\alpha,k_\beta} + \tilde{\mathsf{C}}_{\mathbf{m}_\alpha,k_\alpha} \label{eq:fourier_red_eom_2}
\end{align}
The horizontal strikethrough indicates the terms omitted in the
reduced quasilinear approximation.


\section{Numerical Implementation}
\label{sec:numerics}

In order to compute the reduced basis for a specific, developing flow
case, an ensemble of 24 members 
was run using full-basis quasilinear (FQL) direct numerical
simulations.  Data spread evenly across all times in the flow
evolution and ensemble members is used to compute the POD modes.
Then, the energetically optimized truncated POD basis retaining
specific numbers of modes is used in a quasilinear reduced-order model
(RQL), a Galerkin Projection of the Boussinesq equations (BGP), as
described in equations \ref{eq:fourier_red_eom_1} and
\ref{eq:fourier_red_eom_2}.  For a sampling of truncated-basis sizes,
an ensemble of 12 RQL members are then compared against a reference
ensemble of 24 FQL members.

Because of the novelty of the modeling approach, this initial research
seeks a minimal and simple solver framework.  Harmonic diffusivity is
used.  Advection schemes are centered, second-order.  This caps the
maximum order of interactions at 2, simplifying quasilinearization in
all simulations.  Note that upwind advection schemes that use absolute
values of flow variables cannot be quasilinearized.

\subsection{Full-Basis Simulations}

Full-basis nonlinear (FNL) and quasilinear (FQL) simulations were run
on a modified version of the MITgcm \citep{mitgcm}, a finite-volume
atmospheric and oceanic fluid dynamics solver. Flow variables are
colocated in time and integrated using a second-order Adams-Bashforth
scheme. The surface boundary condition is a rigid lid.

\subsection{Reduced-Basis Determination}

\label{sec:basis}

In order to compute the POD modes from flow data, the flow fields must
be expressed in consistent units.  This is accomplished by means of the
available potential energy (APE) in the flow
\begin{align}
PE =& - \sum_\mathbf{r} g \alpha T\left(\mathbf{r}\right) z \; \Delta V \\
APE =& PE - PE_{min} \label{eq:ape}
\end{align}
where $g$ is the acceleration of gravity at Earth's surface, $\alpha$
is the coefficient of linear thermal expansion of sea water, $\Delta V$ is
the volume of each grid cell, $z<0$ is the fluid depth, and $PE_{min}$
is the potential energy of the temperatures sorted with lower
temperatures at greater depths. This method performed as well or
better than an alternative approach in which the temperature was
converted to a velocity such that the variance of the scaled
temperature was equal to the mean of the variances of the velocity
components. 

\begin{figure}[]
\centering
\includegraphics[width=0.48 \textwidth]{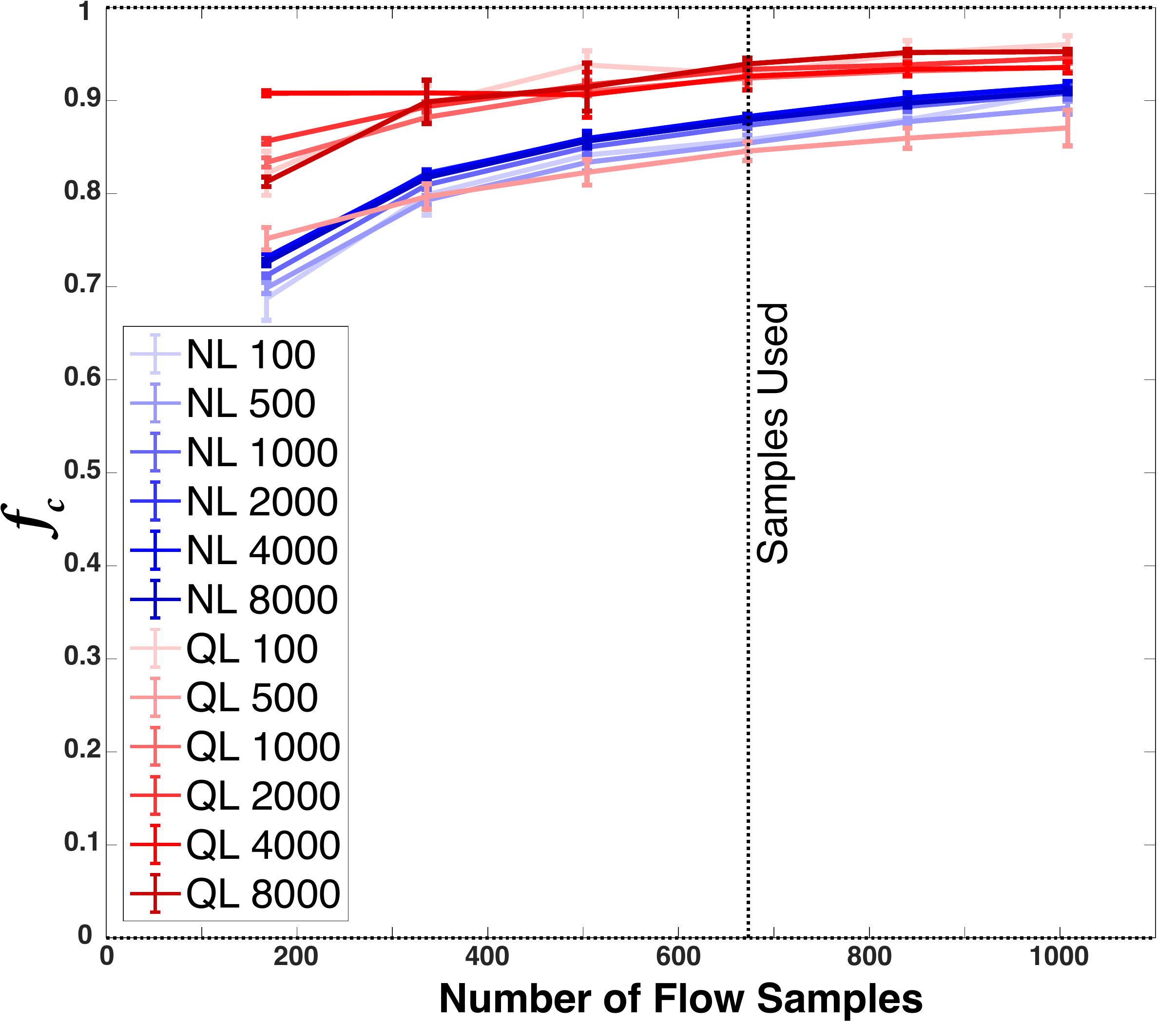}
\caption{\small Basis convergence factor, $f_c$, for the
  thermal-convection case (defined in section
  \ref{sec:systems}\ref{sec:thermal_system}) with a linear free
  surface as a function of the number of flow samples used.  See
  \eqref{eq:basis_convergence_factor} for the definition of $f_c$.  }
\label{fig:thermal_spread_basis_convergence}
\end{figure}

\begin{figure}[h]
\centering
\includegraphics[width=0.48 \textwidth]{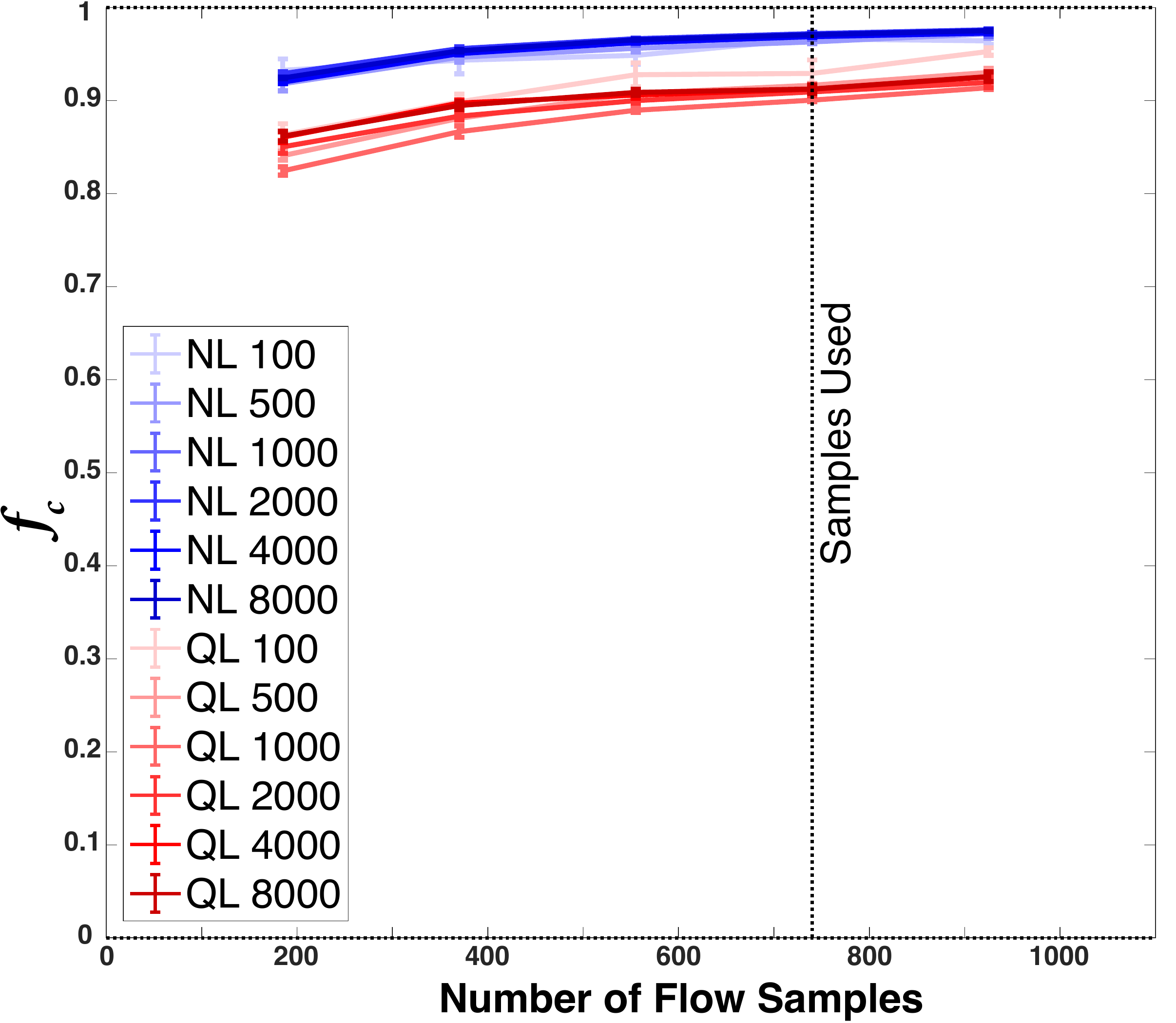}
\caption{\small Basis convergence factore, $f_c$, for the
  Langmuir-turbulence case (defined in section
  \ref{sec:systems}\ref{sec:langmuir_system}) with a linear free
  surface as a function of the number of flow samples used.  See
  \eqref{eq:basis_convergence_factor} for the definition of $f_c$.  }
\label{fig:langmuir_spread_basis_convergence}
\end{figure}

The POD modes are computed separately for each independent horizontal
Fourier mode using the method of snapshots (MOS) \citep{holmes12}.
Because the MITgcm's pressure solver does not yield a
divergenceless flow to numerical precision, flow snapshots are
first projected onto a divergenceless basis to entirely eliminate the
divergent modes from the decomposition.  Enough flow samples are used
to ensure at least 85\% of the truncated basis subspace is
statistically converged (see figures
\ref{fig:thermal_spread_basis_convergence} and
\ref{fig:langmuir_spread_basis_convergence}). Formally, the
convergence is computed using a convergence factor, $f_c$, defined as
the normalized inner product of the subspaces spanned by members of an
ensemble of bases:
\begin{equation}
f_c = \frac{1}{N_b(N_b-1)N_r} \sum_{\alpha \ne \beta} \sum_{i,j} \left| \boldsymbol{\Phi}^\alpha_i \cdot \boldsymbol{\Phi}^\beta_j \right|^2  \label{eq:basis_convergence_factor}
\end{equation}
\begin{equation}
\boldsymbol{\Phi}^\alpha_i \cdot \boldsymbol{\Phi}^\beta_j = \tilde{\delta}_{\mathbf{m}_i-\mathbf{m}_j} \sum_z \Phi^\alpha_{i,z} \left(\Phi^\beta_{j,z}\right)^*
\end{equation}
where $\boldsymbol{\Phi}^\alpha_i$ is the $i$th element of
basis $\alpha$, $N_r$ is the number of elements in the reduced basis,
and $N_b$ is the number of bases in the basis ensemble. 

The cases studied here all involve developing turbulence and momentum
as well as buoyancy entrainment in the oceanic mixed-layer.  Because
the statistics are not converged, samples of flow states at  21 or 37
(depending on the case) evenly spaced times are used to determine the
bases.  Because of this, the bases retain memory of the different
states that the flow may manifest as it develops and excludes states
that should not be part of the flow solution at any time.  It is
useful to think of the basis as encoding part of the dynamics.

\subsection{Reduced-Model Implementation} 

This section serves as an introduction to the piece of software,
Boussinesq Galerkin Projection (BGP), the reduced order model which is
used in the present study and intended for use in subsequent
publications. BGP is a PDE solver that discretizes symbolic equations
of motion up to quadratic order, projects them onto a provided
truncated basis, and integrates the resulting reduced equations
explicitly forward in discrete time.  The discretization is written in
such a way that, when provided with a complete basis (on a necessarily
very small problem), BGP will be in agreement with the MITgcm up to
the precision of the latter's pressure solver. The provided basis must
be divergence-free, which allows for the reduced model to run without a pressure
solver, and defined in Fourier space, which allows for easy
quasilinearization of the quadratic term.

\begin{figure}[]
\centering
\includegraphics[width=0.48 \textwidth]{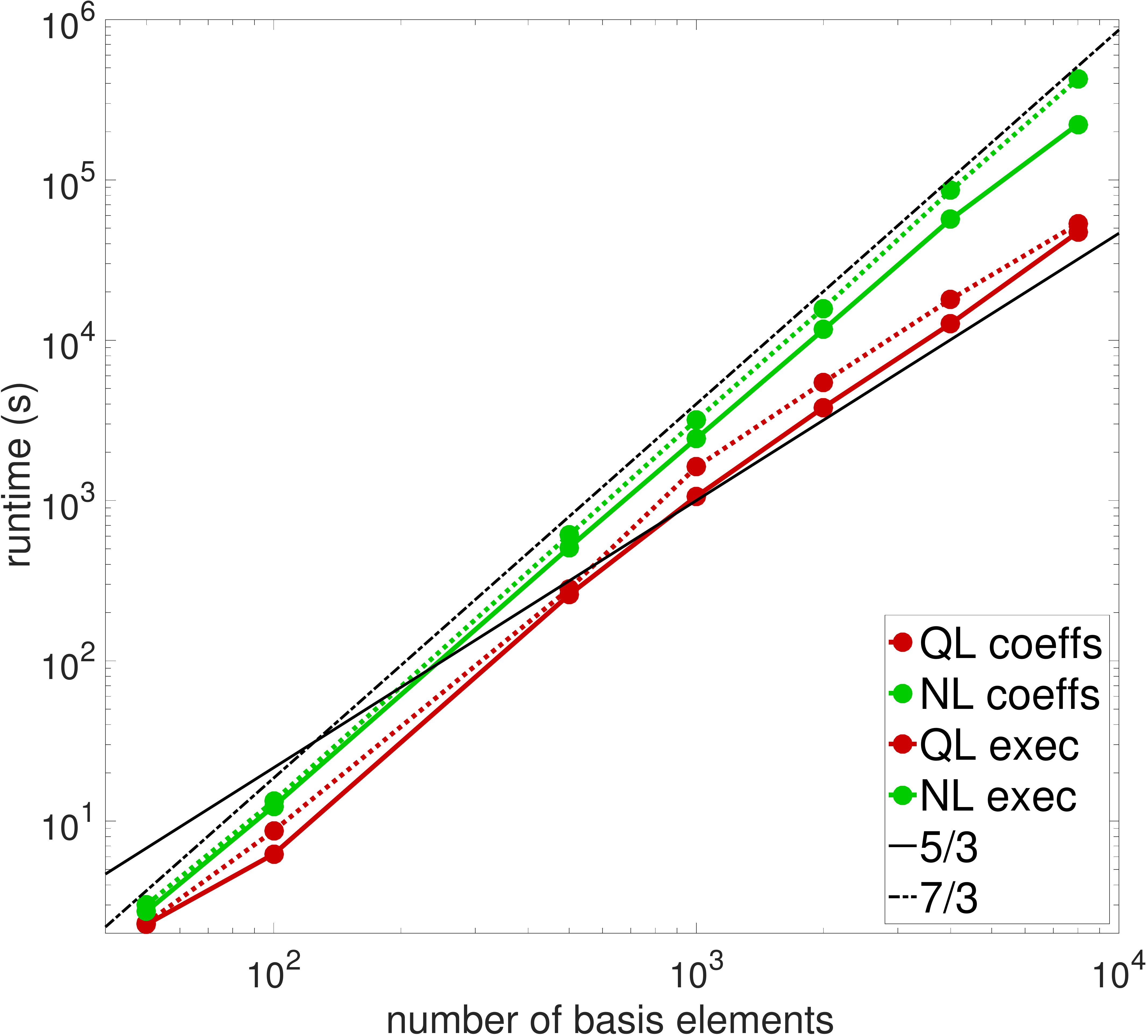}
\caption{\small Runtime scalings of the reduced model (BGP) using the
  Langmuir-turbulence case for both the initial coefficient
  determination (as in equations \ref{eq:gen_quad_coeff} through
  \ref{eq:gen_cst_coeff}) and the simulation execution.  The solid and
  dash-dot black lines indicate the power law scalings that are
  predicted for the RQL and RNL.}
\label{fig:langmuir_scaling}
\end{figure}

The computational complexity of BGP using a truncated basis with
horizontal Fourier modes (POD) may be estimated for both the model
execution and the coefficient computation (which must be computed for
every unique basis and set of flow parameters) for the RNL and RQL
solvers.  Scaling parameters will be the number of retained modes,
$N_r$, the number of vertical discretizations and 3D flow fields in
the underlying full-basis model, $N_{zf}$, and the number of points in
the stencil of the discretization of the quadratic term of the
underlying full-basis model, $N_s$. A key assumption that will be
utilized here is that the distribution function of the truncated flow
representation scales linearly with $N_r$ such that the number of
vertical modes for a given horizontal Fourier mode scales as
$N_r^{\frac{1}{3}}$.  If $N$ is the number of modes in the underlying
full-basis model, then this assumption will break down as $N_r
\rightarrow N$; however, this project is primarily concerned with
model reductions of $N_r \ll N$.  Even for $N_r \ll N$, the assumption
may still break down if the distribution function of the truncated
flow representation changes shape substantially with $N_r$ The
following scalings are verified using runtimes from BGP in figure
\ref{fig:langmuir_scaling}:

\begin{itemize}
\item the coefficient determination for RNL is $\mathcal{O}\left(N_r^{\frac{7}{3}} N_{zf} N_s^2\right)$
\item the coefficient determination for RQL is $\mathcal{O}\left(N_r^{\frac{5}{3}} N_{zf} N_s^2\right)$
\item the execution for RNL is $\mathcal{O}\left(N_r^{\frac{7}{3}}\right)$
\item the execution for RQL is $\mathcal{O}\left(N_r^{\frac{5}{3}}\right)$
\end{itemize}

\section{Turbulence Examples}
\label{sec:systems}

In order to investigate the feasibility and effectiveness of a
quasilinear reduced model for the purpose of modeling SGS
planetary turbulence, two submesoscale or smaller oceanographic flows
are used as case studies: surface-forced thermal convection and
Langmuir turbulence.  The former case is horizontally isotropic, up to
a grid-representation, while the latter is anisotropic.  Crucially,
both cases are formulated to be statistically horizontally
homogeneous, which will permit easy computation of the mean fields and
considerably simplify the POD computations and BGP execution.

The viscosity is $5000\times$ the seawater value, and the diffusivity
is set equal to the viscosity (rather than less by about a factor of
7).  Thus, the Kolmogorov dissipation scale is roughly $600\times$
larger for this simulation than it would be for seawater which permits
representing the flows with a manageable grid and number of model,
i.e., an ``eddy'' viscosity and diffusivity closure is used.  Noting
this, it should be acknowledged that the full-resolution simulations of
these systems that are run on the MITgcm are not, strictly speaking,
direct numerical simulations of seawater.  Boundary layer problems at this resolution are
more typically managed with Large-Eddy-Simulation (LES) closures
\citep[e.g.,][]{sullivan94}, but such an approach introduces
additional quadratic or higher order dependences and qualitatively
similar behavior to the eddy viscosity.  It is preferable, then, to
maintain the linear nature of the dissipation when modeling SGS
turbulent dissipation in QL dynamics. 

The FNL simulation may be considered to be a type of LES
despite the simple linear closure used here because the viscosity
and diffusivity are much larger than the seawater values and the
Prandtl number is one. LES closures generally attempt to dissipate
small-scale energy while preserving the energy cascade and other
dynamics present above the grid scale, when compared with a higher
resolution simulation of the same flow. With QL dynamics, however,
there is no energy cascade, so there can be no direct energy
backscatter; rather, all energy flows through the mean field, and as
long as the flow at the dissipation scales does not affect the
mean-field substantially, there will not be a substantial impact at
intermediate scales.  Therefore, the QL flow solution is not sensitive
to the relative amount of dissipation among different scales provided
the net amount of dissipation is approximately preserved. There is
also some flexibility in the total rate of turbulent energy
dissipation because these cases are not in a steady state and the
tendency of the horizontal mean-field is expected to balance much of
the large-scale forcing.  Roughly speaking, nonlinear LES closures
tend to dissipate energy more weakly at a given scale or at smaller
scales in comparison to harmonic viscosities and diffusivities, or to
be flow-aware selective as to where and when large effects are
needed.  Here, in the FNL case, relatively weak coefficients of
harmonic dissipation are used (gridscale Reynolds and P\'eclet numbers
are $\mathcal{O}(8)$).  It should be stressed, however, that the
primary task here is not having an exhaustively accurate FNL
simulation, but in comparing the much larger differences between the
QL and NL runs and between the reduced-basis and full-basis
runs--particularly with QL runs. At this level of precision, the
differences between the modeling systems are much larger than
differences that result from changing subgrid schemes or parameter
values in the FNL run.

\subsection{Thermal Convection}
\label{sec:thermal_system}

\begin{figure}[]
\centering
\includegraphics[width=0.48 \textwidth]{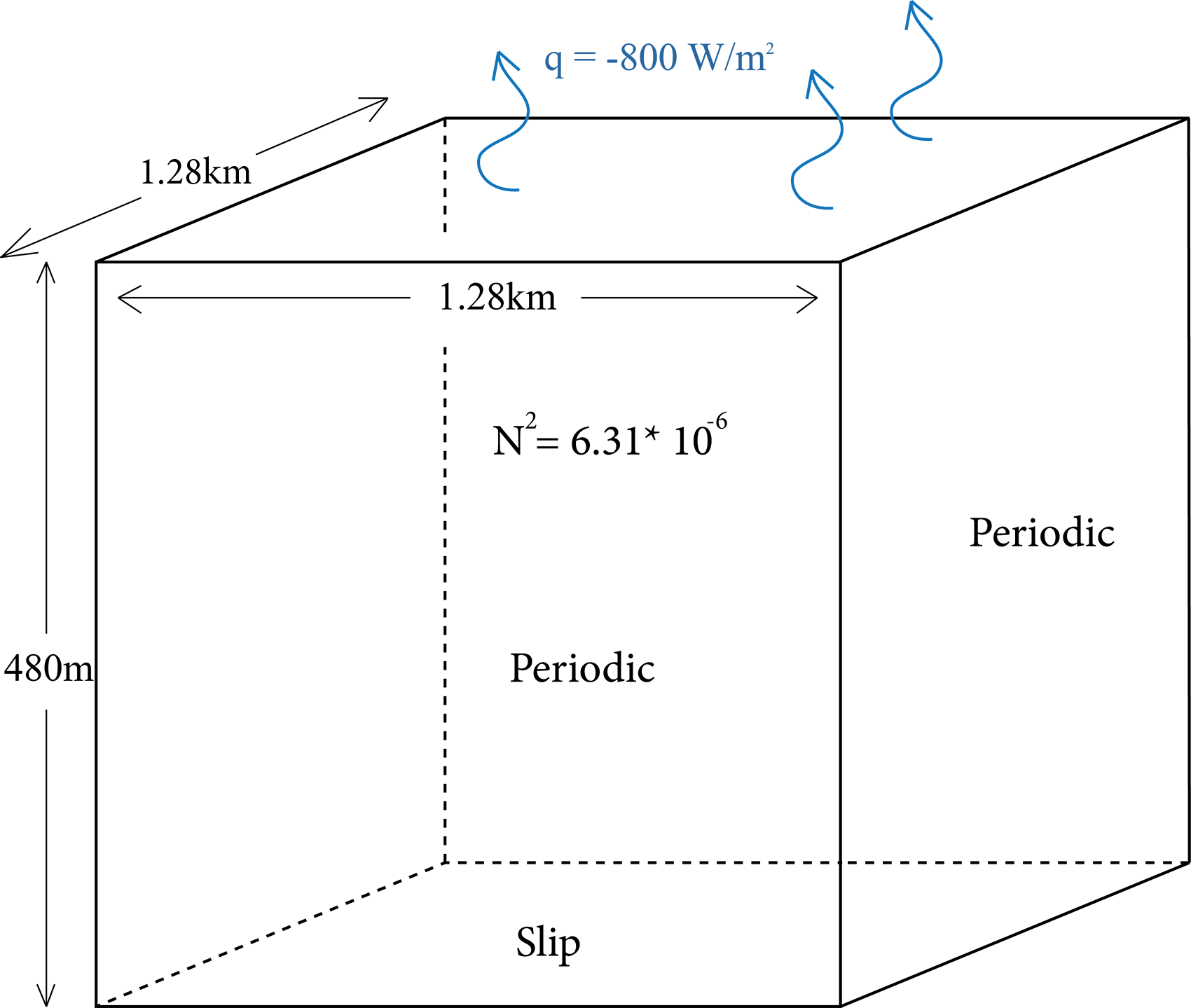}
\caption{\small thermal-convection case}
\label{fig:thermal_diagram}
\end{figure}

The surface-forced-thermal-convection case (figure
\ref{fig:thermal_diagram}) is formulated with an initial uniform
stable stratification that is cooled from the surface
\citep{Chaalal:2016jx}.  This causes the mixed layer to deepen.  The
nonhydrostatic Boussinesq equations that govern this case are
\begin{align}
\boldsymbol{\nabla} \cdot \mathbf{u} &= 0 \\ \partial_t \mathbf{u} +
\left(\mathbf{u} \cdot \boldsymbol{\nabla}\right)\mathbf{u} +
\mathbf{f} \times \mathbf{u} &= - \frac{1}{\rho_0} \boldsymbol{\nabla}
p + \mathbf{g} \alpha \left(T - T_0\right) \nonumber\\
& \qquad \qquad \qquad + \nu \nabla^2 \mathbf{u}
\\ \partial_t T + \left(\mathbf{u} \cdot \boldsymbol{\nabla}\right)T
&= \frac{Q}{c_p \rho_0} + \kappa \nabla^2 T
\end{align}
where $\mathbf{u} = \left\{u,v,w\right\}$ is the 3-dimensional
velocity vector, $T$ is the temperature field, $t$ is time,
$\mathbf{f} = f \hat{\mathbf{z}}$ is the Coriolis parameter, $\rho_0$
is the background density, $p$ is the pressure, $\mathbf{g} = g
\hat{\mathbf{z}}$ is the local acceleration of gravity, $\alpha$ is
the coefficient of thermal expansion, $T_0$ is the background
temperature, $\nu$ is the harmonic kinematic viscosity, $\kappa$ is
the coefficient of thermal diffusivity, $Q$ is the volumetric heating,
and $c_p$ is the specific heat of sea water at constant pressure.
This case is defined on a regular, horizontally periodic Cartesian
grid with the number of cells and grid spacing per dimension
represented as $\left\{N_x,N_y,N_z\right\}$ and
$\left\{\Delta_x,\Delta_y,\Delta_z\right\}$, respectively. The domain
floor and surface have rigid, slip boundary conditions.  In the discrete model, $Q = \delta_{k,0}
\frac{q}{\Delta_z}$, where $k$ is the vertical layer index and $q$ is
the surface heat flux.

Introducing the following non-dimensional parameters
\begin{align}
Ra^* &= \frac{g \alpha q L_{m}^4}{\nu \kappa^2 \rho_0 c_p} \\
Pr &= \frac{\nu}{\kappa}
\end{align}
where $Ra^*$ is the modified Rayleigh Number, appropriate for a
prescribed heat flux as opposed to a prescribed temperature at the
boundaries, $Re$ is the Reynolds Number, $Pr$ is the Prandtl Number,
and $L_m$ is the mixed-layer depth.  Velocity and temperature scales
may be introduced as in \cite{schmidt89}, allowing for the definition
of the Reynolds Number, a convective Froude Number, and the
conventional Rayleigh Number based on a characteristic temperature
difference, $T_*$:
\begin{align}
w_* &= \sqrt[\leftroot{0}\uproot{6}\textrm{\small $3$}]{\frac{\alpha g q L_m}{c_p \rho_0}} \\
T_* &= \sqrt[\leftroot{0}\uproot{6}\textrm{\small $3$}]{\frac{q^2}{c_p^2 \rho_0^2 \alpha g L_m}} \\
Fr &= \frac{w_*}{N L_m} \\
Ra &= \frac{\alpha g T_* L_m^3}{\nu \kappa} \\
Re &= \frac{w_* L_{m}}{\nu} 
\end{align}
where $N = \sqrt{\alpha g \frac{\partial \overline{T}}{\partial z}}$
is the Brunt-V\"{a}is\"{a}l\"{a} frequency, and $\overline{T}$ is the
background temperature profile.  Using these definitions, the case
studied has the following parameters:
\begin{figure}[]
\centering
\includegraphics[width=0.48 \textwidth]{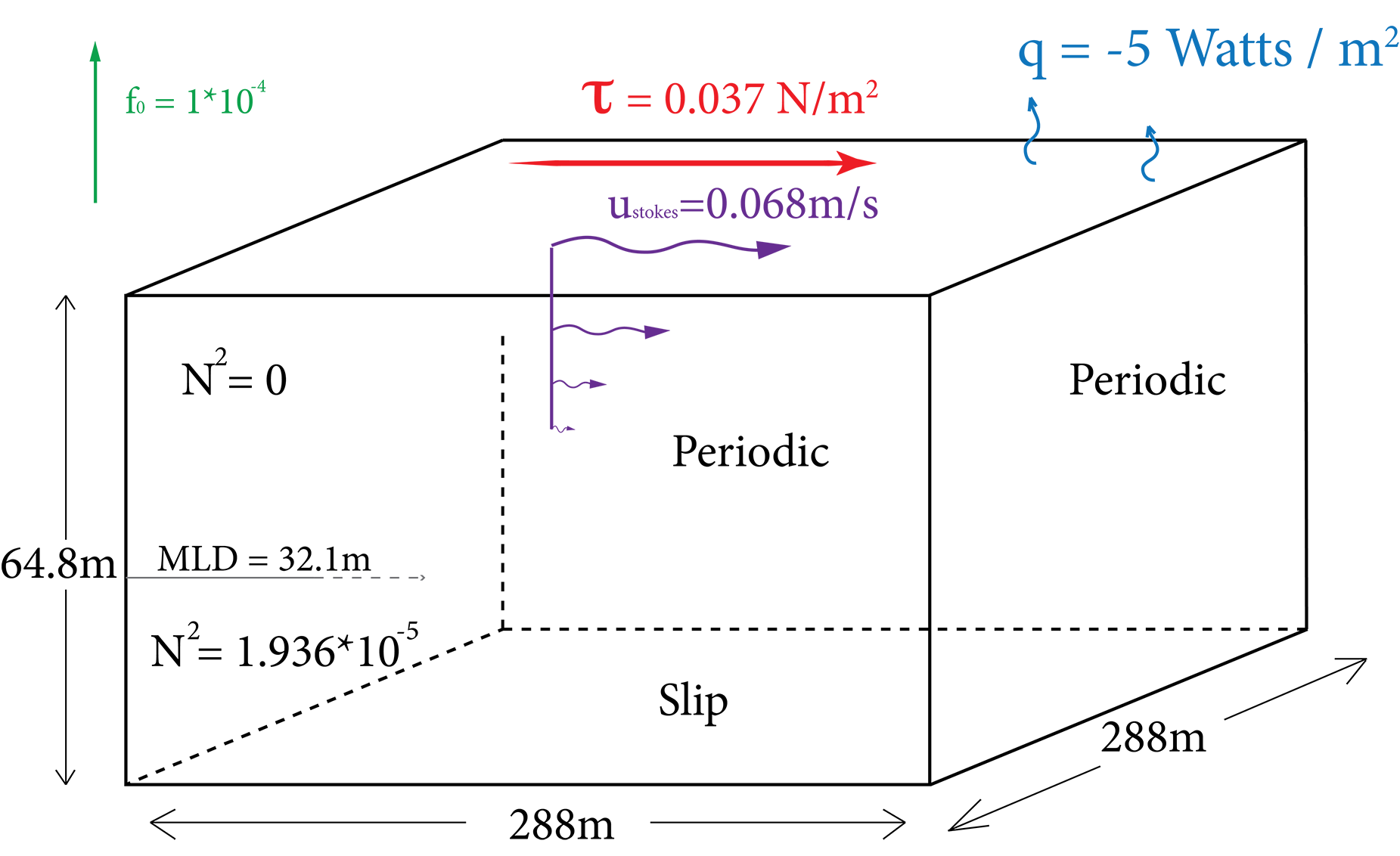}
\caption{\small Langmuir-turbulence case}
\label{fig:langmuir_diagram}
\end{figure}
\begin{alignat}{2}
&\left\{N_x,N_y,N_z\right\} &&= \left\{128,128,48\right\} \nonumber \\
&\left\{\Delta_x,\Delta_y,\Delta_z\right\} &&= \left\{10,10,10\right\} \textrm{m} \nonumber \\
&f &&= 0 \nonumber \\
&\rho_0 &&= 1035 \; \textrm{kg m}^{-3} \nonumber \\
&g &&= 9.81 \; \textrm{m s}^{-2} \nonumber \\
&\alpha &&= 2 \times 10^{-4} \; \textrm{C}^{-1} \nonumber \\
&T_0 &&= 20 \; \textrm{C} \nonumber \\
&\nu &&= 5 \times 10^{-2} \; \textrm{m}^2 \textrm{s}^{-1} \nonumber \\
&\kappa &&= 5 \times 10^{-2} \;\textrm{m}^2 \textrm{s}^{-1} \nonumber \\
&c_p &&= 3994 \; \textrm{J kg}^{-1} \textrm{C}^{-1} \nonumber \\
&q &&= -800 \; \textrm{W m}^{-2} \nonumber \\
&N^2 && = 5.45 \times 10^{-7} \; \textrm{s}^{-2} \nonumber \\
&w_* &&= 5.67 \times 10^{-2} \; \textrm{m s}^{-1} \nonumber \\
&T_* &&= 3.41 \times 10^{-2} \; \textrm{C} \nonumber \\
&Ra^* &&= 1.61 \times 10^8 \nonumber \\
&Ra &&= 2.96 \times 10^5 \nonumber \\
&Pr &&= 1 \nonumber \\
&Fr &&= 0.160 \nonumber \\
&Re &&= 544 \nonumber
\end{alignat}
Here, $N$ is the Brunt-V\"{a}is\"{a}l\"{a} frequency.  Using these
parameters, the mixed layer depth, initially zero, will reach the
bottom of the domain at approximately $t = 40$ hours, at which point
the execution is stopped.  

\subsection{Langmuir Turbulence}
\label{sec:langmuir_system}

The Langmuir-turbulence forcing is based on \cite{mcwilliams97}.  The
domain size and resolution have been reduced in this work, while most
other case parameters are identical (see figure
\ref{fig:langmuir_diagram}).  Langmuir circulation is modeled using
the phase-averaged Craik-Leibovich equations \citep{craik77}.  The
nonhydrostatic equations governing this case are
\begin{align}
&\boldsymbol{\nabla} \cdot \mathbf{u} = 0 \\
&\partial_t \mathbf{u} + \left(\mathbf{u} \cdot \boldsymbol{\nabla}\right)\mathbf{u} + \mathbf{f} \times \left( \mathbf{u} + \mathbf{\textcolor{black}{u_s}}\right) = \nonumber \\
& \qquad \qquad \qquad \qquad - \frac{1}{\rho_0} \boldsymbol{\nabla}\left( p + \textcolor{black}{\frac{\rho_0}{2} \left(\left|\mathbf{u} + \mathbf{u_s}\right|^2 - \left|\mathbf{u}\right|^2 \right)}\right) \nonumber \\
& \qquad \qquad \qquad \qquad + \nu_h \nabla_h^2  \mathbf{u} + \nu_z \nabla_z^2  \mathbf{u} \nonumber \\
& \qquad \qquad \qquad \qquad + \textcolor{black}{\mathbf{u_s}\times \boldsymbol{\nabla} \times \mathbf{u}} + \textcolor{black}{\frac{\boldsymbol{\tau}}{\rho_0}} \nonumber \\
& \qquad \qquad \qquad \qquad + \mathbf{g} \alpha \left(T - T_0\right) \\
& \partial_t T + \left((\mathbf{u} + \textcolor{black}{\mathbf{u_s}})\cdot \boldsymbol{\nabla}\right)T = \frac{Q}{c_p \rho_0} + \kappa_h \nabla_h^2 T + \kappa_z \nabla_z^2 T \\
& \textcolor{black}{\mathbf{u_s} = u_s e^{2 k_s z}\hat{\mathbf{x}}} 
\end{align}
where $\mathbf{u}_s$ is the Stokes drift velocity, $\boldsymbol{\tau}
= \tau \hat{\mathbf{x}}$ is the surface wind stress, $k_s$ is the
Stokes wavenumber that sets the depth scale of the effect,
$\nabla_h^2$ are the horizontal components of the vector Laplacian,
$\nabla_z^2$ is the vertical component of the vector Laplacian,
$\nu_h$ and $\nu_z$ are the horizontal and vertical kinematic
viscosities, and $\kappa_h$ and $\kappa_z$ are the horizontal and
vertical thermal diffusivities.

The parameters used for this case are
\begin{alignat}{2}
&\left\{N_x,N_y,N_z\right\} &&= \left\{96,96,72\right\} \nonumber \\
&\left\{\Delta_x,\Delta_y,\Delta_z\right\} &&= \left\{3,3,0.9\right\} \; \textrm{m} \nonumber \\
&f &&= 1 \times 10^{-4} \; \textrm{s}^{-1} \nonumber \\
&\rho_0 &&= 1035 \; \textrm{kg m}^{-3}\nonumber \\
&g &&= 9.81 \; \textrm{m s}^{-2} \nonumber \\
&\alpha &&= 2 \times 10^{-4} \; \textrm{C}^{-1} \nonumber \\
&T_0 &&= 20 \; \textrm{C} \nonumber \\
&\nu_h &&= 2 \times 10^{-3} \; \textrm{m}^2 \textrm{s}^{-1} \nonumber \\
&\kappa_h &&= 2 \times 10^{-3} \; \textrm{m}^2 \textrm{s}^{-1} \nonumber \\
&\nu_z &&= 1 \times 10^{-3} \; \textrm{m}^2 \textrm{s}^{-1} \nonumber \\
&\kappa_z &&= 1 \times 10^{-3} \; \textrm{m}^2 \textrm{s}^{-1} \nonumber \\
&c_p &&= 3994 \; \textrm{J kg}^{-1} \textrm{C}^{-1} \nonumber \\
&q &&= -5 \; \textrm{W m}^{-2} \nonumber \\
&N^2 && = 1.936 \times 10^{-5} \; \textrm{s}^{-2} \nonumber \\
&\tau &&= 0.037 \; \textrm{N m}^{-2} \nonumber \\
&u_s &&= 0.068 \; \textrm{m s}^{-1} \nonumber \\
&k_s &&= 0.105 \; \textrm{m}^{-1} \nonumber \\
&L_m && = 32.1 \; \textrm{m} \nonumber \\
&u_* && = 6.0 \times 10^{-3} \; \textrm{m} \textrm{s}^{-1}  \nonumber \\
&Re_* && = 191.9 \nonumber \\
&La &&= 0.2965 \nonumber 
\end{alignat}
Here, $L_m$ is the initial mixed-layer depth, above which the fluid
temperature is initially uniform, $u_*= \tau^{\frac{1}{2}}\rho_0^{\frac{1}{2}}$ is the friction velocity, Re$_* = \frac{u_* L_m}{\nu_z}$ and the Langmuir number is defined as
\begin{equation}
La = \sqrt{\frac{\tau^{\frac{1}{2}}}{\rho_0^{\frac{1}{2}}u_s}}
\end{equation}
 The surface cooling is included for
consistency with \cite{mcwilliams97} where it is used to help
destabilize the flow; it does not have a significant impact on the
dynamics otherwise, so the Rayleigh and Prandtl number are not important.  The simulation is stopped at 72 hours.

\begin{figure*}[]
\centering
\includegraphics[width=1.0 \textwidth]{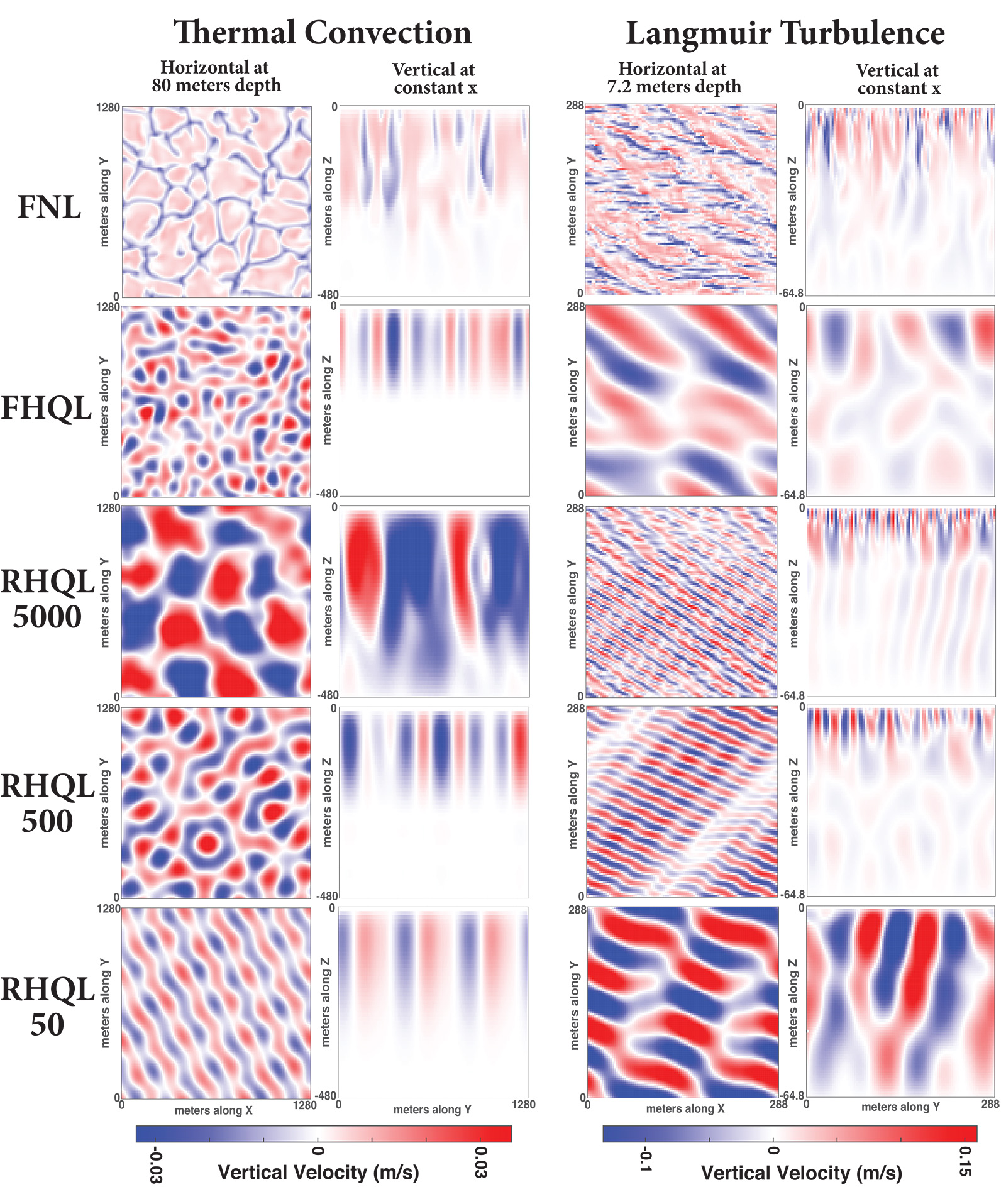}
\caption{\small Cross-sections of vertical velocity fields from an
  ensemble member of the thermal-convection case at $t = 10$ hours
  and from an ensemble member of the Langmuir-turbulence case at $t =
  54$ hours.  Horizontal cross sections (top half of figure) are
  at 80 meters depth in the thermal convection case and 7.2 meters
  depth in the Langmuir case.  By incompressibility, $\overline{w}=0$, so $w = w'$.}
\label{fig:flow_fields}
\end{figure*}

\begin{figure*}[]
\centering
\includegraphics[width=1.0 \textwidth]{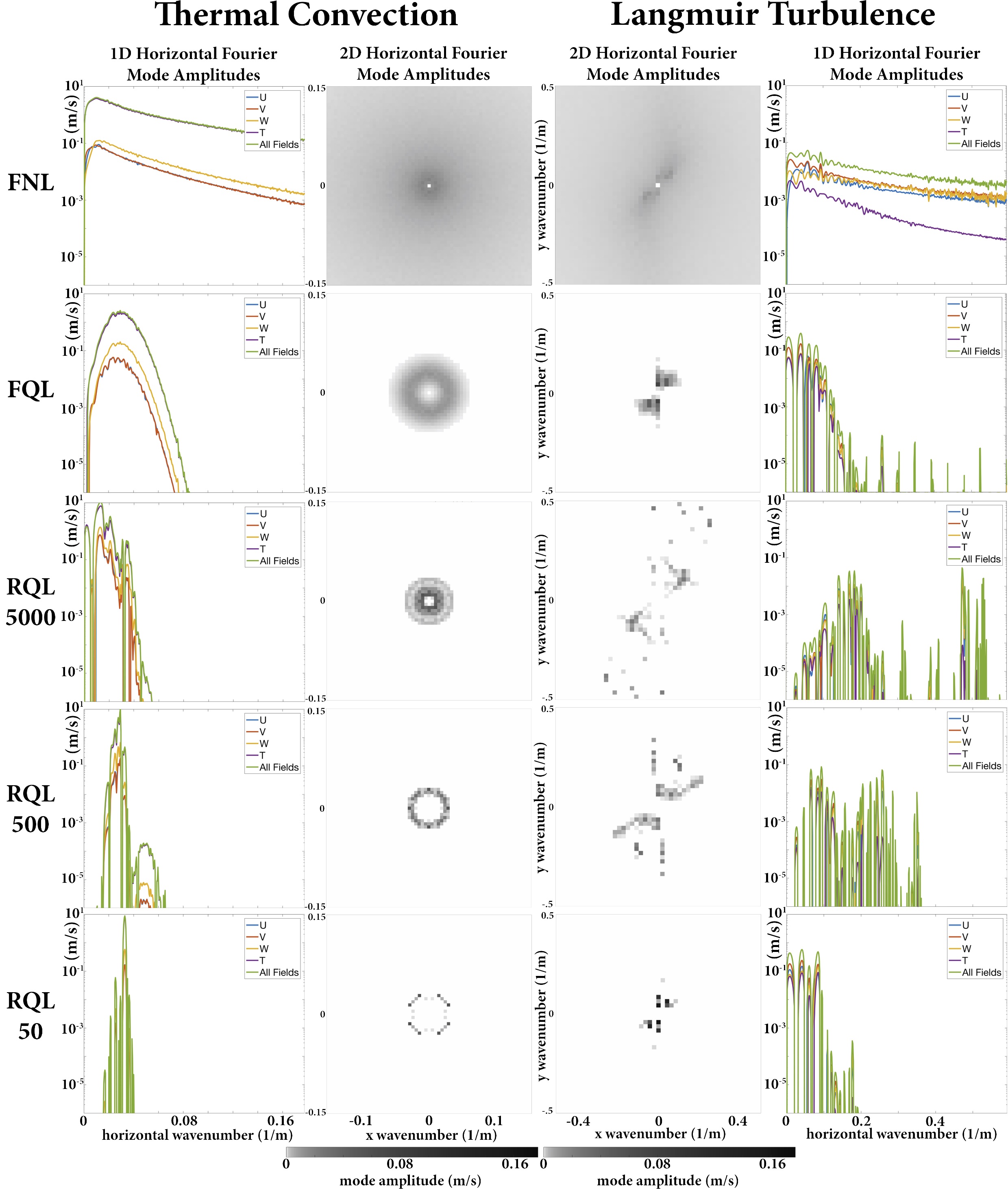}
\caption{\small Center columns: The 2D horizontal mode amplitude of
  the thermal-convection fields at 20 hours and the
  Langmuir-turbulence fields at 54 hours for different types of
  dynamics and basis sizes. These are computed from an ensemble
  average of 12 reduced-model fields and 24 full-basis fields.  The
  wavenumber axes show half of the modes accessible in the full-basis
  representation.  Outer columns: The 1D horizontal mode amplitudes at
  the same times.  Temperature is scaled as described in section
  \ref{sec:numerics}\ref{sec:basis}. }
\label{fig:mode_density}
\end{figure*}

\begin{figure*}[h!]
\centering
\begin{subfigure}{0.28\textwidth}
\centering
\begin{subfigure}{\textwidth}
\centering
\includegraphics[width=\textwidth]{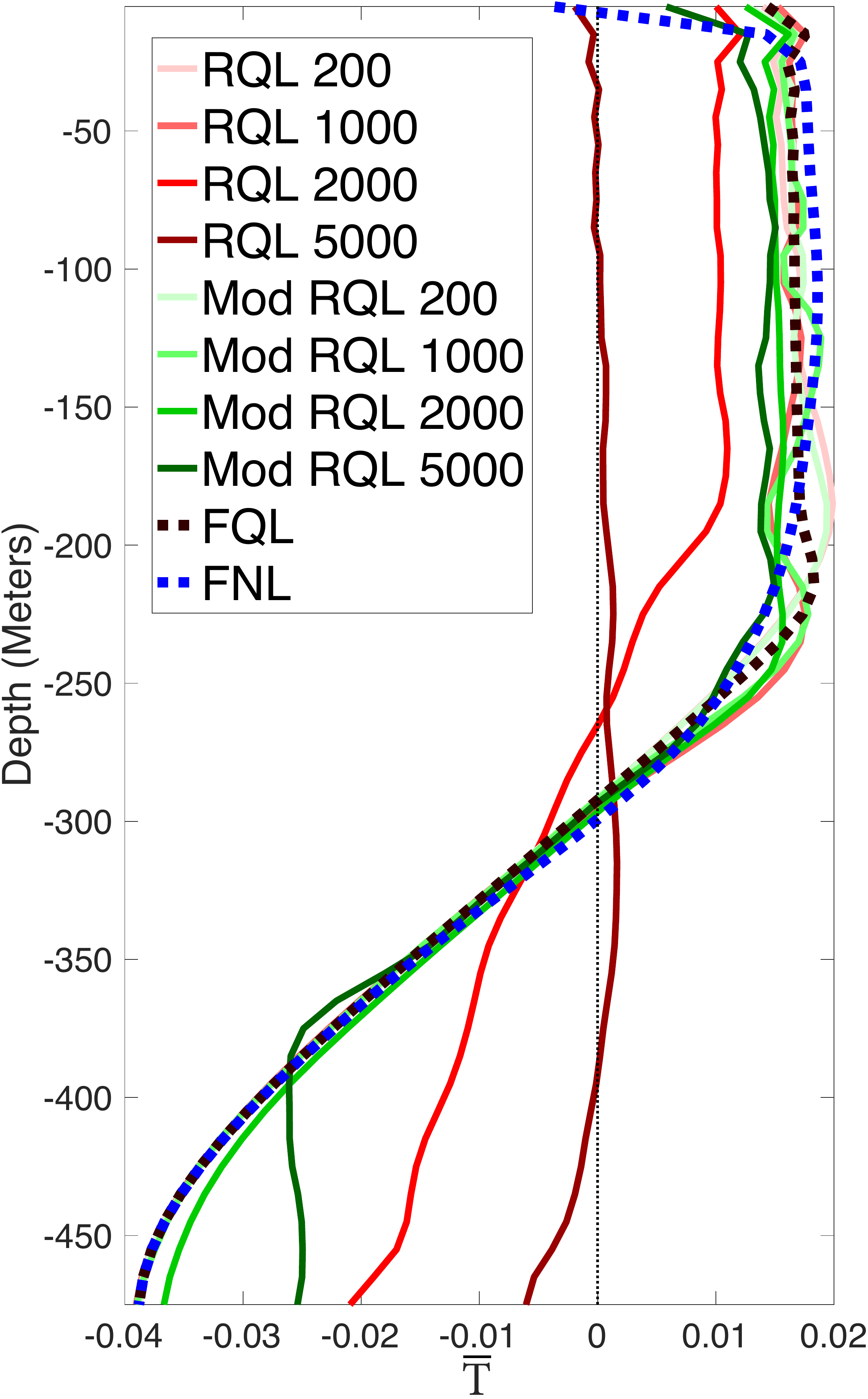}
\label{fig:thermal_T}
\end{subfigure}

\begin{subfigure}{\textwidth}
\centering
\includegraphics[width=\textwidth]{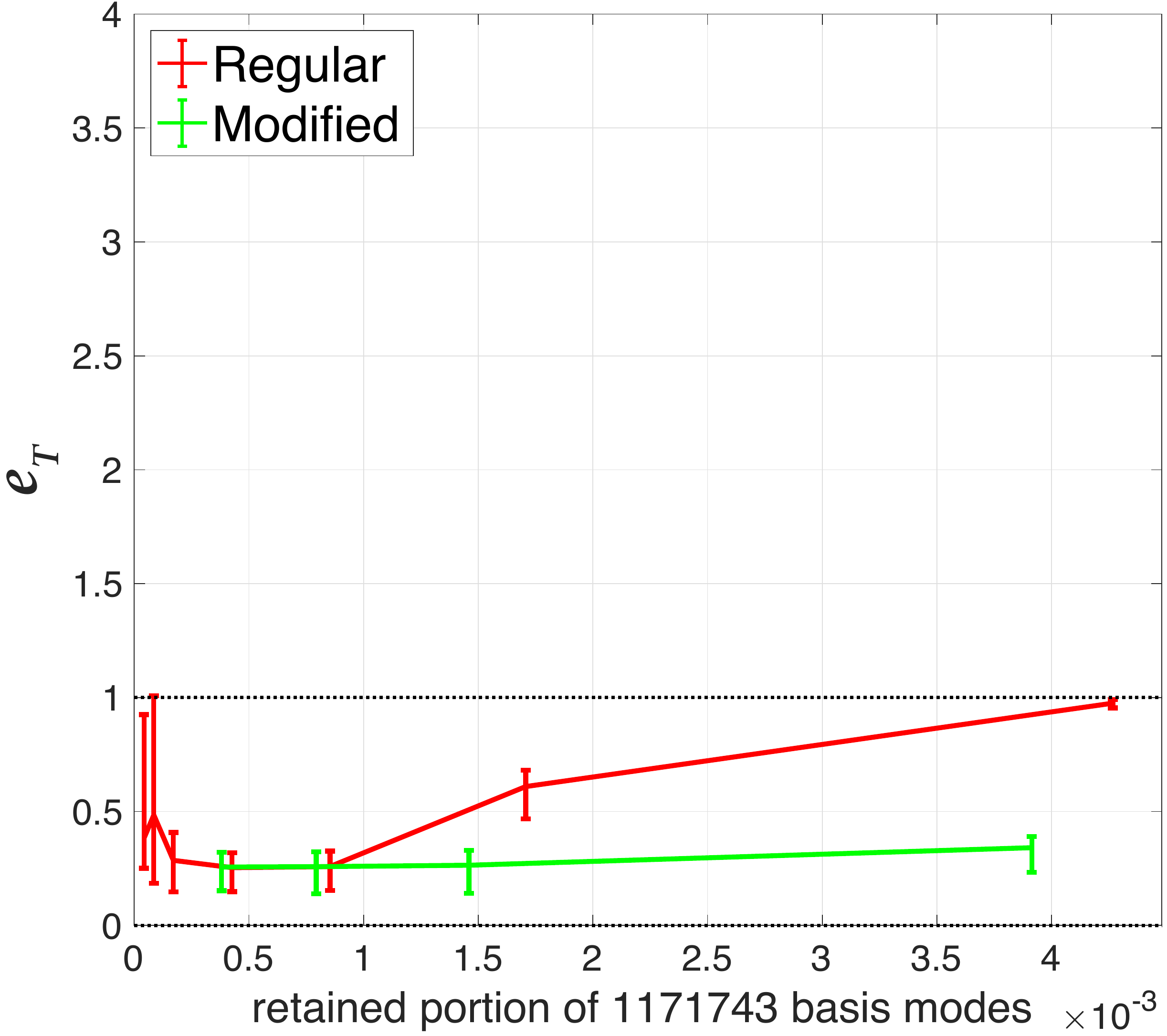}
\end{subfigure}
\caption{Thermal Convection}
\label{fig:thermal_mean_error}
\end{subfigure}
\hspace{0.25cm}
\rulesep
\hspace{0.25cm}
\begin{subfigure}{0.56\textwidth}
\centering
\begin{subfigure}{0.49\textwidth}
\centering
\includegraphics[width=\textwidth]{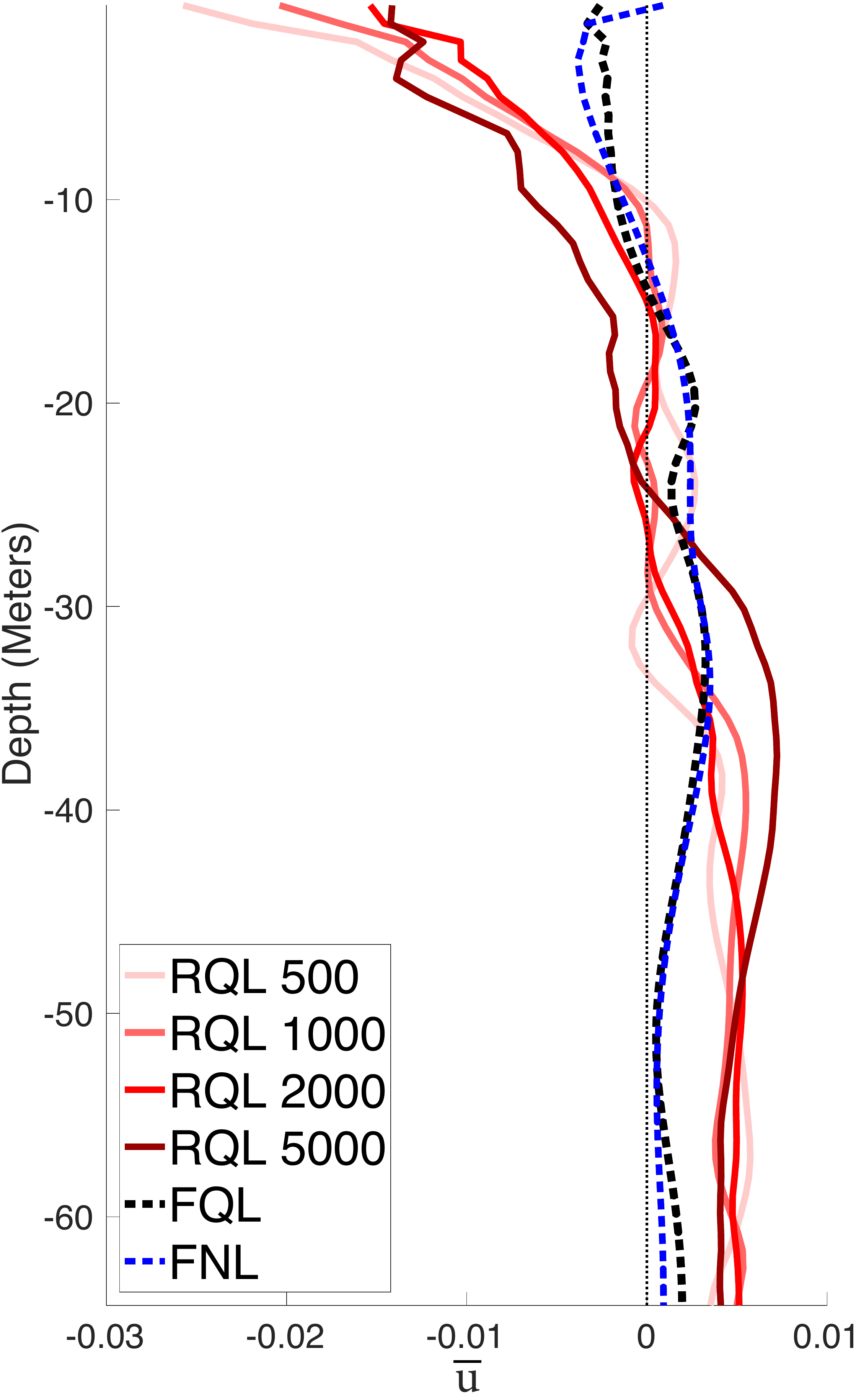}
\label{fig:langmuir_U}
\end{subfigure}
\begin{subfigure}{0.49\textwidth}
\centering
\includegraphics[width=\textwidth]{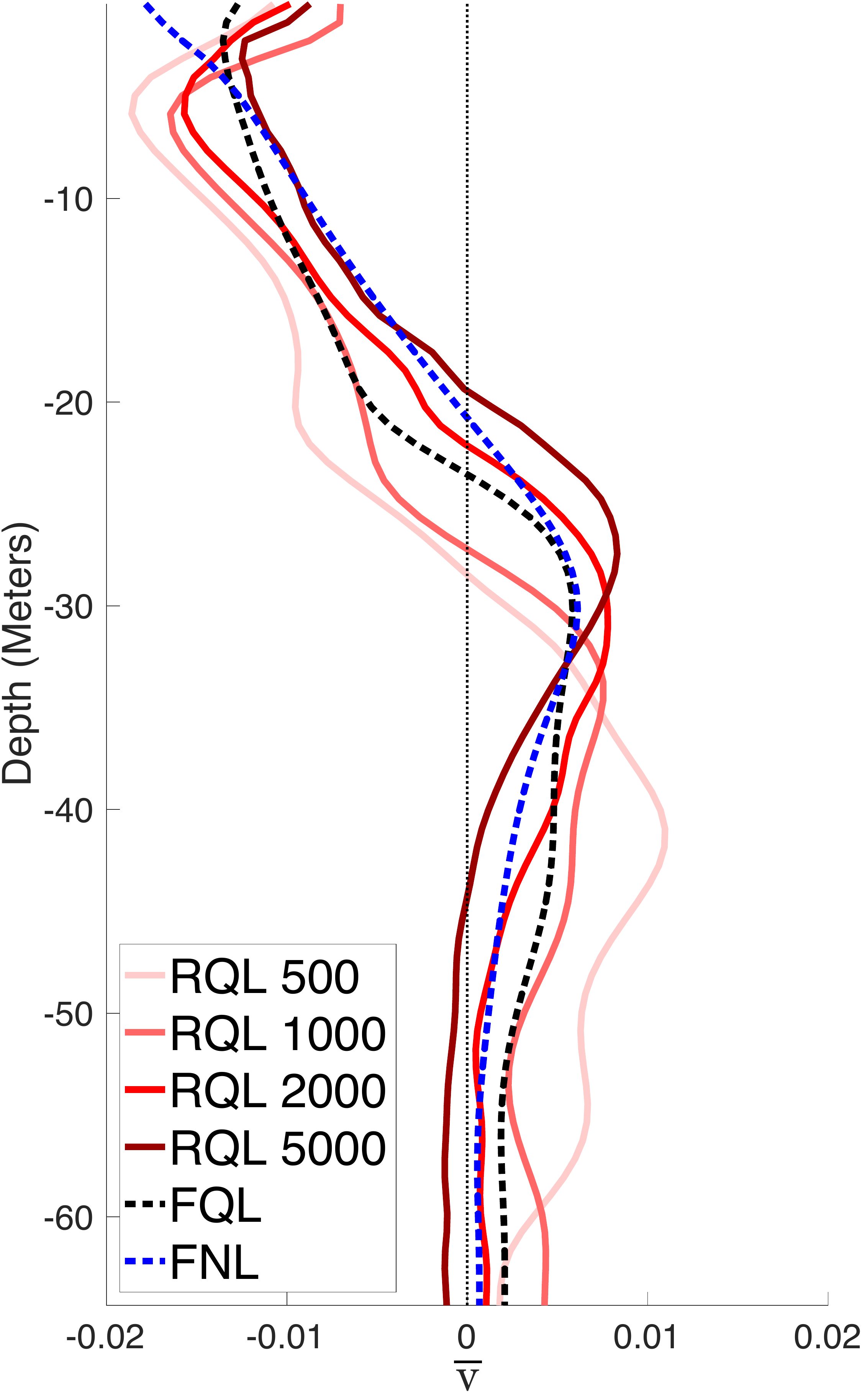}
\label{fig:langmuir_V}
\end{subfigure}

\begin{subfigure}{0.525\textwidth}
\centering
\includegraphics[width=\textwidth]{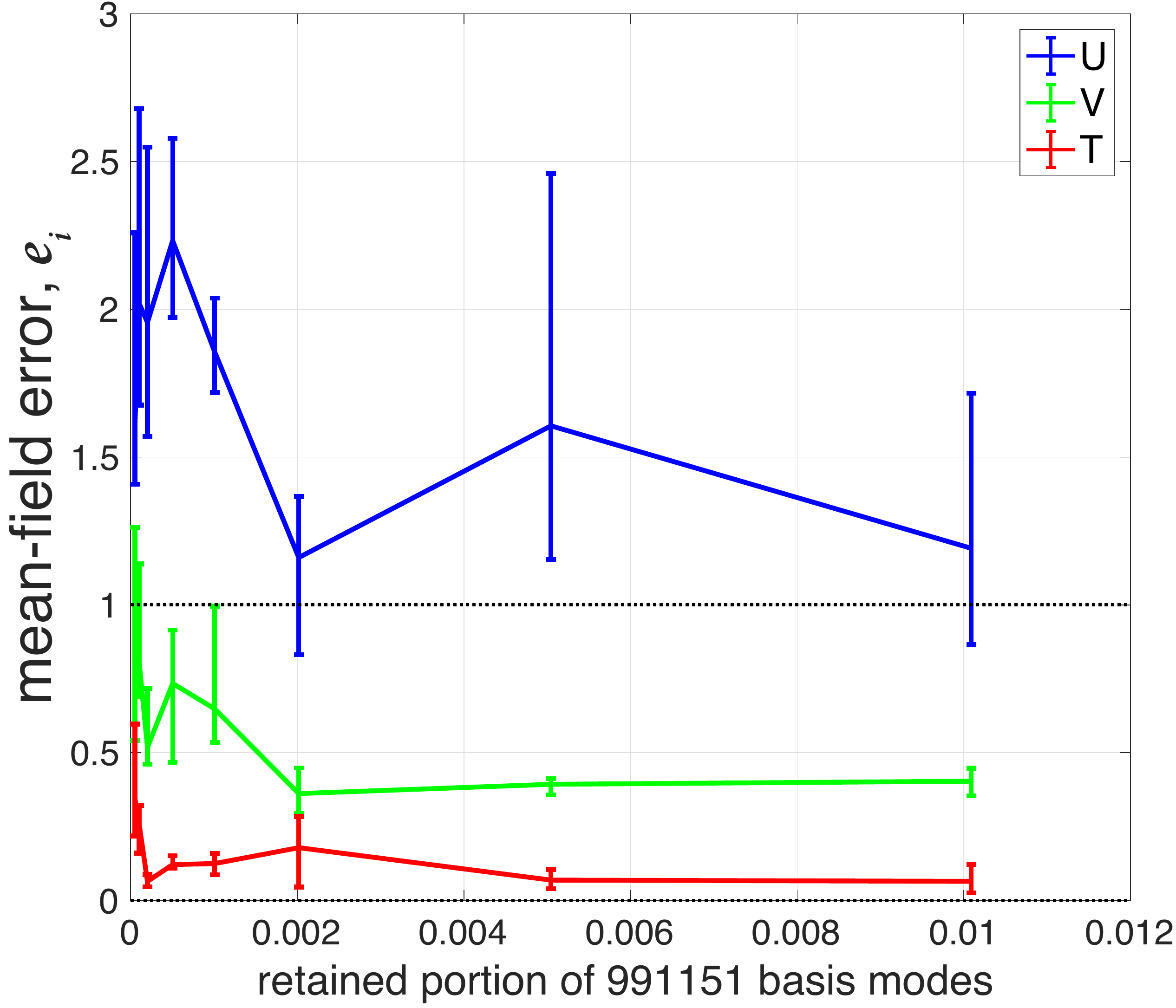}
\caption{Langmuir Turbulence}
\label{fig:langmuir_mean_error}
\end{subfigure}
\end{subfigure}
\centering \caption{Top Row: Ensemble-averaged mean-field vertical
  profiles for the thermal-convection case at $t=10$ hours and for the
  Langmuir-turbulence case at $t=18$ hours. Bottom Row: error of
  ensemble averaged mean-field profiles as a function of truncated
  basis size, as defined in equation \ref{eq:mean_error_def}.  For the thermal-convection profiles, both the regular 5000-mode truncated POD basis and a modified version omitting some large-scale horizontal modes (defined in section \ref{sec:results}\ref{sec:displacement} and depicted in figure \ref{fig:basis_trim}) are shown.}
\label{fig:mean_stats}
\end{figure*}

\begin{figure*}[]
\centering
\begin{subfigure}{0.30\textwidth}
\centering
\includegraphics[width=\textwidth]{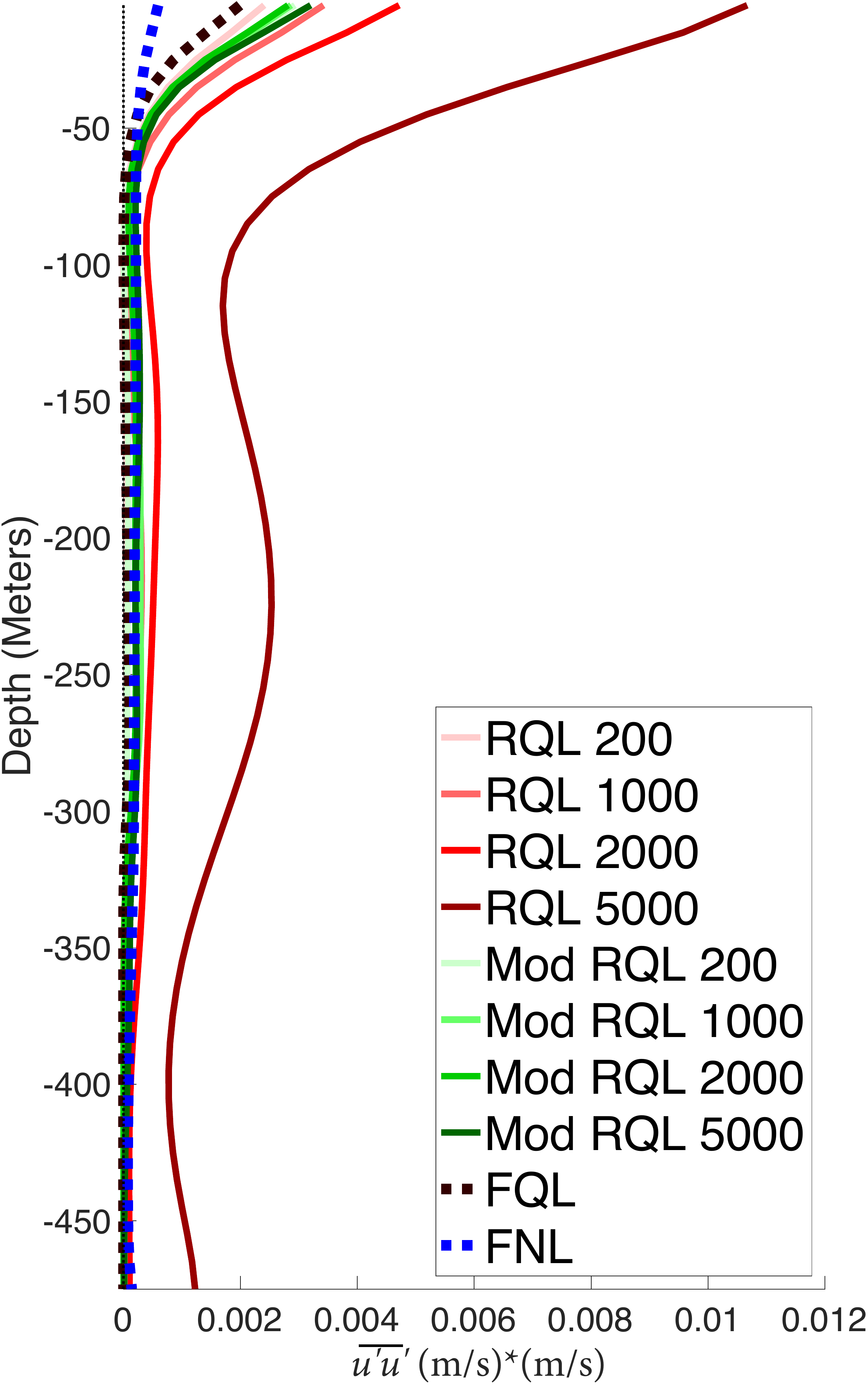}
\end{subfigure}
\begin{subfigure}{0.30\textwidth}
\centering
\includegraphics[width=\textwidth]{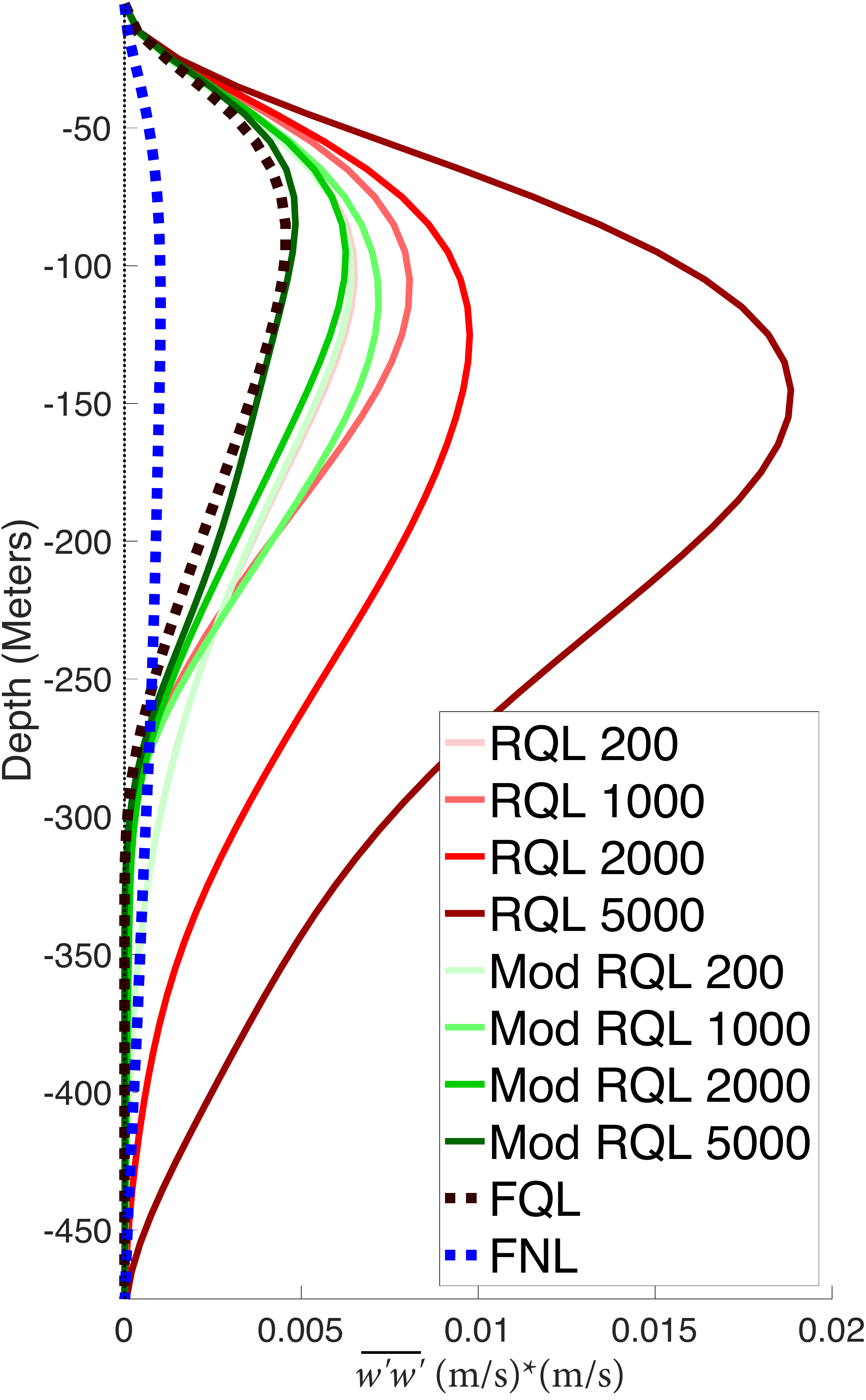}
\end{subfigure}
\begin{subfigure}{0.30\textwidth}
\centering
\includegraphics[width=\textwidth]{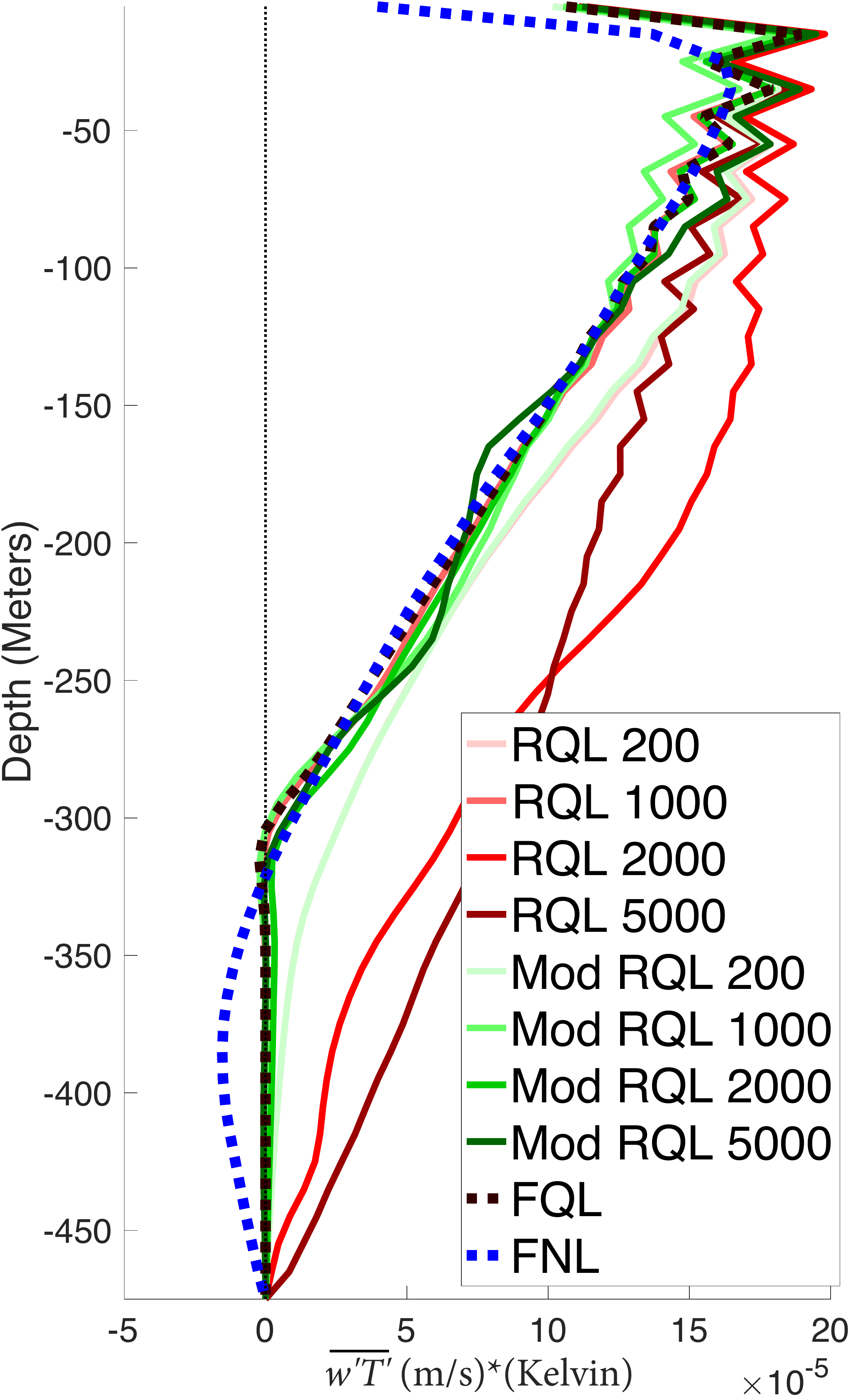}
\end{subfigure}

\begin{subfigure}{0.30\textwidth}
\centering
\includegraphics[width=\textwidth]{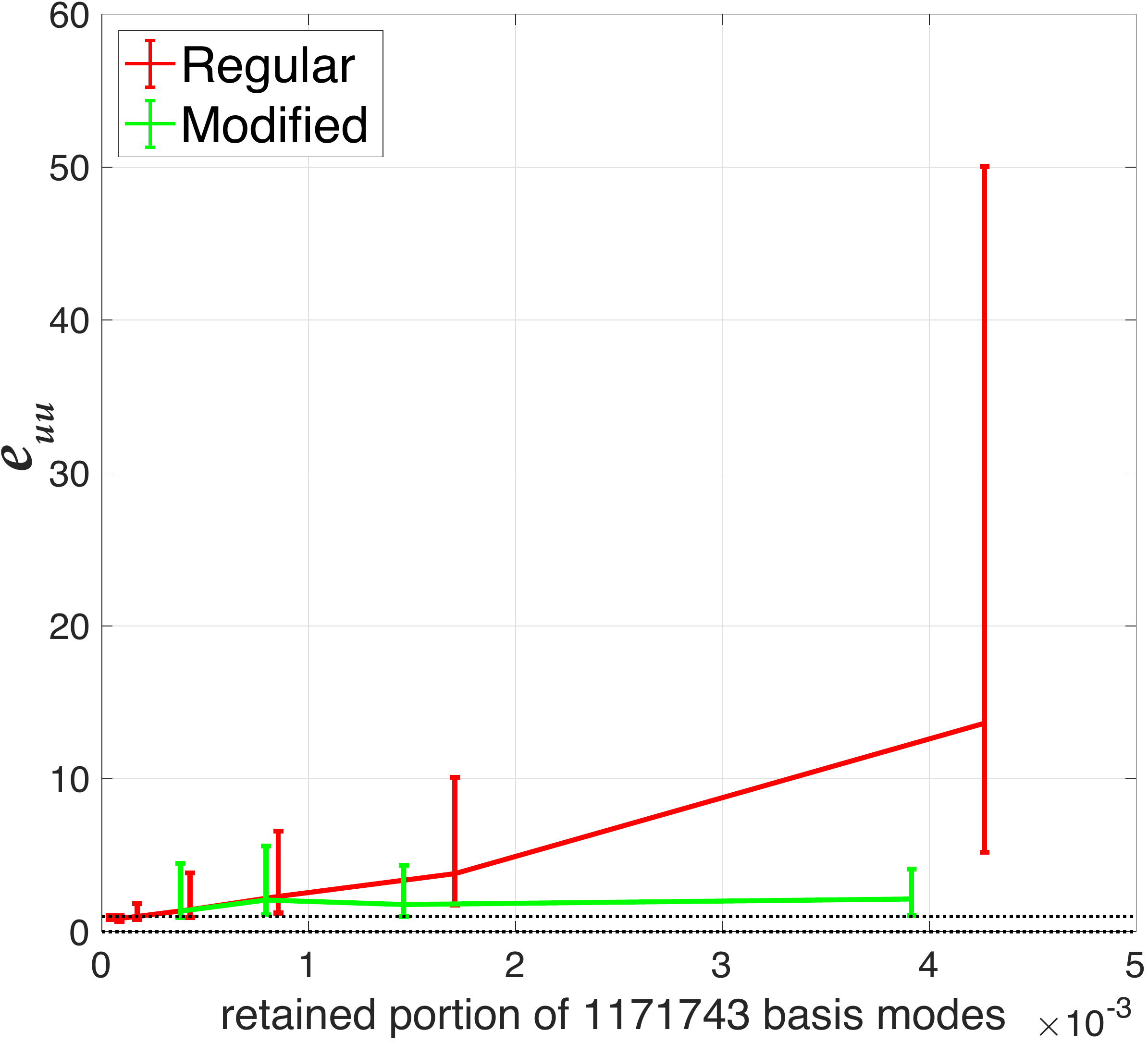}
\caption{\normalsize $\overline{u'u'}$}
\label{fig:thermal_uu_error}
\end{subfigure}
\begin{subfigure}{0.30\textwidth}
\centering
\includegraphics[width=\textwidth]{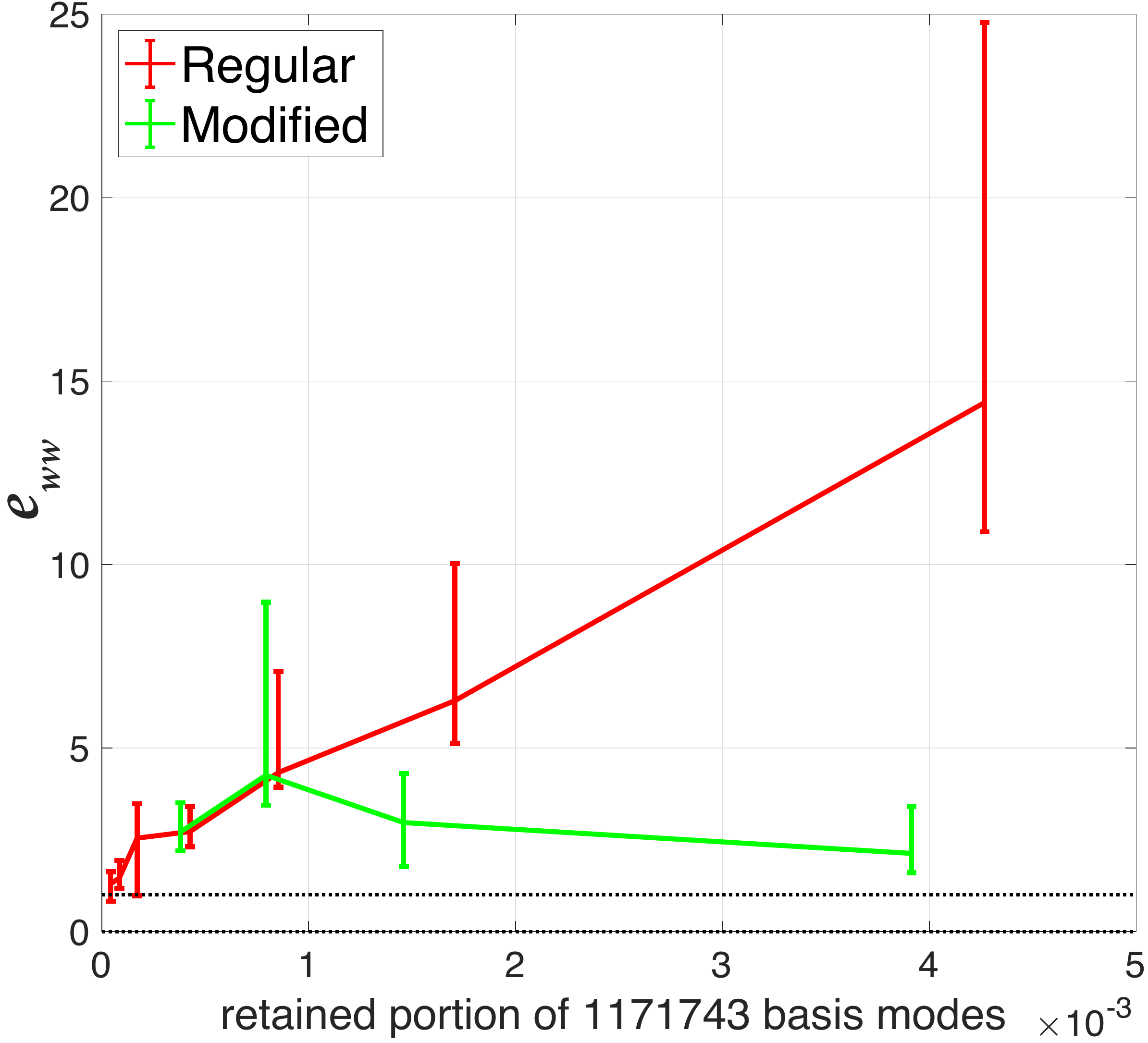}
\caption{\normalsize $\overline{w'w'}$}
\label{fig:thermal_ww_error}
\end{subfigure}
\begin{subfigure}{0.30\textwidth}
\centering
\includegraphics[width=\textwidth]{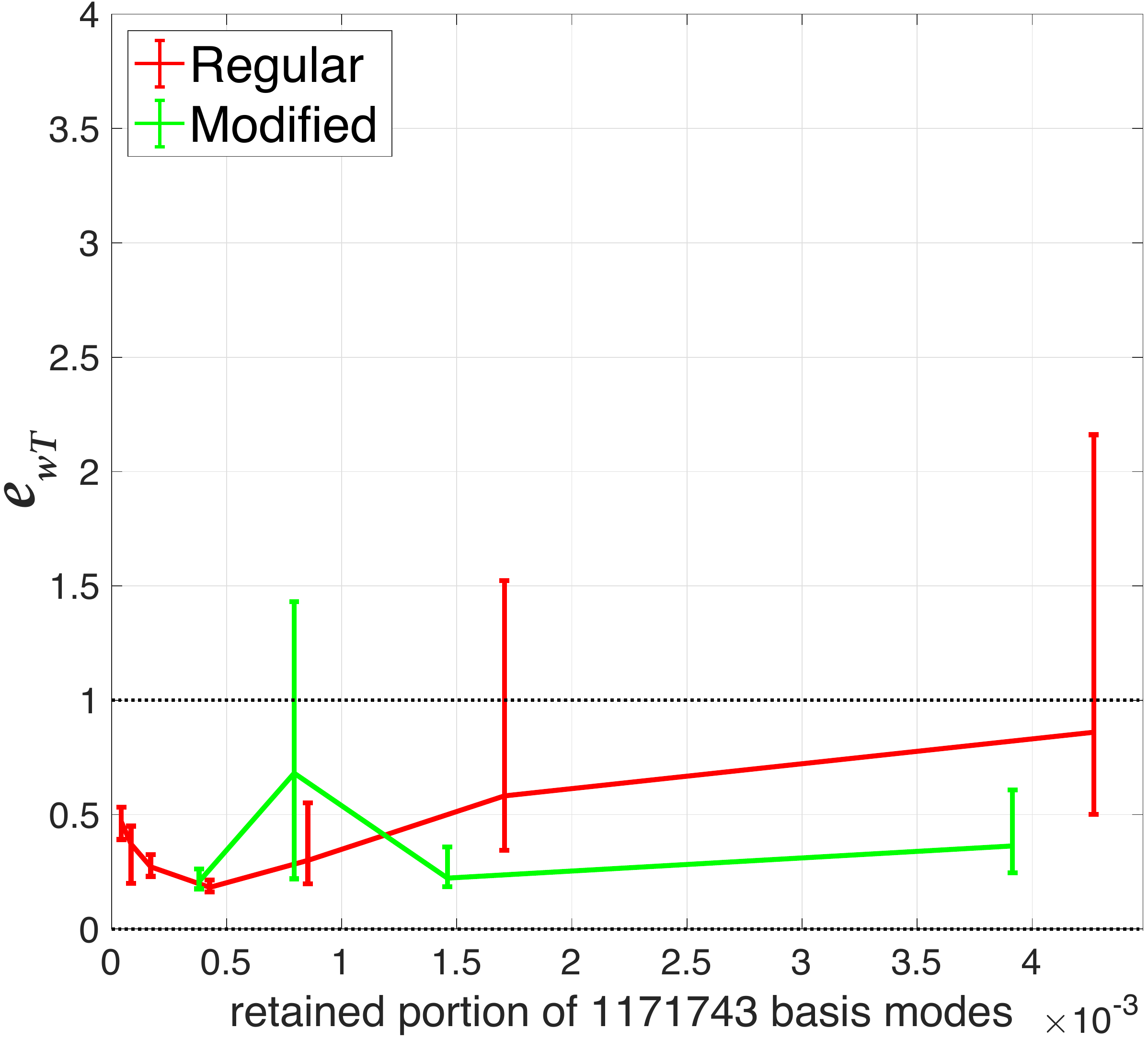}
\caption{\normalsize $\overline{w'T'}$}
\label{fig:thermal_wT_error}
\end{subfigure}
\centering \caption{Top Row: Vertical profiles of turbulent transports
  and energies of the thermal-convection cases at $t = 20$ hours.
  Bottom Row: Vertical-profile error, $e_{ij}$, defined in equation
  \ref{eq:error_def}, vs retained basis size for different statistical
  profiles of the thermal-convection cases.  The modified version of the truncated bases is defined in section \ref{sec:results}\ref{sec:displacement} and depicted in figure \ref{fig:basis_trim}.  Note that $e_{ij}$
  accounts for the profiles that are evenly sampled throughout the
  flow evolution.  Although there must be a general trend for the RQL
  statistical profiles to converge on the FQL profiles, this
  convergence will not generally be uniform, as is seen in the poor
  representation of the 5000-mode profiles in particular.  Convergence
  properties will be explored more thoroughly in section
  \ref{sec:results}\ref{sec:small}.}
\label{fig:thermal_stat}
\end{figure*}

\begin{figure*}[]
\centering
\begin{subfigure}{0.30\textwidth}
\centering
\includegraphics[width=\textwidth]{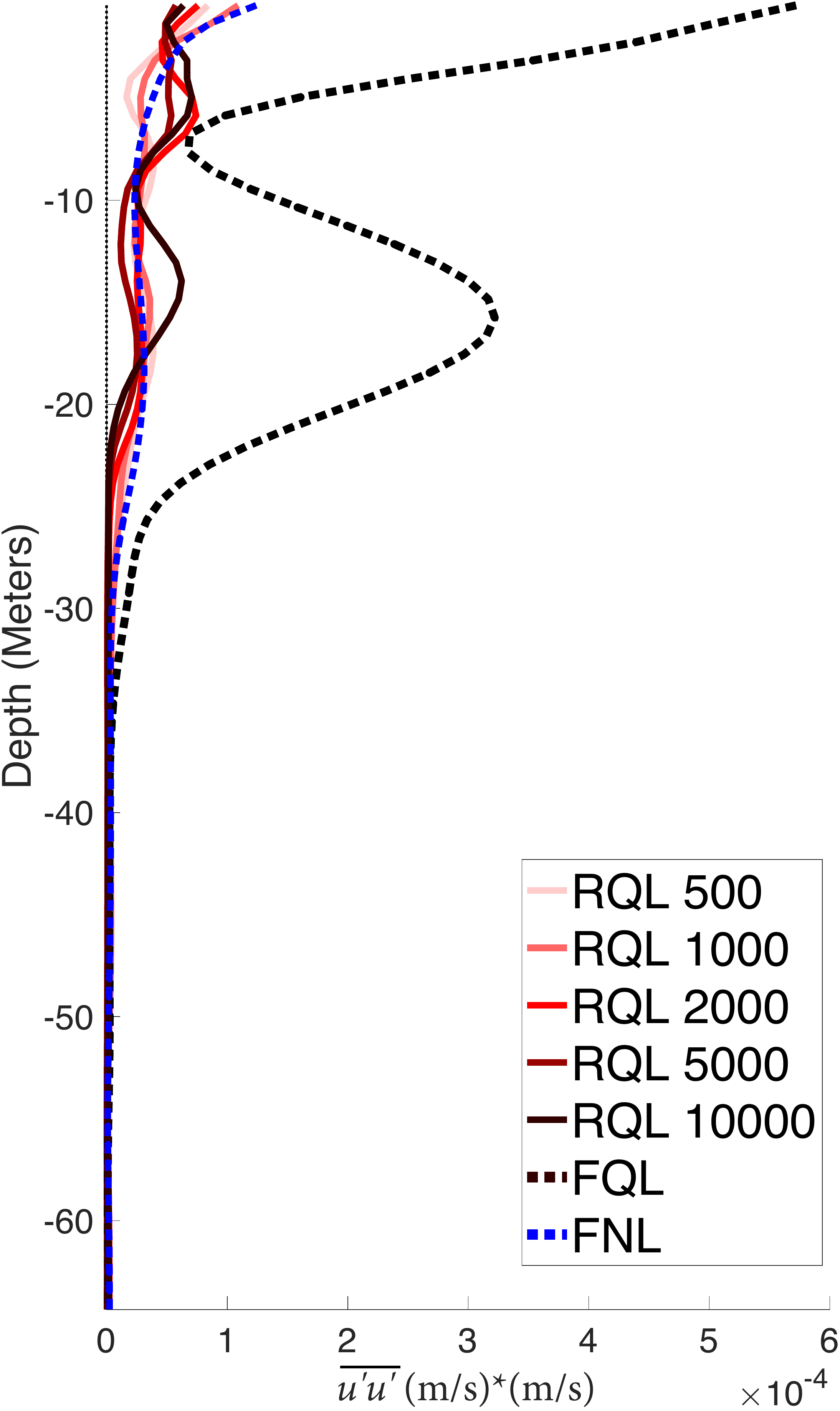}
\end{subfigure}
\begin{subfigure}{0.30\textwidth}
\centering
\includegraphics[width=\textwidth]{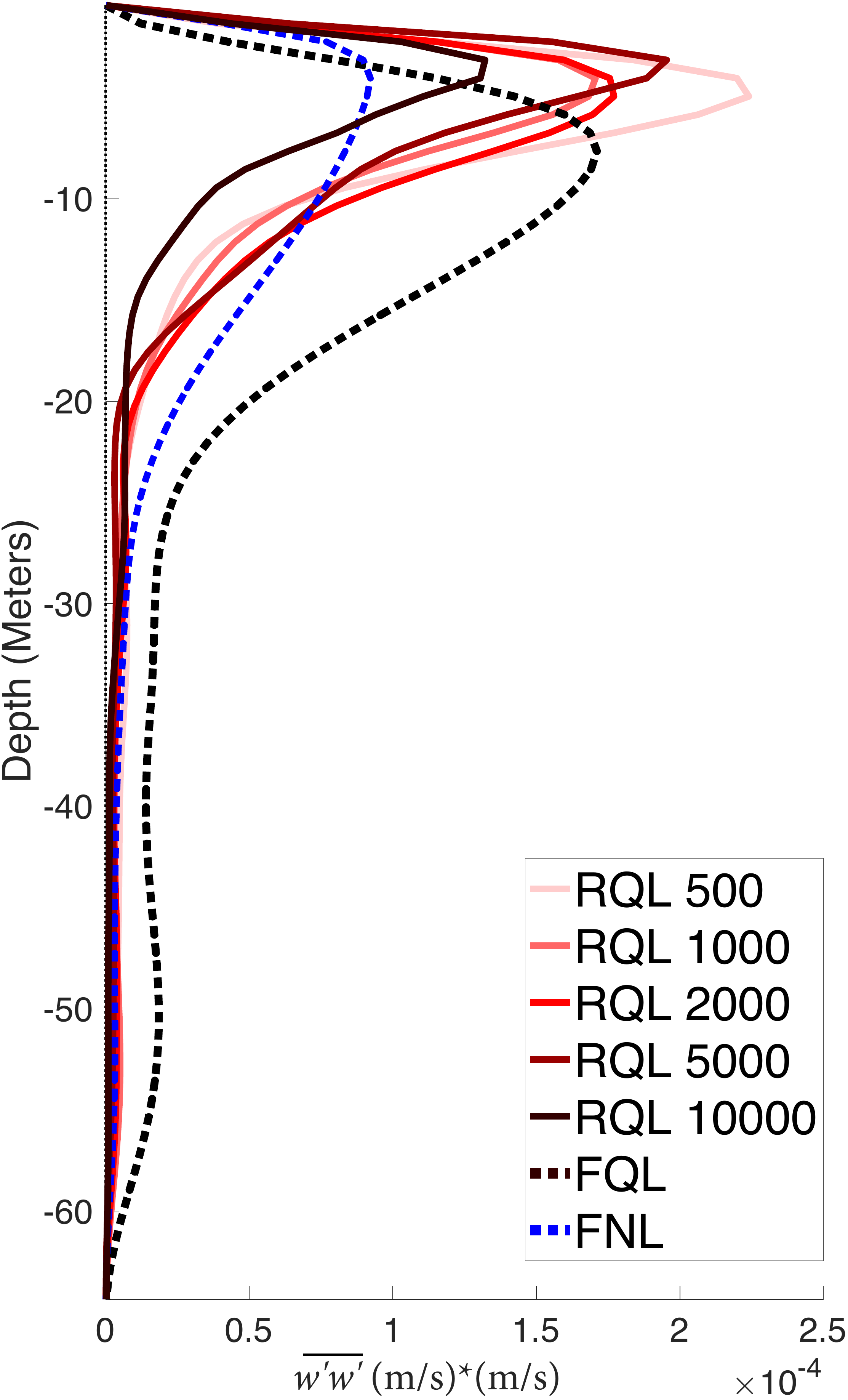}
\end{subfigure}
\begin{subfigure}{0.30\textwidth}
\centering
\includegraphics[width=\textwidth]{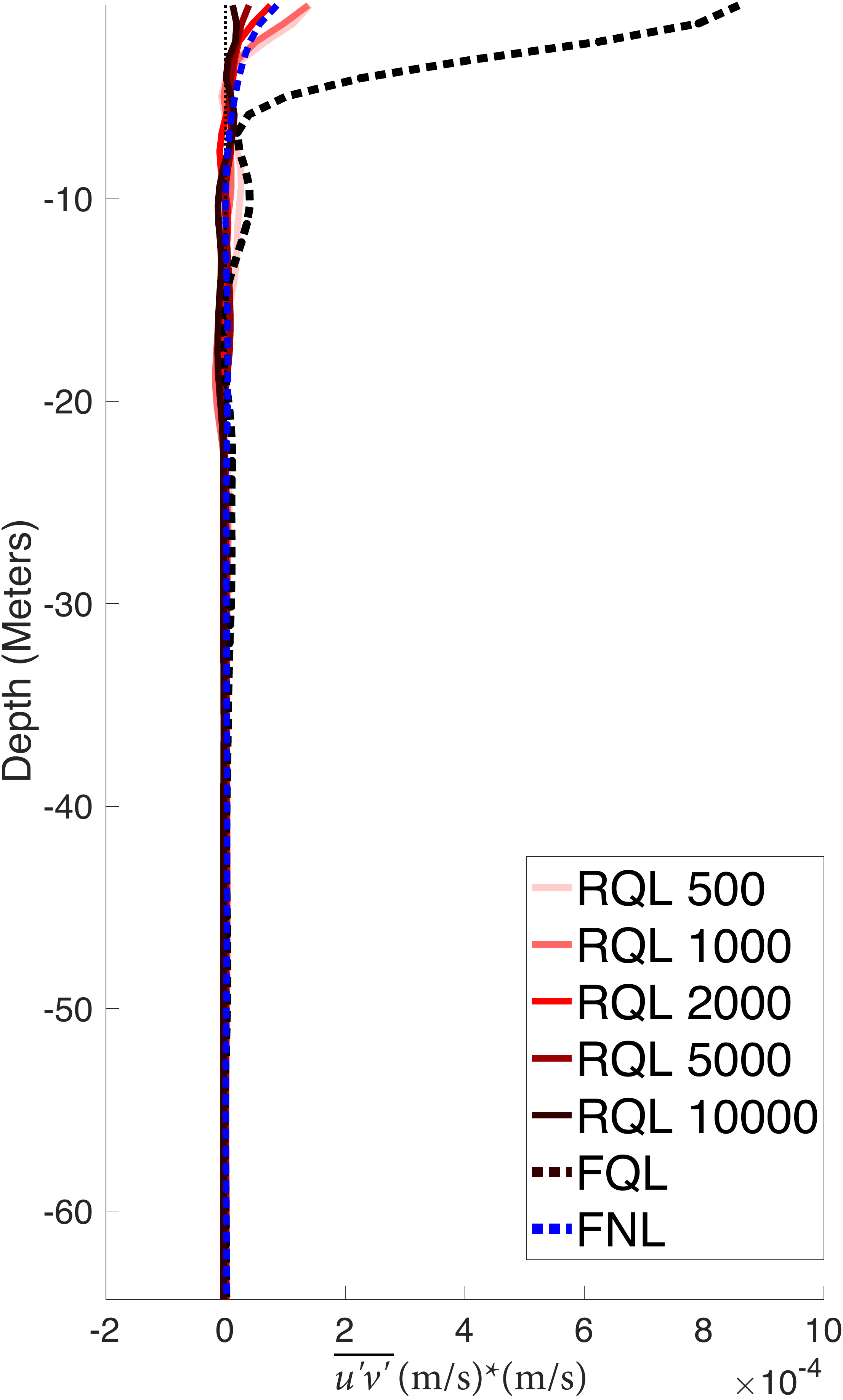}
\end{subfigure}

\begin{subfigure}{0.30\textwidth}
\centering
\includegraphics[width=\textwidth]{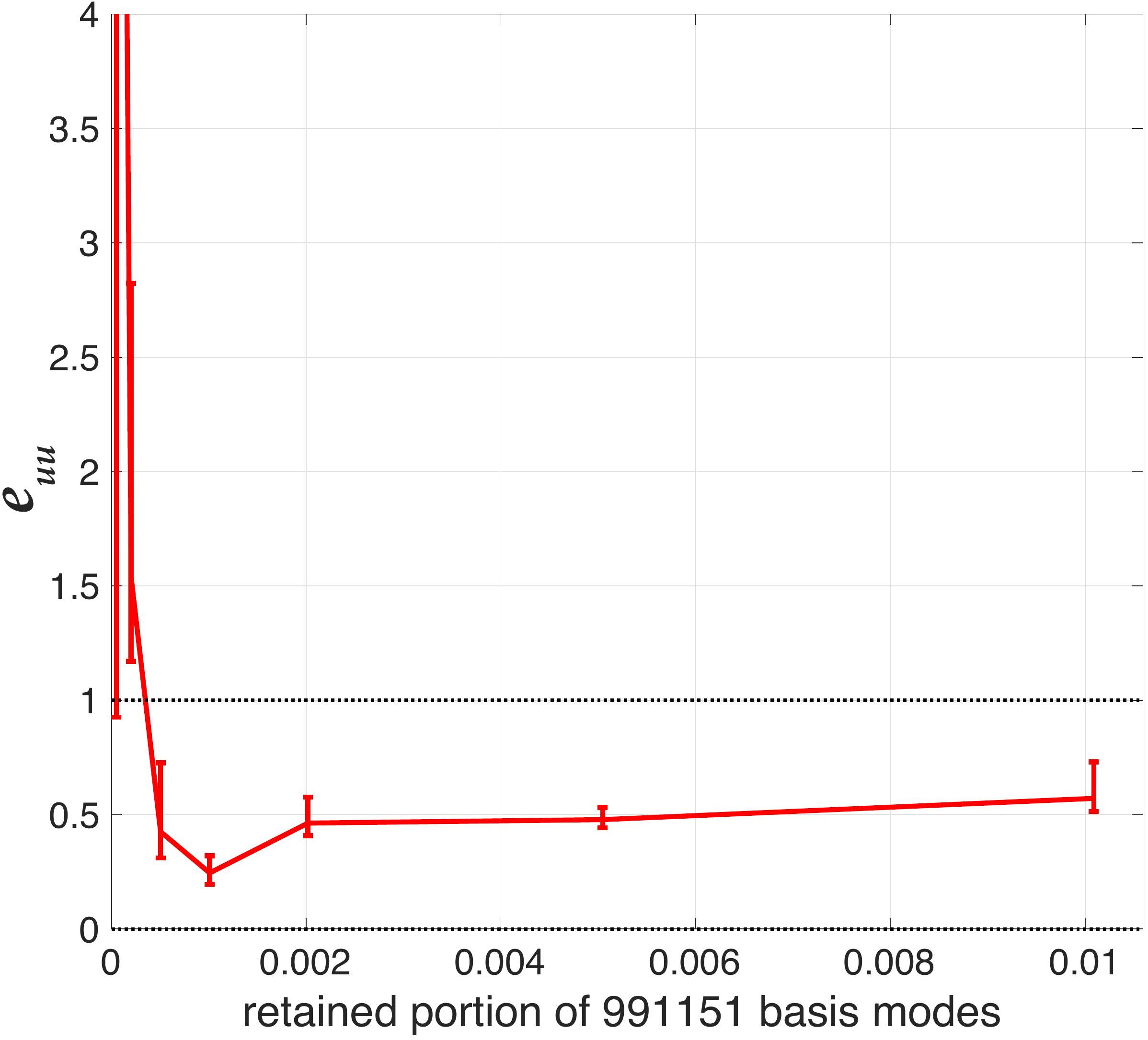}
\caption{\normalsize $\overline{u'u'}$}
\label{fig:langmuir_uu_error}
\end{subfigure}
\begin{subfigure}{0.30\textwidth}
\centering
\includegraphics[width=\textwidth]{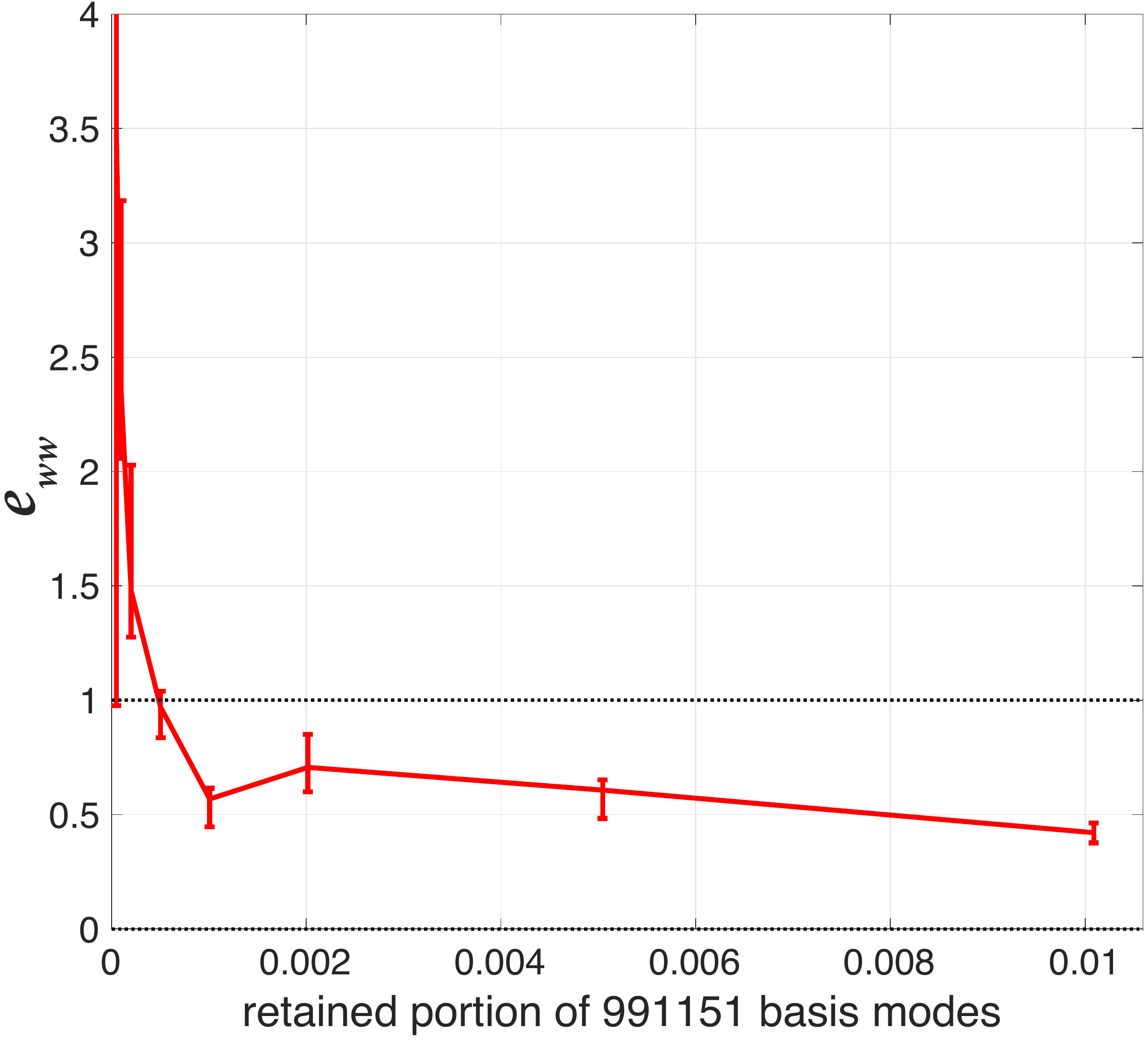}
\caption{\normalsize $\overline{w'w'}$}
\label{fig:langmuir_ww_error}
\end{subfigure}
\begin{subfigure}{0.30\textwidth}
\centering
\includegraphics[width=\textwidth]{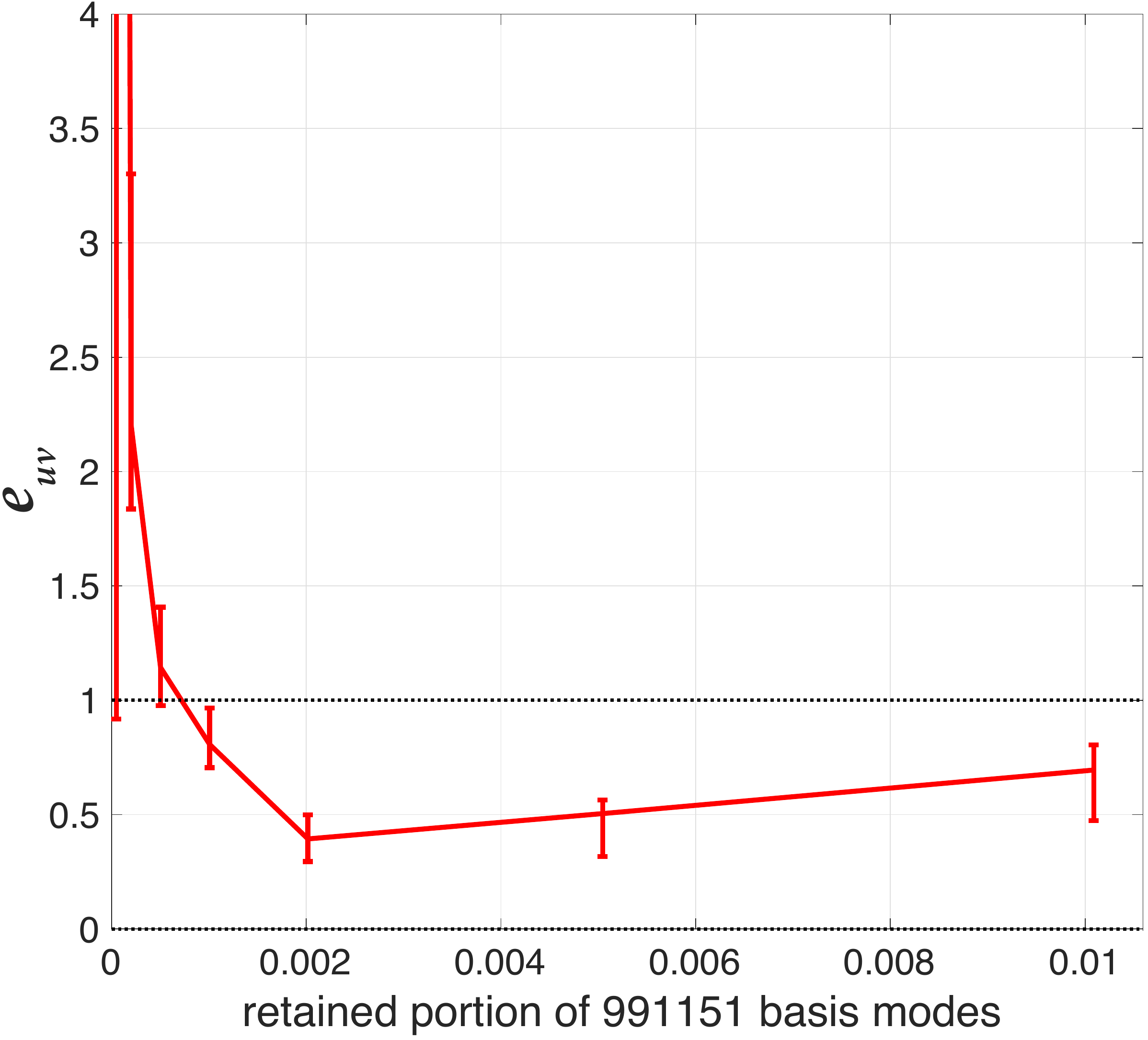}
\caption{\normalsize $\overline{u'v'}$}
\label{fig:langmuir_uv_error}
\end{subfigure}
\centering \caption{Top Row: Vertical profiles of turbulent transports
  and energies of the Langmuir-turbulence cases at $t = 36$ hours.
  Bottom Row: Vertical-profile error, $e_{ij}$, defined in equation
  \ref{eq:error_def}, vs retained basis size for different statistical
  profiles of the Langmuir-turbulence cases.  Note that $e_{ij}$
  accounts for the profiles that are evenly sampled throughout the
  flow evolution.}
\label{fig:langmuir_stat_1}
\end{figure*}

\begin{figure*}[]
\centering
\begin{subfigure}{0.30\textwidth}
\centering
\includegraphics[width=\textwidth]{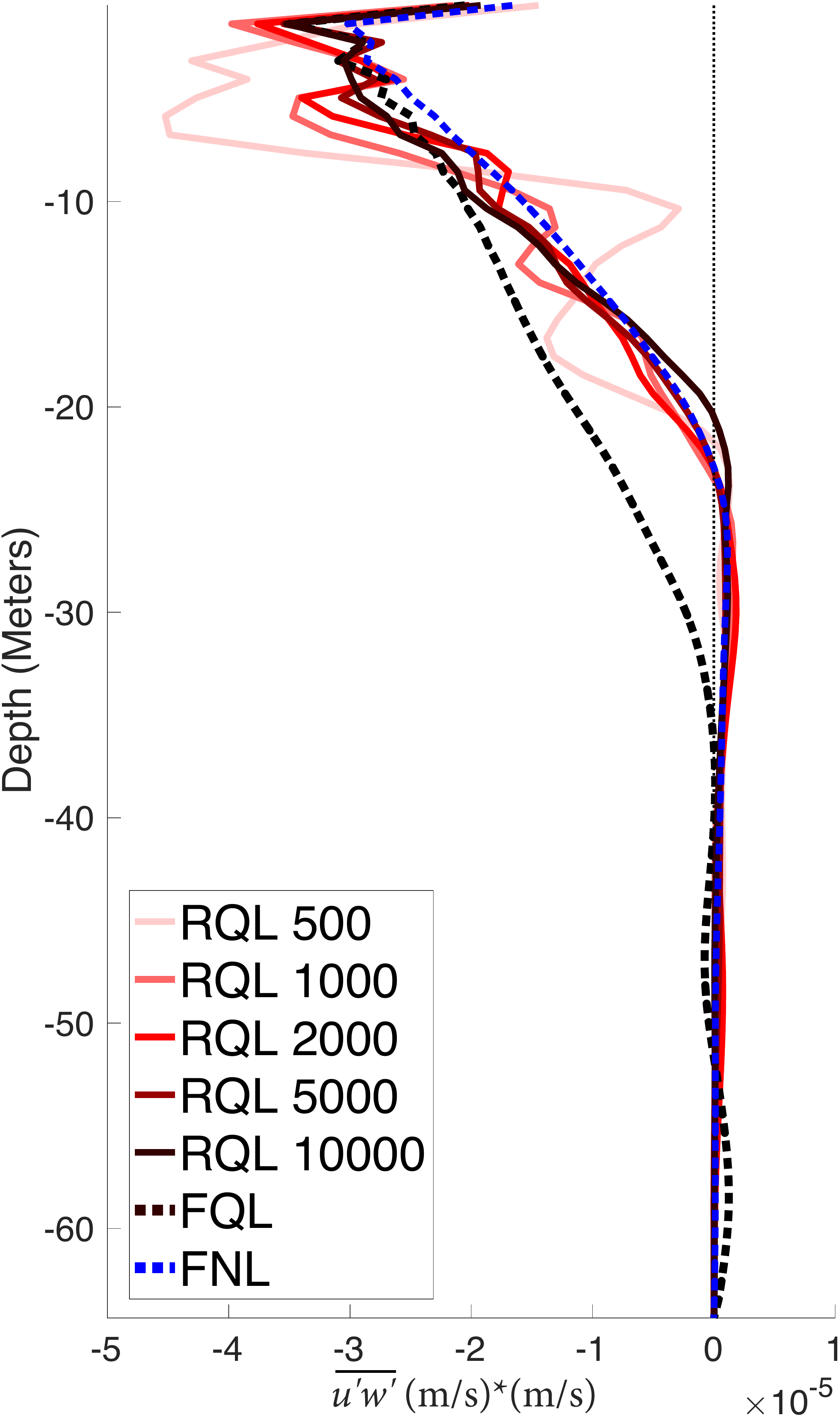}
\end{subfigure}
\begin{subfigure}{0.30\textwidth}
\centering
\includegraphics[width=\textwidth]{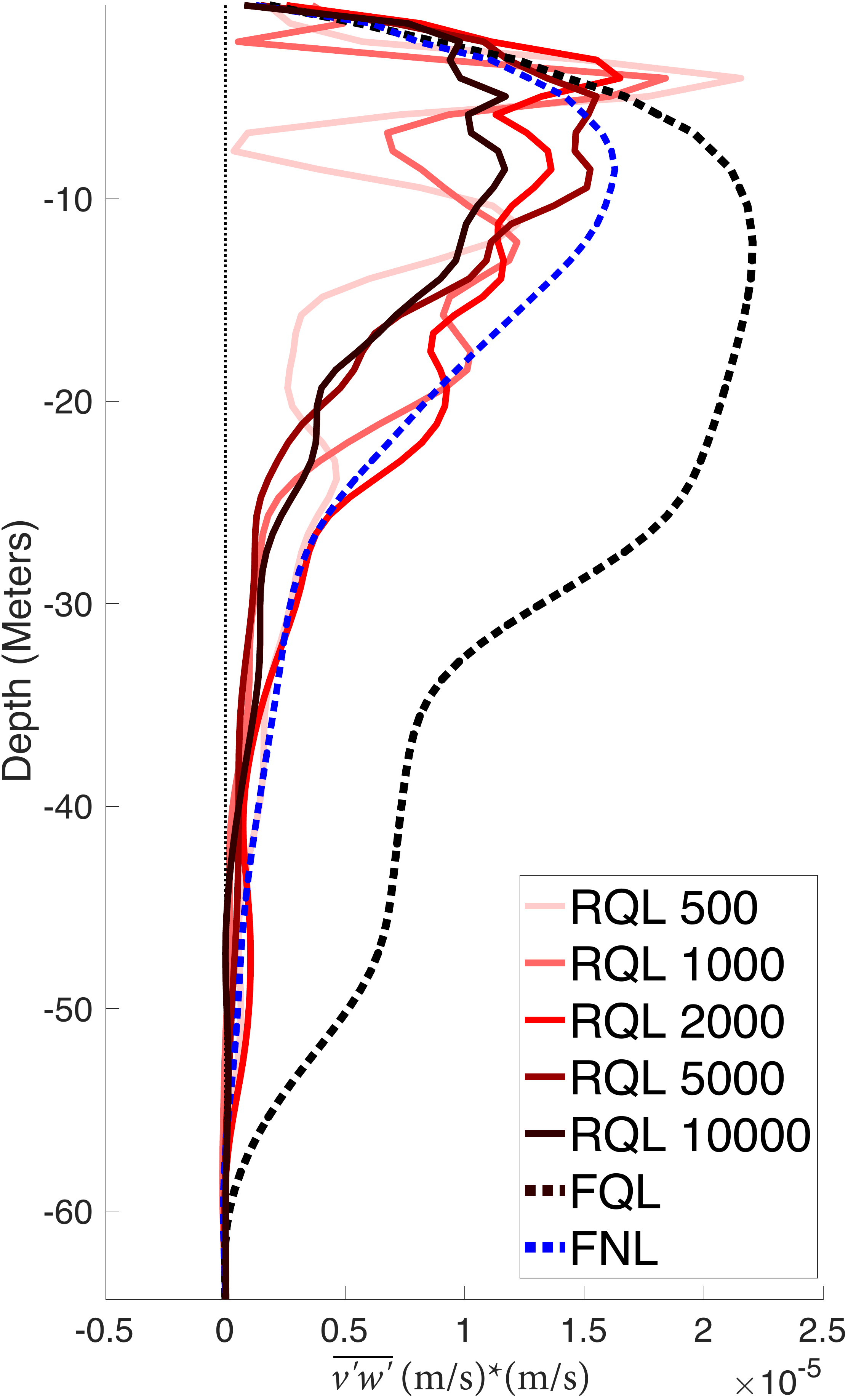}
\end{subfigure}
\begin{subfigure}{0.30\textwidth}
\centering
\includegraphics[width=\textwidth]{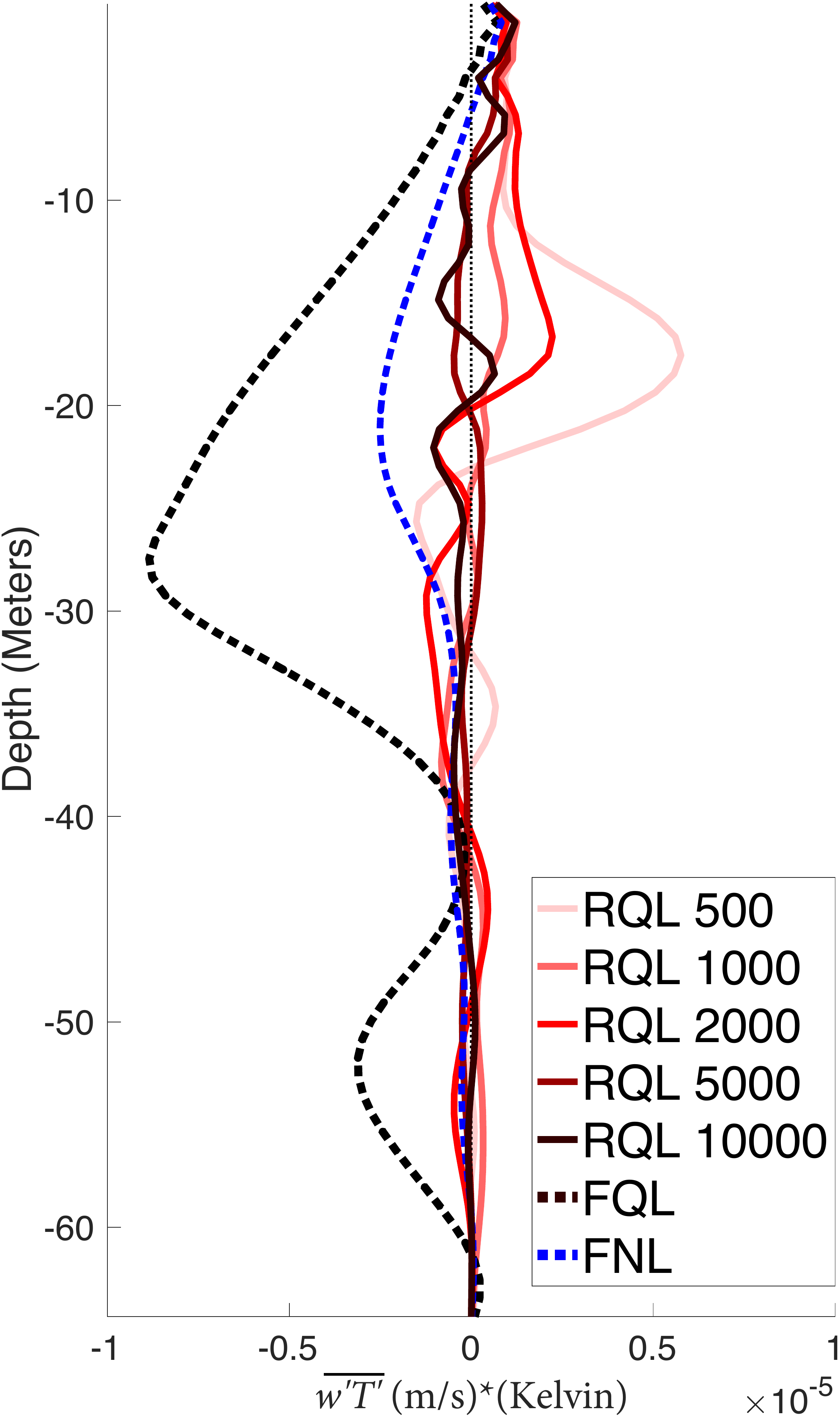}
\end{subfigure}

\begin{subfigure}{0.30\textwidth}
\centering
\includegraphics[width=\textwidth]{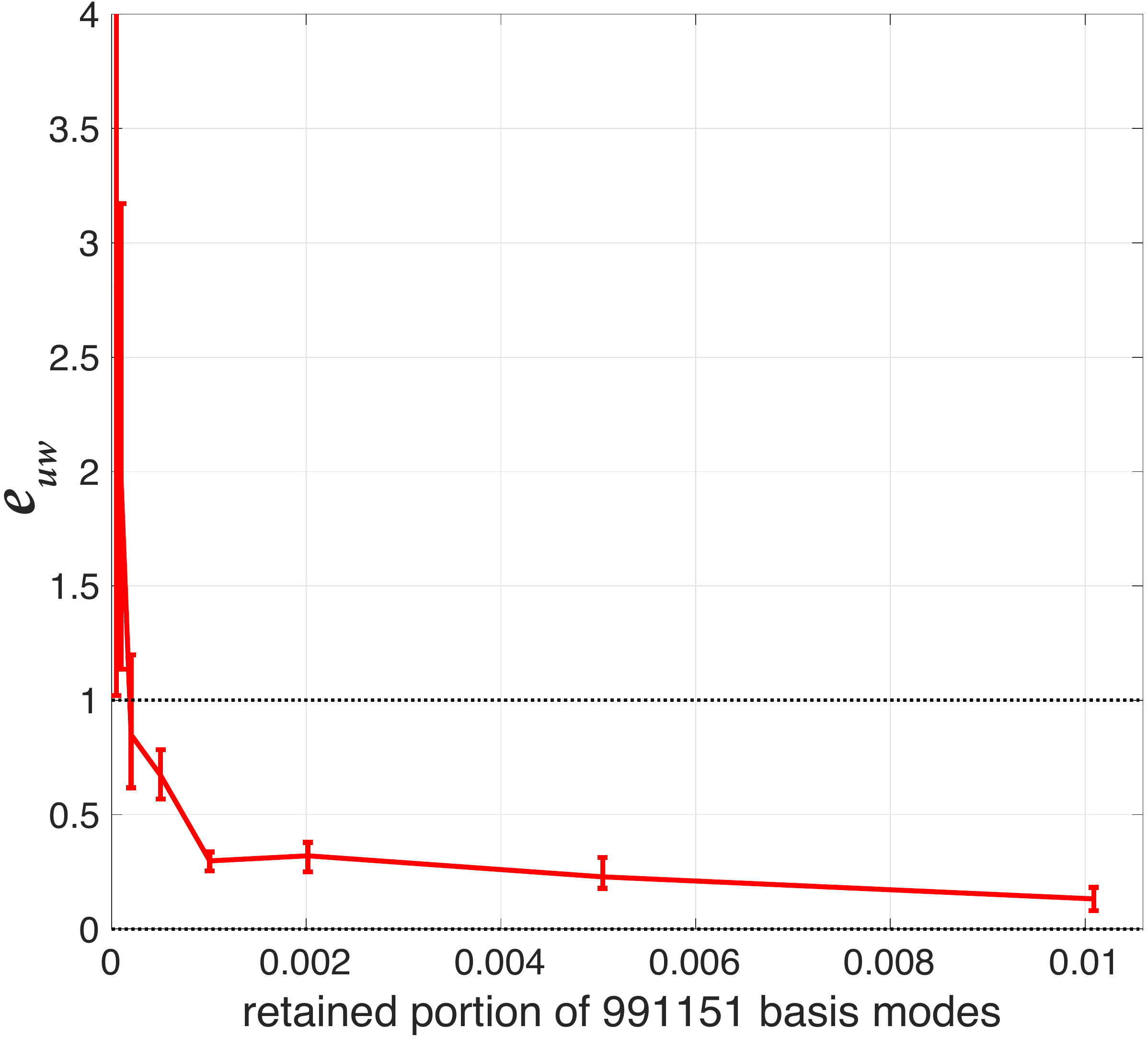}
\caption{\normalsize $\overline{u'w'}$}
\label{fig:langmuir_uw_error}
\end{subfigure}
\begin{subfigure}{0.30\textwidth}
\centering
\includegraphics[width=\textwidth]{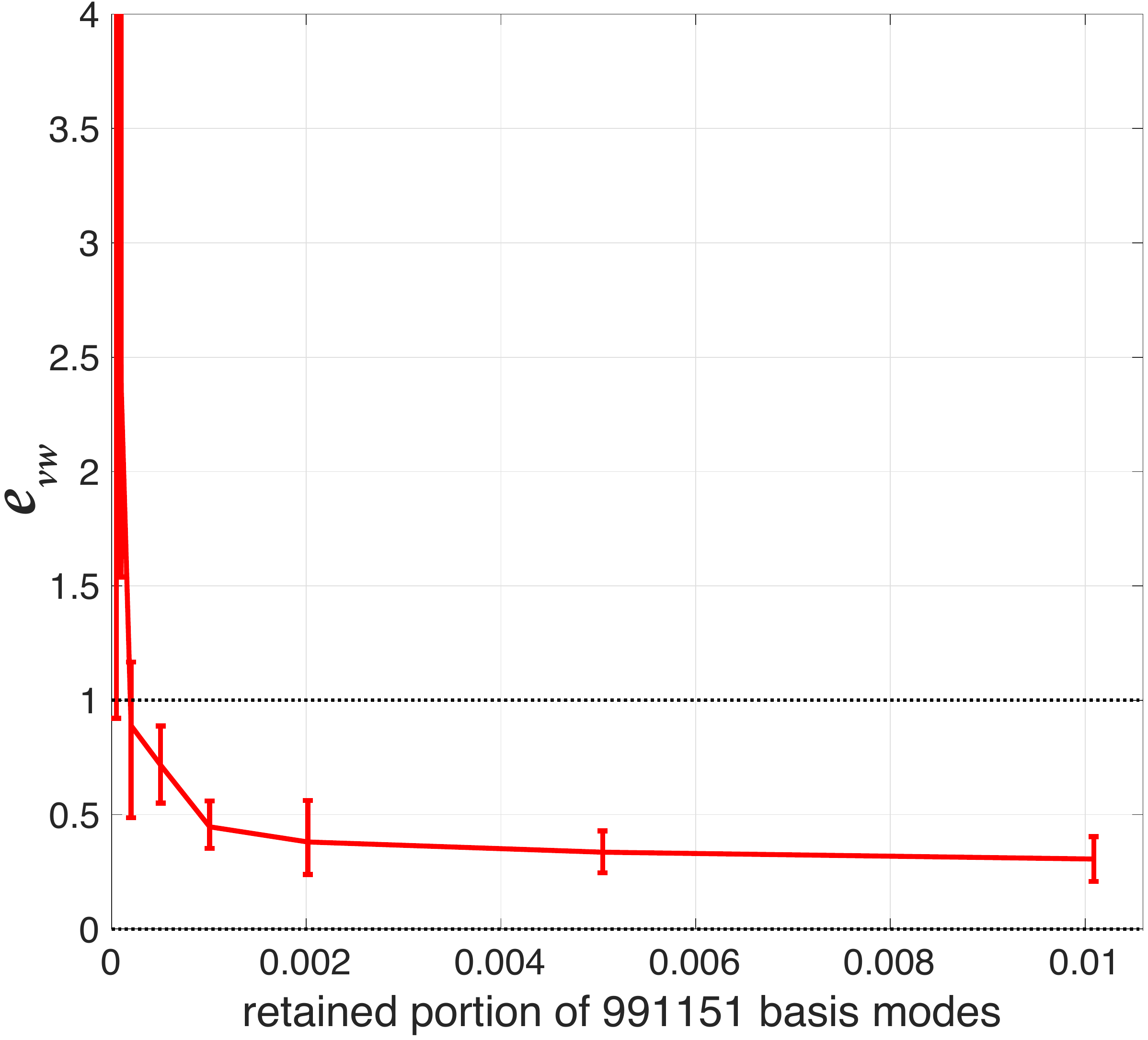}
\caption{\normalsize $\overline{v'w'}$}
\label{fig:langmuir_vw_error}
\end{subfigure}
\begin{subfigure}{0.30\textwidth}
\centering
\includegraphics[width=\textwidth]{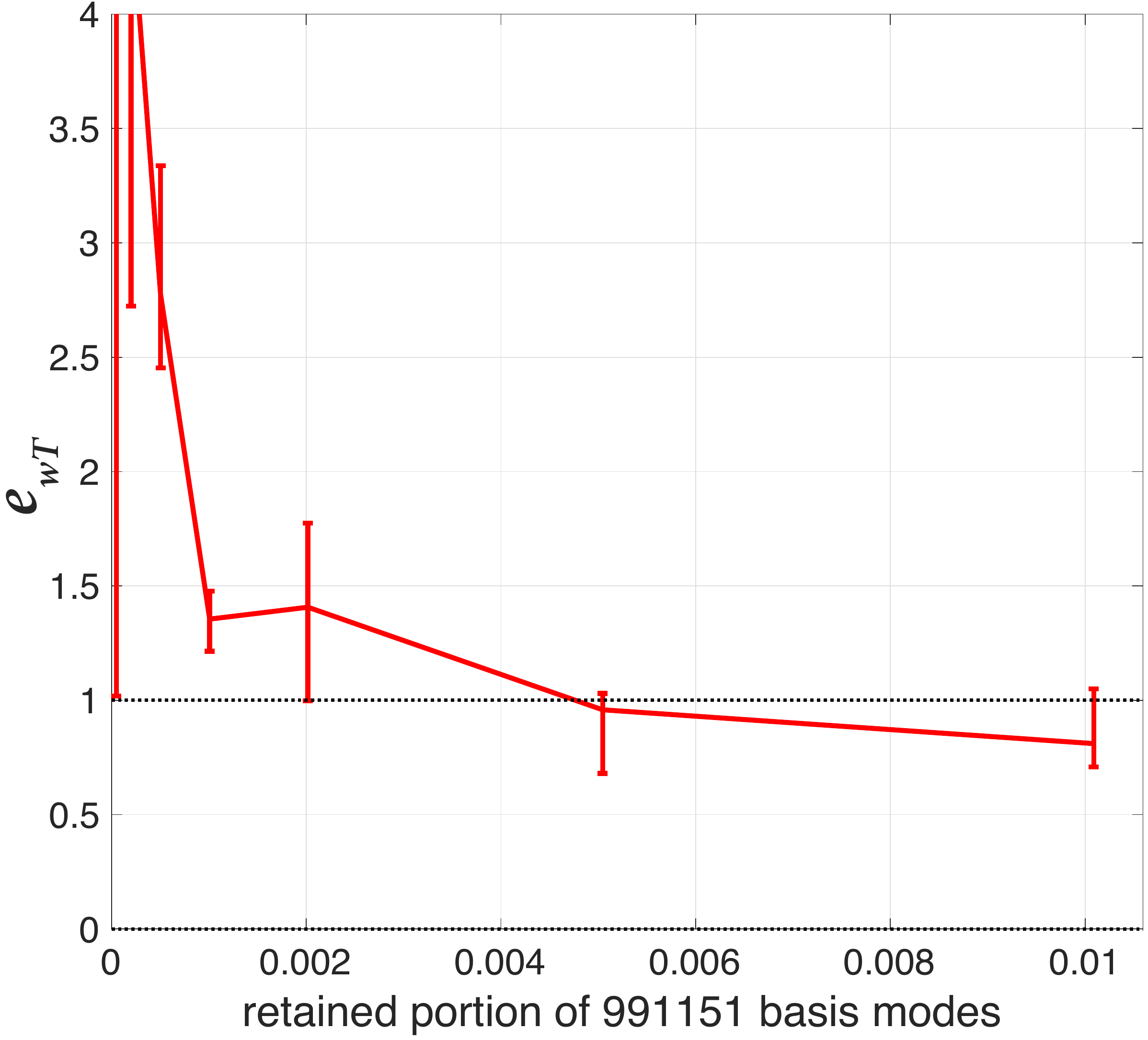}
\caption{\normalsize $\overline{w'T'}$}
\label{fig:langmuir_wT_error}
\end{subfigure}
\centering \caption{Top Row: Vertical profiles of turbulent transports
  and energies of the Langmuir-turbulence cases at $t = 36$ hours.
  Bottom Row: Vertical-profile error, $e_{ij}$, defined in equation
  \ref{eq:error_def}, vs retained basis size for different statistical
  profiles of the Langmuir-turbulence cases.  Note that $e_{ij}$
  accounts for the profiles that are evenly sampled throughout the
  flow evolution.}
\label{fig:langmuir_stat_2}
\end{figure*}

\begin{figure*}[h!]
\centering
\begin{subfigure}{0.4\textwidth}
\centering
\includegraphics[width=\textwidth]{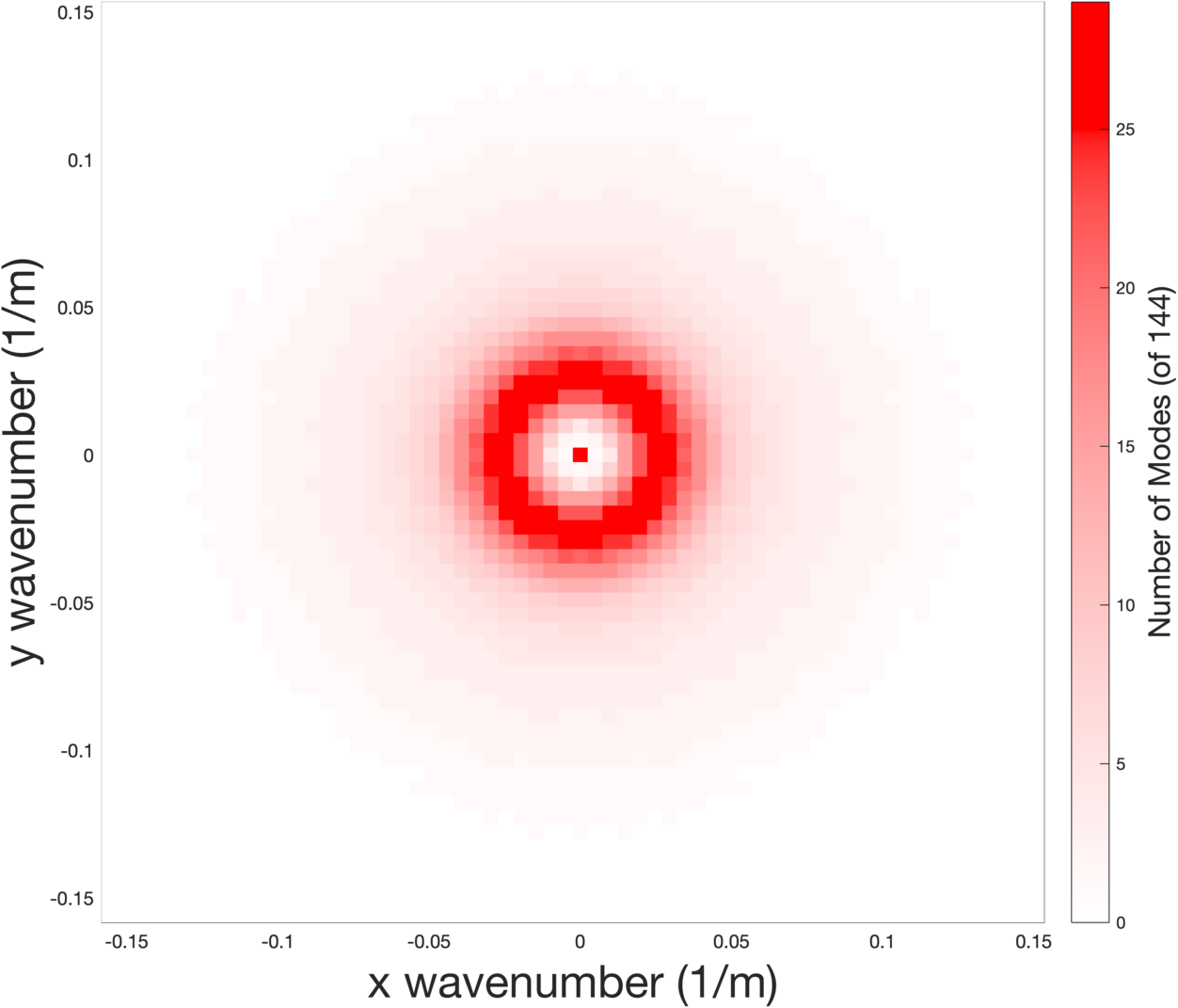}
\label{fig:thermal_T}
\end{subfigure}
\begin{subfigure}{0.4\textwidth}
\centering
\includegraphics[width=\textwidth]{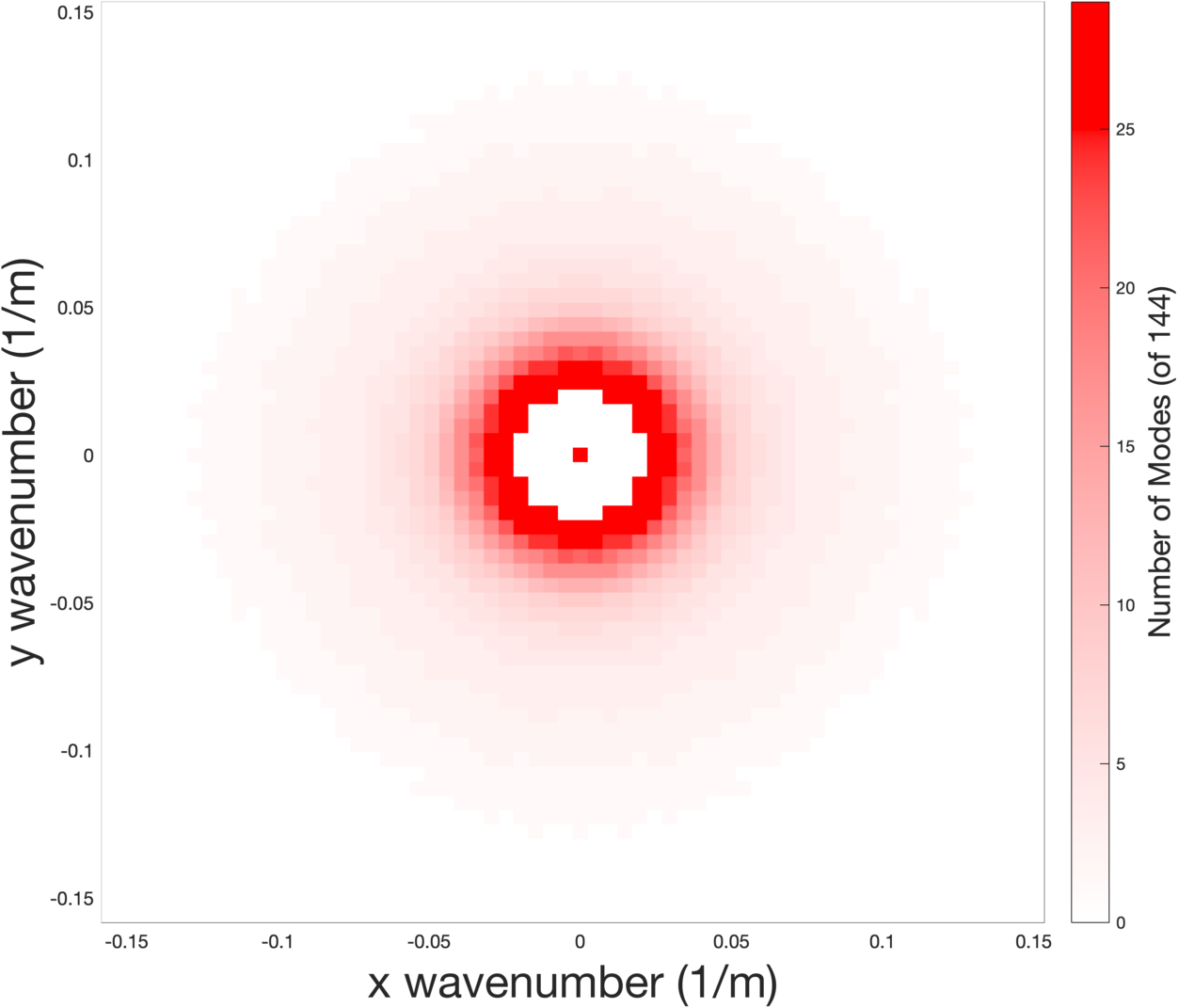}
\label{fig:thermal_T}
\end{subfigure}
\centering \caption{The horizontal Fourier mode density within the 5000-mode truncated basis: (left) the regular POD truncation (right) the "modified" version of this basis in which horizontal Fourier modes with less than 85\% of the peak representation mode and with larger horizontal scales than that mode are omitted.  The omitted large-scale modes are associated with plumes that extend beyond the mixed-layer early in the flow evolution.}
\label{fig:basis_trim}
\end{figure*}

\begin{figure*}[]
\centering
\begin{subfigure}{0.30\textwidth}
\centering
\includegraphics[width=\textwidth]{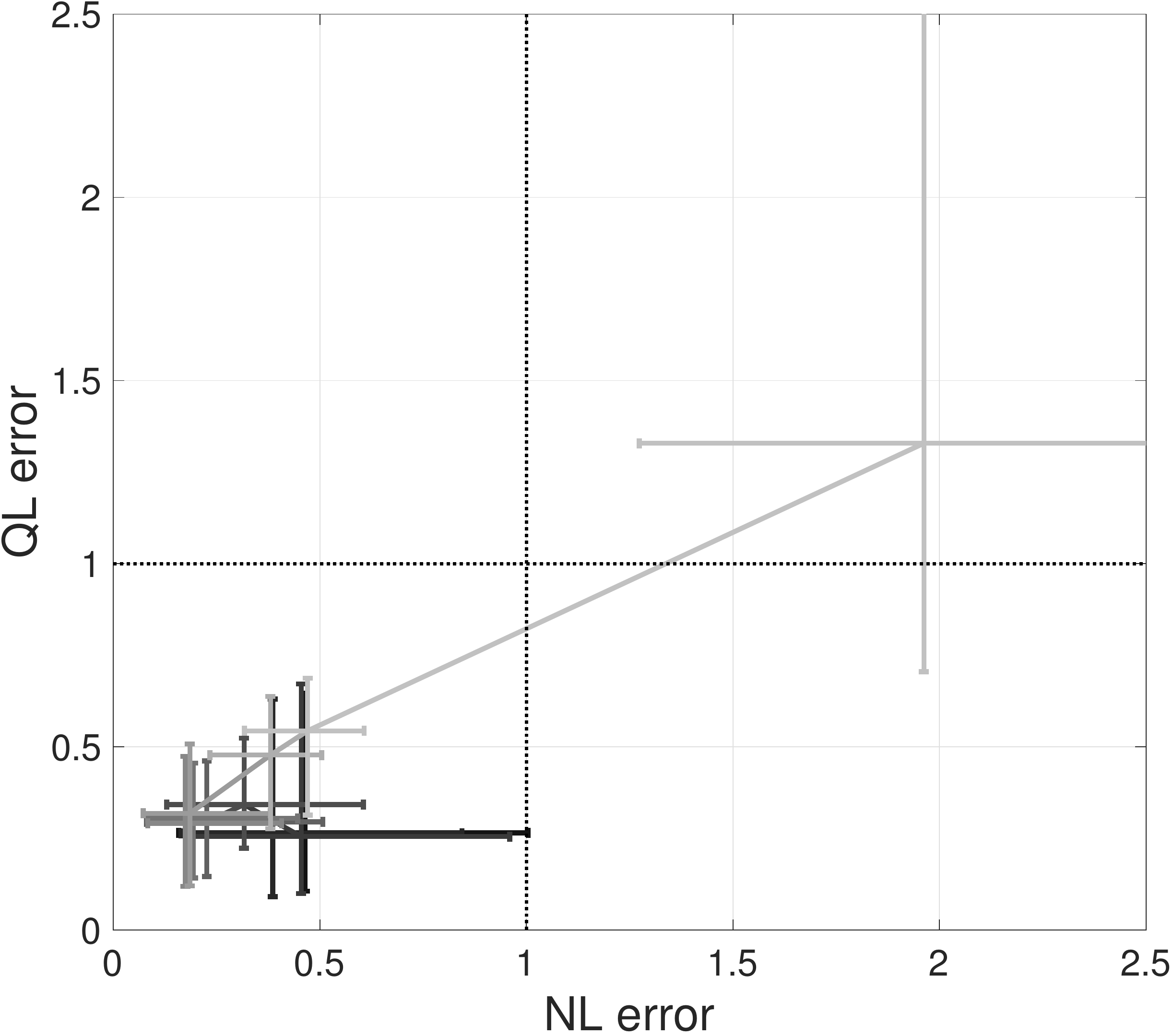}
\label{fig:small_13_nl_ql}
\end{subfigure}
\begin{subfigure}{0.30\textwidth}
\centering
\includegraphics[width=\textwidth]{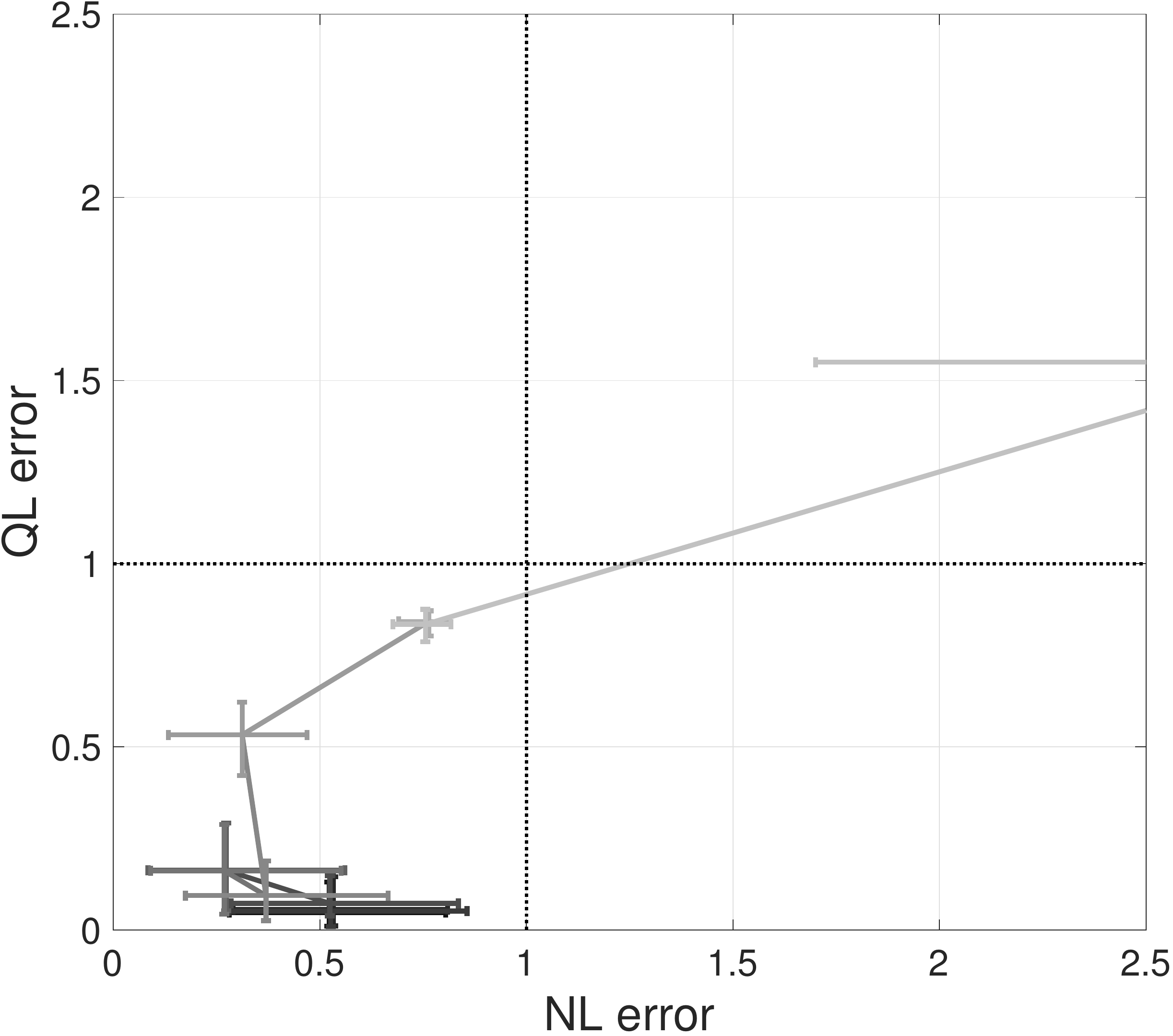}
\label{fig:small_33_nl_ql}
\end{subfigure}
\begin{subfigure}{0.30\textwidth}
\centering
\includegraphics[width=\textwidth]{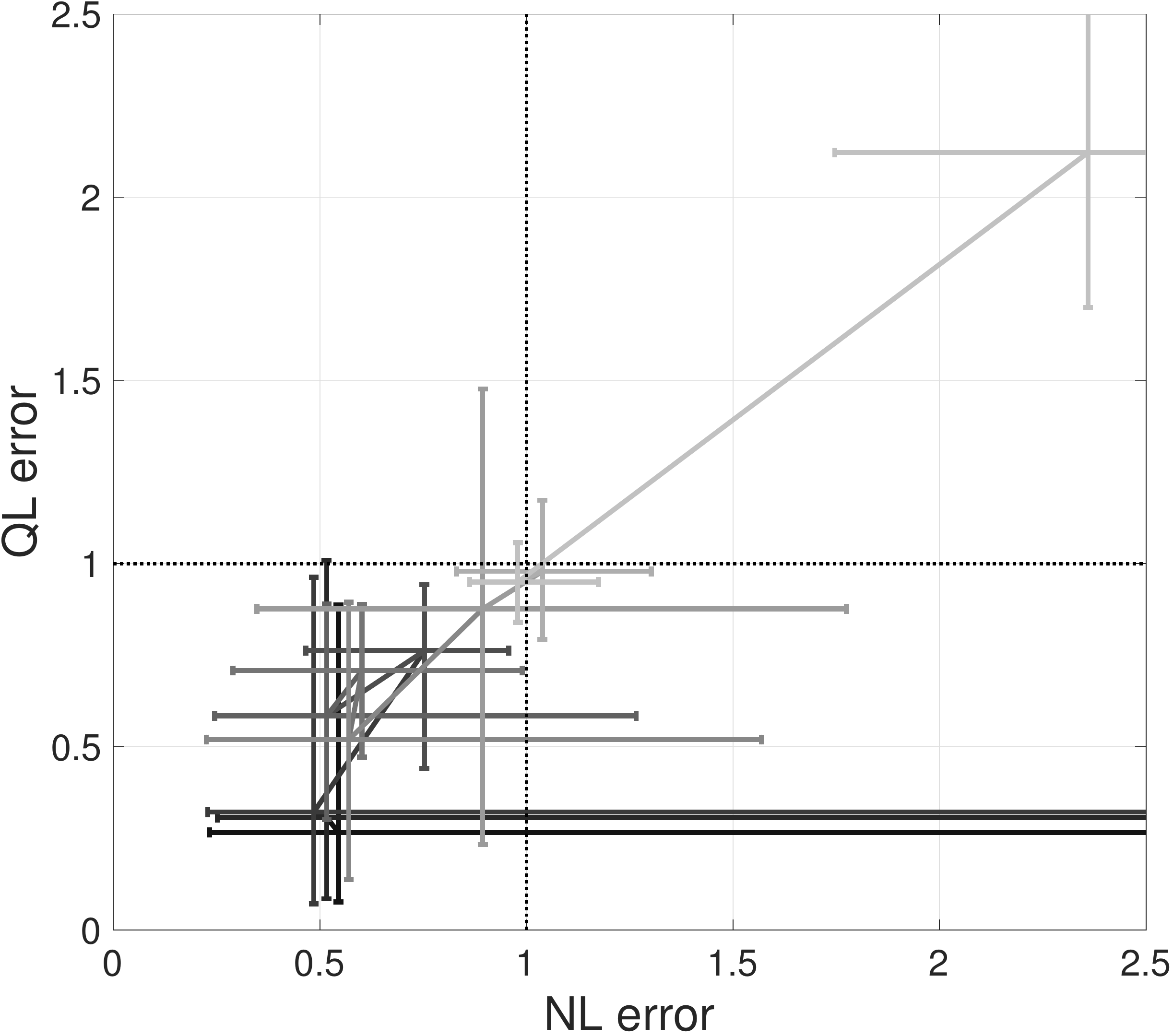}
\label{fig:small_23_nl_ql}
\end{subfigure}

\begin{subfigure}{0.30\textwidth}
\centering
\includegraphics[width=\textwidth]{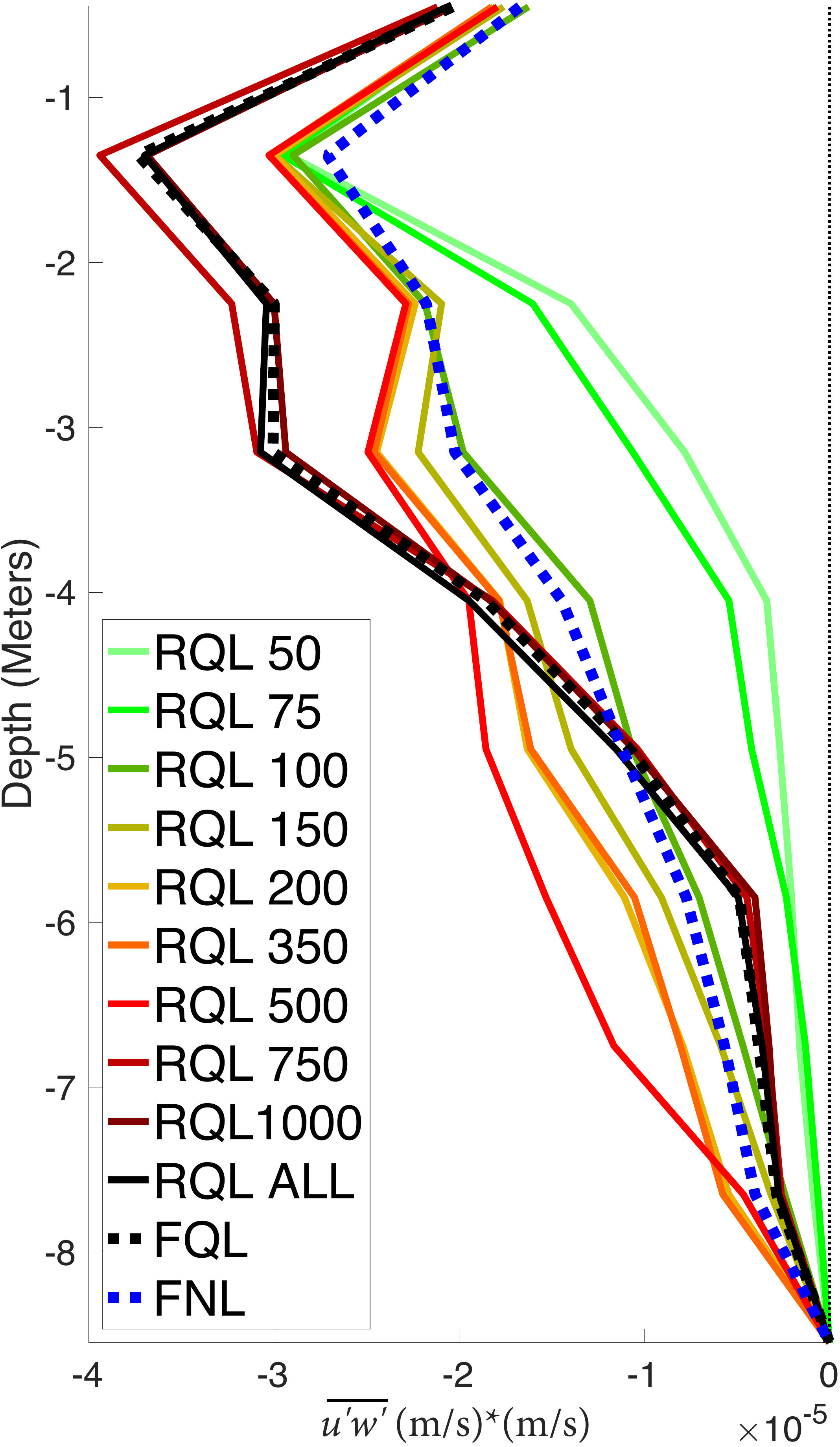}
\caption{\normalsize $\overline{u'w'}$}
\label{fig:small_13_profiles}
\end{subfigure}
\begin{subfigure}{0.30\textwidth}
\centering
\includegraphics[width=\textwidth]{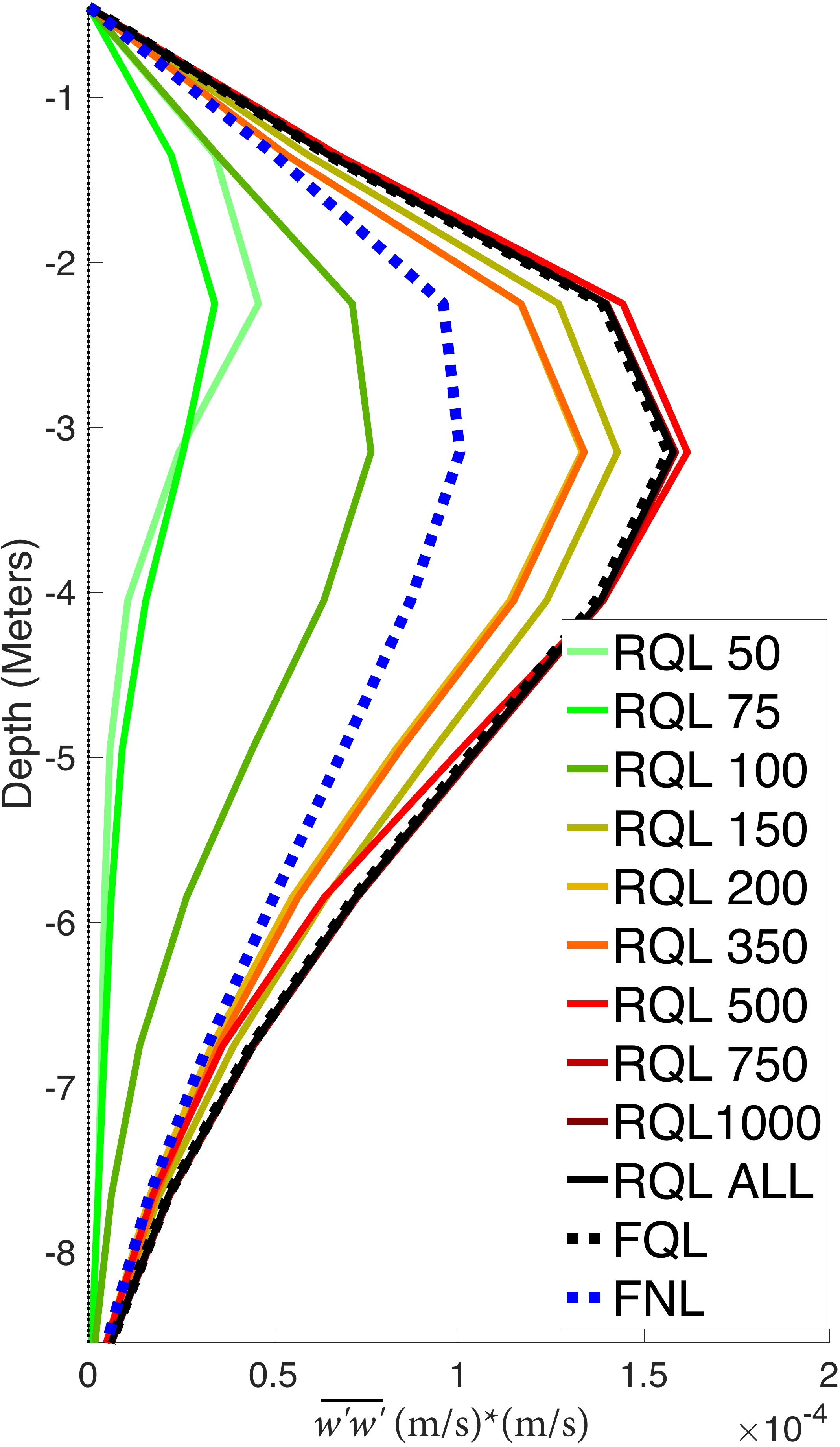}
\caption{\normalsize $\overline{w'w'}$}
\label{fig:small_33_profiles}
\end{subfigure}
\begin{subfigure}{0.30\textwidth}
\centering
\includegraphics[width=\textwidth]{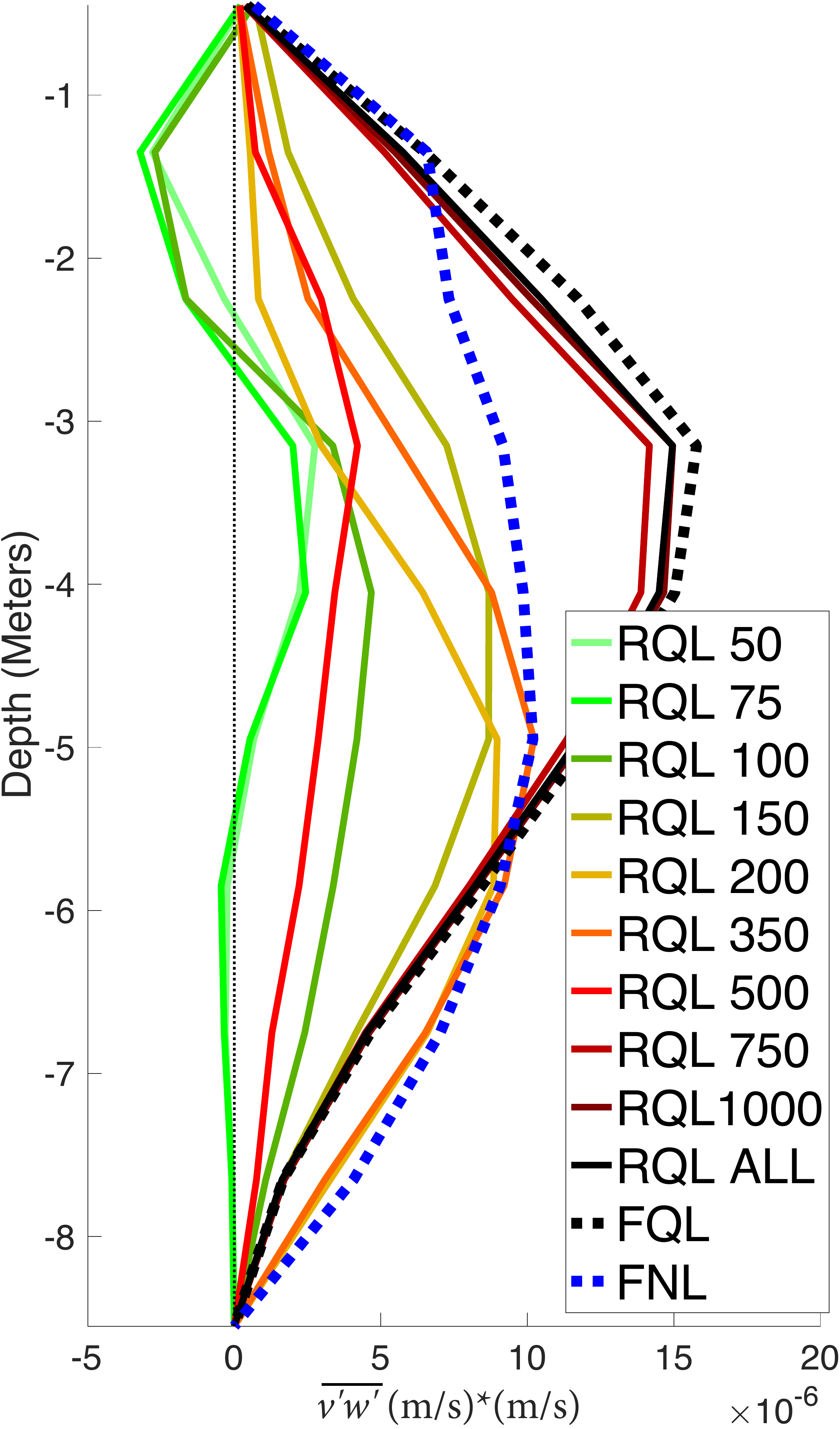}
\caption{\normalsize $\overline{v'w'}$}
\label{fig:small_23_profiles}
\end{subfigure}
\centering \caption{For a 10 $\times$ 10 $\times$ 10 miniature version
  of the Langmuir-turbulence model. Top Row: error of RQL turbulent
  transport and energy vertical profiles in representing FNL vs FQL
  profiles.  Darker lines indicate greater basis retention, with black
  being complete representation.  Bottom Row: turbulent statistical
  profiles as a function of depth for different truncated basis sizes.
  Some of the RQL turbulent profiles, such as $\overline{u'w'}$ after
  120 minutes of flow evolution, are more representative of FNL than
  FQL, something observed in the larger Langmuir-turbulence case at
  incomplete basis representation (see figures
  \ref{fig:langmuir_stat_1} and \ref{fig:langmuir_stat_2}.)  Other RQL
  turbulent transports and energies, such as $\overline{w'w'}$ at 90
  minutes, sweep through the space of accessible profiles and appear
  to represent the FNL profiles well at certain retention rates
  seemingly by chance.  Finally, other RQL statistical profiles, such
  as $\overline{v'w'}$ at 120 minutes, do not approximate the FNL
  profiles better than the FQL profiles at any basis size.}
\label{fig:small}
\end{figure*}


\section{Results}
\label{sec:results}

\subsection{Full-Basis Results}
\label{sec:full_basis_results}

Figure \ref{fig:flow_fields} shows horizontal cross sections of specific instances of the vertical velocity field of both the thermal-convection case (TC) and the Langmuir-turbulence case (LT) for FNL, FQL, and various truncations of RQL.  Figure \ref{fig:mode_density} compliments this with ensemble-averaged horizontal Fourier spectra of the flow fields taken at the same times.  The full-basis results in the top two rows of these figures will first be addressed.  

For TC vertical-velocity fields (top-left of figure \ref{fig:flow_fields}), FQL has different horizontal scales and shapes than FNL.  FNL has geometric patterns of upwelling delineated by relatively intense, spoke-like plumes.  FQL, on the other hand, has very regular, horizontally wave-like plumes that are almost homogeneous in the vertical direction within the mixed layer.   The amplitude of the velocities is also observed to be substantially exaggerated in the FQL case.   Additionally, there is an apparent sign-reversal symmetry within the vertical velocity field of FQL.  This is because the Fourier modes have phases that are independent of one another, as described in section \ref{sec:theory}.  For a given wavenumber amplitude, there are many such independent Fourier modes on or near an annulus in wavenumber space of that radius; these must be approximately equally excited due to horizontal isotropy, creating an apparent sign-inversion symmetry within a single snapshot of the flow. 

For LT FNL vertical-velocity fields (top-right of figure \ref{fig:flow_fields}), alternating bands of upwelling and downwelling are observed, broken up by small-scale turbulent features.  Like with TC, although less noticeably, the downwelling part of the structures are narrower and more concentrated.  The rolls have been rotated away from the direction of the wind and waves by the Coriolis force; the angle of rotation is dependent on depth.  FQL looks a bit like it could be a filtered version of the FNL solution (unlike TC); however, again, there is an apparent sign-inversion symmetry that is not present in FNL.  As with TC, the vertical velocity field is overly strong, likely due to a lack of small-scale turbulence acting on the primary instability.  

The Fourier spectra in the top two rows of figure \ref{fig:mode_density} reveal that in FQL, for both cases, the modes that are populated are near those which are the most energetic in FNL.  Outside of this concentrated region of wavenumbers the energy falloff is steep.  This reflects the lack of a cascade to spectrally local modes.  

The top row of figure \ref{fig:mean_stats} shows vertical profiles of horizontal and ensemble-averaged mean fields for FNL and FQL, (as well as several truncations of RQL, to be discussed in section \ref{sec:results}\ref{sec:reduced_basis_statistics}).  FQL is able to reproduce the FNL fairly well.  The horizontal mean fields are particularly important in the quasilinear approximation as they are the only way turbulent energy can move among horizontal Fourier modes, so it is expected that the rough agreement between FQL and FNL mean profiles is an important foundation for making further approximations in the form of model reduction. 

In contrast to the mean-field profiles (figure \ref{fig:mean_stats}), the second-order FQL profiles (figures \ref{fig:thermal_stat}, \ref{fig:langmuir_stat_1}, and \ref{fig:langmuir_stat_2}) generally do not exhibit good agreement with the FNL profiles.  For the thermal-convection case, the only nonlinear profile that is well represented by FQL is the vertical turbulent temperature transport, $\overline{w'T'}$, figure \ref{fig:thermal_wT_error}.  This is likely because the transport of temperature is what is driving the flow; a mean-field temperature anomaly builds at the surface due to the prescribed heat flux and is preferentially transported downwards due to the eddy field.  This term dominates the turbulent-kinetic-energy production, which is also accurately represented although not depicted.  Note that other off-diagonal Reynolds stresses are exactly zero in the limit of a large ensemble average because of the horizontal isotropy of this case, so there are relatively few non-trivial second-order vertical statistical profiles to observe in TC.  For LT, none of the second-order FQL profiles are well-represented.

\subsection{Reduced-Model Fields and Spectra}
\label{sec:reduced_basis_fields_and_spectra}

Reduced-basis fields and horizontal Fourier spectra for TC and LT are provided in figures \ref{fig:flow_fields} and \ref{fig:mode_density}, respectively, along with those of the full bases.    At a glance, the behavior of TC and LT are substantially different under the model reduction. In the former case, modes are populated at slightly smaller wavenumbers than FQL while in LT at truncations larger than 50, energy tends to be in much larger wavenumbers than FQL.  The RQL flow solution and statistics must eventually converge upon those of FQL as the basis size is further increased.  The convergence of the Fourier energy footprint is therefore nonuniform, where smaller basis retentions can actually look more like FQL than the larger basis retentions.  It will turn out to be the case that this nonuniform convergence can be advantageous in that it allows for better representation of the FNL statistical profiles.   The nonuniform convergence of the RQL flow statistics on those of FQL is explored more thoroughly in a small test case in section \ref{sec:results}\ref{sec:small}. 

In TC RQL 50, the observed rows of plumes are due to an insufficient number of retained eddy-modes on the dominant wavenumber (see bottom left of figure \ref{fig:mode_density}).  The basis was selected in a manner that observes discrete horizontal isotropy exactly, so the non-zero modes show an eight-fold symmetry; however, with fifty modes, the truncation is so severe that the annulus is not complete and modes along the axes are omitted.  Because so few modes are excitable near this wavenumber, their amplitudes do not average out to form smooth structures that reflect the symmetry of the problem, rather, one or two may randomly dominate at a given time, giving rise in some instances to rows, in others to regular triangular patterns.  Despite the qualitative shortcomings, the vertical profile resembles the FQL solution, exaggerating the plume depth slightly.  In LT RQL 50, the breaking of horizontal isotropy allows for 50 modes to better capture the essential shape of the windrows than with TC; however, the velocity field is too strong and penetrates through the stratified layer.  

At 500 modes, both TC and LT show improvement in the vertical cross sections of the flow fields (figure \ref{fig:flow_fields}), a reflection of the larger number of accessible modes, as seen in the 2D Fourier spectrum in figure \ref{fig:mode_density}.  The horizontal cross sections of both cases reflect RQL's shift to different horizontal scales compared with FQL; in TC RQL 500, the solution has similar horizontal structure as FQL except with slightly larger length scales, while in LT RQL 500, the windrows are much more narrow.   This trend is more pronounced at 5000 modes.  The displacement of the excitation of horizontal Fourier modes at intermediate basis truncations in RQL is explored more thoroughly in section \ref{sec:results}\ref{sec:displacement}.

\subsection{Reduced-Model Statistical Profiles}
\label{sec:reduced_basis_statistics}

The mean-field profiles are depicted in figure \ref{fig:mean_stats} along with their representation error:
\begin{equation}
e_{i} = \frac{\left(\left<\overline{q}_{i, FNL}\right> - \left<\overline{q}_{i, RQL}\right>\right) \cdot \left<\overline{q}_{i, FNL}\right>}{\left<\overline{q}_{i, FNL}\right> \cdot \left<\overline{q}_{i, FNL}\right>} \label{eq:mean_error_def}
\end{equation}
Here, the mean field is $\left<\overline{q}\left(z,t\right)\right>$, $\left<\right>$ indicates an ensemble average, the subscript runs over the fields $\left\{u,v,w,T\right\}$, $t$ runs over four evenly sampled times throughout the flow evolution, $z$ runs over the vertical position, and the dot product is defined over both $z$ and $t$.  The top-rows of figures \ref{fig:thermal_stat}, \ref{fig:langmuir_stat_1}, and \ref{fig:langmuir_stat_2} depict vertical profiles of horizontal and ensemble averaged turbulent transports and energy terms, while the bottom rows show the RQL model's representation error, $e_{ij}$ of the FNL statistical profiles, computed as a normalized inner product of the Reynolds stresses $\left<\overline{q_i q_j}\right> = \boldsymbol{\tau}_{ij} = \tau_{ij}\left(z,t\right) $:
\begin{equation}
e_{ij} = \frac{\left(\boldsymbol{\tau}_{ij, FNL} - \boldsymbol{\tau}_{ij, RQL}\right) \cdot \boldsymbol{\tau}_{ij, FNL}}{\boldsymbol{\tau}_{ij, FNL} \cdot \boldsymbol{\tau}_{ij, FNL}} \label{eq:error_def}
\end{equation}
Note that errors less than $0.3$ usually indicate a good qualitative representation of the vertical profile. 

At small basis sizes (200 and 1000), the mean-field thermal-convection temperature profile, depicted on the left of figure \ref{fig:thermal_mean_error}, as well as the vertical temperature transport, $\overline{w'T'}$ (right of figure \ref{fig:thermal_stat}), are represented well by RQL.  At 5000 modes, however, unrealistically strong and deep profiles of $\overline{w'w'}$ and $\overline{w'T'}$ in figures \ref{fig:thermal_ww_error} and \ref{fig:thermal_wT_error}, respectively, have mixed the domain entirely by the time of 20 hours, demonstrating a deviation from uniform convergence towards the FQL statistics.  An explanation and solution for this anomalous mixing using a modified POD basis is given in section \ref{sec:results}\ref{sec:displacement}.

Unlike the thermal-convection case, the Langmuir-case momentum mean field profiles in figure \ref{fig:langmuir_mean_error} do not exhibit a trend of worse representation at larger basis sizes. The FQL $v$-velocity profile appears qualitatively consistent with the FNL profiles; however, the $u$-velocity component exhibits errors on the order of the magnitude of the FNL mean field (meaning $e_u \approx 1$). Note that for none of the mean-field profiles does RQL outperform FQL at reproducing the FNL mean fields.  This may be unsurprising given the degree of reduction of the model size; however, it will turn out that RQL will sometimes perform better than FQL for second-order profiles. 

The Langmuir-turbulence case, on the other hand, exhibits good representation of several turbulent transport and energy profiles with RQL; however, it requires a larger portion of retained basis modes to do this, about $0.2\%$ of modes, or about a factor of 10 more than the thermal-convection case. Several of the RQL turbulent transport and energy profiles (specifically $\overline{u'u'}$, $\overline{w'w'}$, $\overline{u'v'}$ and $\overline{v'w'}$), appear to initially converge on the FNL profiles (as opposed to FQL), something which was not observed in the mean-field profiles in figure \ref{fig:mean_stats}.  The potential of RQL to reproduce second-order FNL statistics would be a huge boon to sub-grid-scale modeling within the ocean-surface boundary layer, so the mechanism for this convergence is scrutinized in section \ref{sec:results}\ref{sec:displacement}.  It can also be noted that the relatively good performance of RQL at reproducing second-order FNL profiles compared with first-order mean profiles indicates the mean fields may not be as essential as would have been expected given their important role as an energy pathway in the quasilinear approximation.

\subsection{Reduced Basis Excitation Displacement}
\label{sec:displacement}

In order to understand the convergence of RQL on FNL statistics in LT as well as the degradation of performance of RQL TC at intermediate basis truncations, it is helpful to first identify the cause and impact of the change in excited horizontal Fourier wavenumbers that occurs in RQL relative to FQL (see figure \ref{fig:mode_density}).  Recall that relative to FQL, RQL 5000 prefers smaller wavenumbers in TC and quite a bit larger in LT. To probe the mechanism for this spectral excitation displacement, simulations were run to determine the growth rate of quasilinear instabilities around a fixed FQL mean field for different basis truncations. Based on these numerical experiments, the peak quasilinear instability is displaced to different horizontal scales because the truncation of these excited horizontal modes prevents the redistribution of energy to different vertical profiles within that horizontal mode. This then prevents the draining of energy that would otherwise occur in the full-basis solution. In the LT fixed-mean-field experiments, high-wavenumber modes are driven by the linear Stokes instability. In the full-basis flow solution, the vertical profile of this mode evolves through an interaction with the mean field, and backscatter and vertical viscosity both act to dissipate this energy such that the mode never becomes substantially energized. In the 5000-mode version, however, the representation of these horizontal Fourier modes is limited and energy is not able to vertically redistribute in order to drain through these mechanisms. Similarly, in TC the smaller-wavenumber modes are excited via scattering off of the mean field and efficiently dissipated via backscatter, viscosity, and thermal diffusion in the full-basis solution, with the truncated basis (1000+) suppressing this pathway.  

Having identified the cause of these shifted peak excitations to be an insufficient representation of the associated horizontal Fourier modes, next consider their impact on the mean-field and vertical structure of the flow.  Recall that the POD bases are computed using snapshots of FQL flow solutions.  In FQL, boundary layer structures are simple, spectrally focused, and uniform throughout the mixed layer (relative to FNL).  The horizontal dimension of these flow structures scales monotonically with the mixed-layer depth.  When the basis truncation of a horizontal Fourier mode is very limited, such as at the horizontal wavenumbers of the displaced peak excitations in LT and TC, the included vertical modes will allow for only a small variation of mixed-layer depths around the preferred depth of the original FQL solution.  Because of this, in LT the displaced peak-excitation Fourier modes of a high-wavenumber create structures of a shallower profile than the overall MLD (which is controlled by some of the less energetic and smaller wavenumber modes).  The interior of the mixed layer becomes structurally less homogeneous, and therefore more FNL-like (bottom-right of figure \ref{fig:flow_fields}).  For TC, the displaced peak-excitation horizontal Fourier modes artificially push the mixed-layer beyond its correct size, causing the RQL profiles to be a much worse representation of the FQL results (Fig.~\ref{fig:thermal_stat}).  

The reason that TC and not LT exhibits displaced mode excitation at larger scales is because in TC the representation of small-wavenumber modes gradually trails off at small wavenumbers in the FQL solution while for LT they are continuously represented up to a sharp cutoff, missing only a single discrete wavenumber next to the mean field (see figure \ref{fig:mode_density}.)  A means of addressing the inaccuracies associated with large-scale displaced RQL mode excitation while preserving the benefits of the small-scale modes is to allow the reduced-basis solver to use only those partially represented modes that are approximately of equal or larger horizontal wavenumber than the peak excitations. This basis modification could be efficient to implement because it can take effect at runtime, which stands in contrast to modifications to the POD formulation, such as different energy scalings of the temperature, which must be configured in advance of basis determination.

To test this idea, the reduced basis was further truncated at poorly represented wavenumbers that are smaller than the modes of peak representation (see figure \ref{fig:basis_trim}). Poorly represented is taken as 85\% or less of the peak represented mode. This was done with the understanding that a dynamically adaptive solver would perform as well or better than this test with similar computational expense. The results of this test, shown in figures \ref{fig:mean_stats} and \ref{fig:thermal_stat}, show that the technique can avoid the most severe inaccuracies of the truncated large-scale horizontal modes and also preserve the approximately accurate evolution of the deepening mixed layer.

\subsection{Miniature Langmuir-Case Convergence Study}
\label{sec:small}

In order to examine more closely the convergence properties of RQL on FQL, a miniature test system containing 10 $\times$ 10 $\times$ 10 grid cells was simulated that could be run using the Galerkin Projection of the equations of motion all the way up to complete resolution of the underlying basis.  The system parameters are exactly the same as the Langmuir turbulence case (see section \ref{sec:langmuir_system}), with the exception of the mixed-layer depth, which is initialized at 5 meters.  The grid spacing is the same in both cases, meaning the domain only extends to 9 meters depth instead of 72 meters depth for the larger case.  The resulting flow field is far from realistic and the mixed layer quickly reaches the bottom of the domain; however, the nonlinear and quasilinear flow solutions do have substantially different vertical statistical profiles prior to that, during the first 3 hours of the flow evolution.  384 ensemble members were run of FNL, FQL, and RQL; the simulations were run until 10 hours into the flow evolution. 

The miniature test system indicates that for some turbulent transport profiles, the RQL model trends towards the FNL profiles at intermediate basis sizes, for instance, the vertical turbulent x-component velocity transport $\overline{u'w'}$ (left side of figure \ref{fig:small}).  This quantity is responsible for transporting much of the turbulent energy in the flow introduced near the surface to greater depths and Monin-obukhov similarity theory constrains this quantity heavily in this case.  Perhaps it is for these reasons that $\overline{u'w'}$ is represented relatively well in both the miniature test case and in the larger Langmuir-turbulence case. 

Other RQL turbulent profiles are representative of the FNL profiles at intermediate basis retention because the former incidentally sweeps through the FNL profile as with $\overline{w'w'}$ in the center column of figure \ref{fig:small}.  Finally, some profiles do not exhibit agreement between the FNL and RQL at intermediate basis retention, as in $\overline{v'w'}$ in the right column where the RQL reproduces only part of the FNL profile at a given truncation size. The magnitude of this transport is considerably smaller than $\overline{u'w'}$, perhaps suggesting why it might not be accurately represented.  Many of these turbulent profiles do not exhibit uniform convergence towards the FQL versions.  This can be seen in $\overline{v'w'}$ in the right column of figure \ref{fig:small} most clearly among the three statistical quantities depicted.

Overall, the miniature Langmuir-turbulence test case demonstrates that the intermediate convergence exhibited by many of the statistical profiles in the larger Langmuir case towards the FNL profile (see figures \ref{fig:langmuir_stat_1} and \ref{fig:langmuir_stat_2}) as well as the nonuniform convergence found in all statistical profiles in the thermal-convection case in figure \ref{fig:thermal_stat} occurs in other cases and is not an unusual feature or error.


\section{Discussion}
\label{sec:discussion}

\begin{figure}[]
\centering
\begin{subfigure}{0.45\textwidth}
\centering
\includegraphics[width=\textwidth]{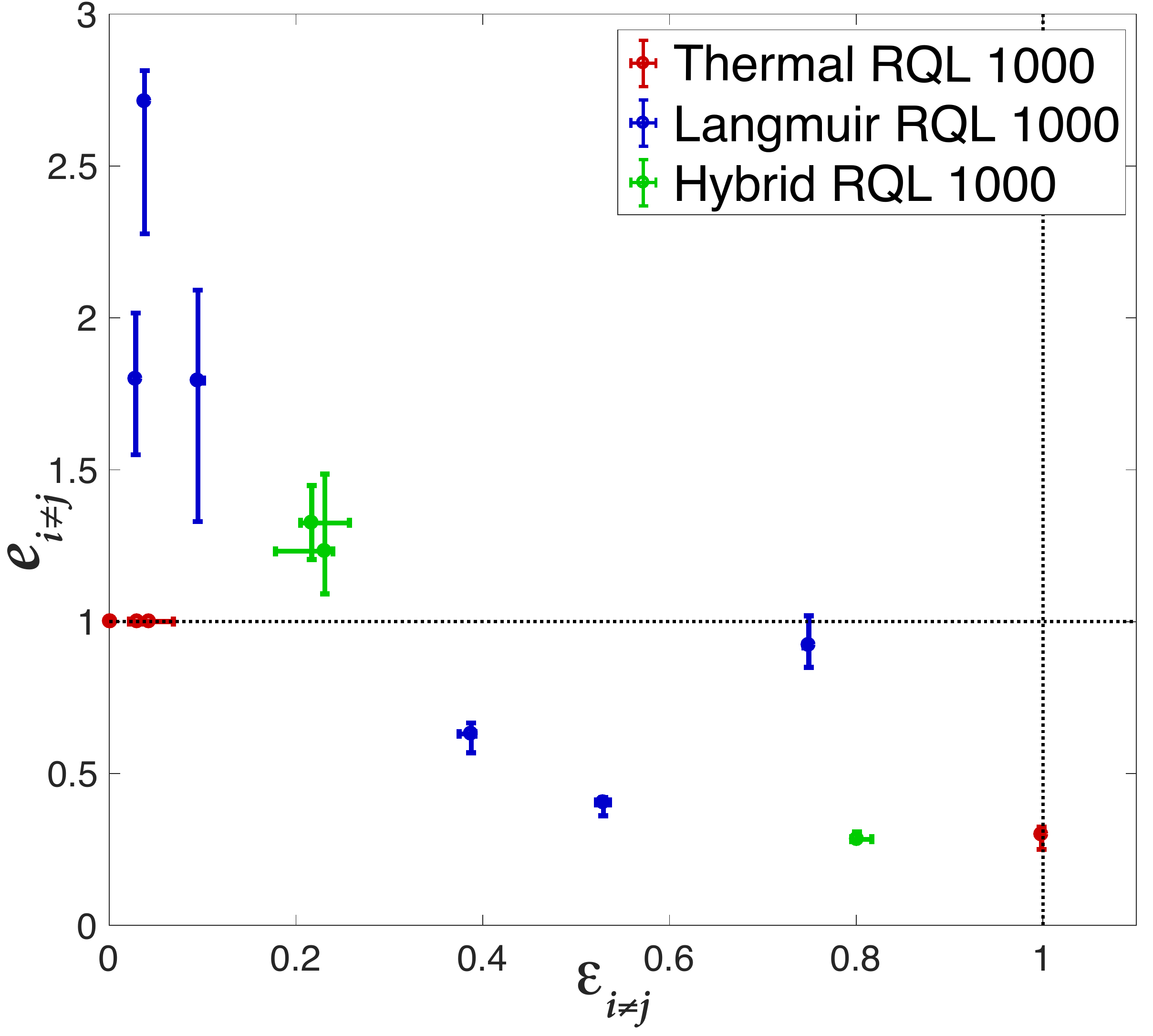}
\label{fig:off_diag_stat_importance}
\end{subfigure}

\begin{subfigure}{0.45\textwidth}
\centering
\includegraphics[width=\textwidth]{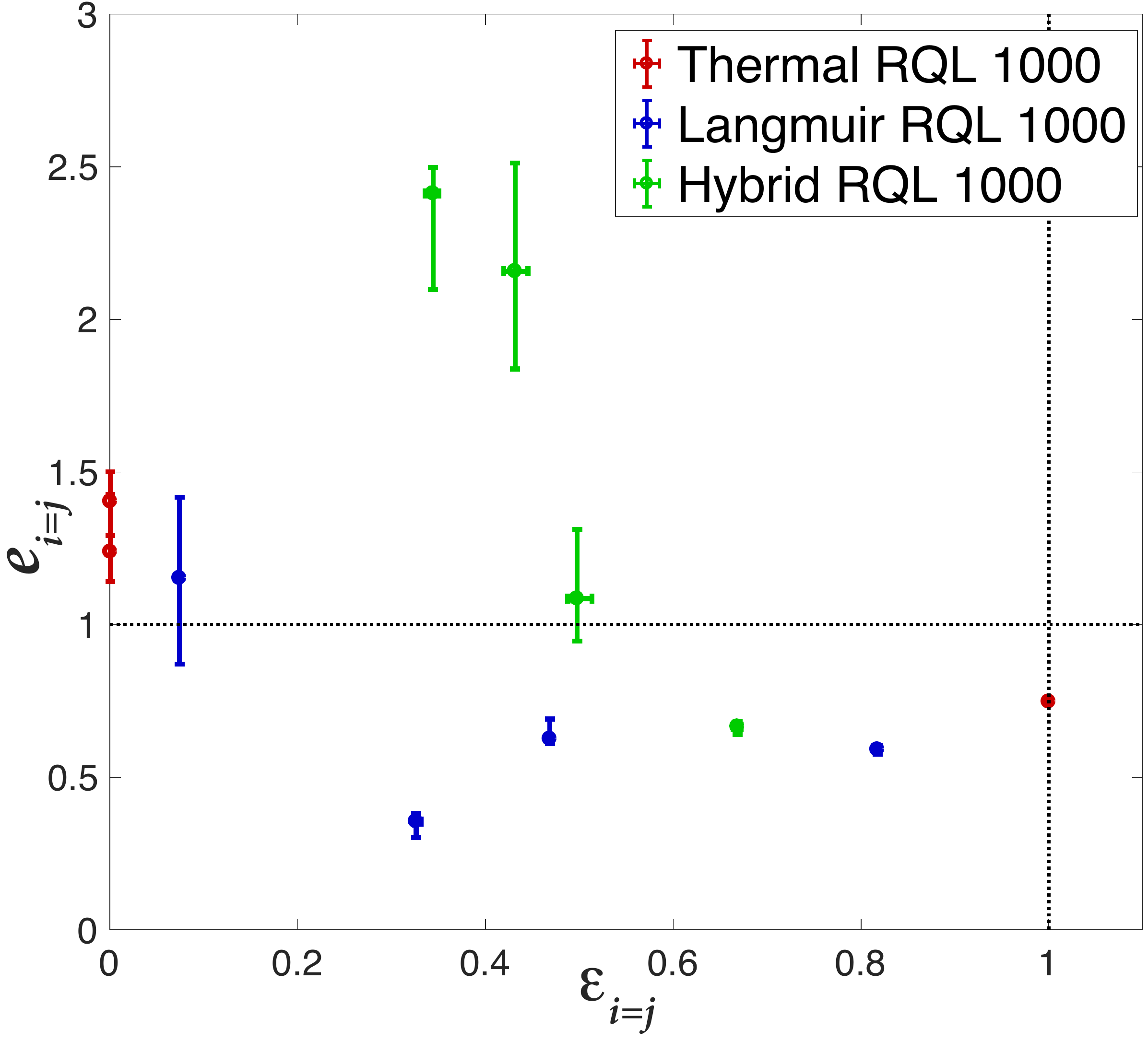}
\label{fig:diag_stat_importance}
\end{subfigure}
\centering \caption{Comparison of second-order turbulent statistical
  profile representation errors (equation \ref{eq:error_def}) with
  their variance-normalized amplitudes (equation \ref{eq:norm_amp})
  from the FNL solutions.  This is done after temperatures have been scaled to have
  units of velocities using the available-potential energy technique
  described in section \ref{sec:basis}.  The ``Hybrid''- case is a
  mixture of the thermal-convection case and the Langmuir-turbulence
  case formulated so that the scaled turbulent temperature fluxes and
  turbulent velocity fluxes are similar in magnitude. When the
  turbulent energies (diagonal terms in the Reynolds-stress tensor, bottom)
  and transports (off-diagonal terms, top) are treated separately,
  variance-normalized amplitudes of these statistics are correlated
  with their representation error.}
\label{fig:stat_importance}
\end{figure}

One takeaway from this work is that some turbulent-transport and
energy profiles are well represented in the reduced models, while
others are not.  The ones that are well represented seem to be those
which are most important for transporting energy from externally
forced boundary conditions into the flow interior.  It would be
useful to be able to predict which of these transports would
be most accurately represented without necessarily referring back to
an ensemble of full-basis simulations.  Monin-Obukhov similarity
theory or \citep{KrausTurner67} budgets might be revealing, but more
simulations with both convective and stress forcing are needed to
evaluate the robustness of this hypothesis.

As an idea for what might indicate the importance of second-order
statistical transports, a 4-dimensional Reynolds-stress tensor was
computed using the temperature scaling described in section
\ref{sec:numerics}\ref{sec:basis} as well as the APE definition in
equation \ref{eq:ape} and normalized using the variances of all the
elements at different depths and times:
\begin{equation}
\epsilon_{ij} = \frac{\sum_{z,t} \overline{q_i'q_j'\left(z,t\right)}^2}{\sum_{\left\{ij\right\}}\sum_{z,t} \overline{q_i'q_j'\left(z,t\right)}^2} \label{eq:norm_amp}
\end{equation} 
where $\left\{ij\right\}$ is the set of pairs of velocity-temperature
components in consideration (energy and transport elements are treated
separately for best results, so this will be either $i=j$ or $i\ne j$).
This was then plotted against the representation error, $e_{ij}$ (see
equation \ref{eq:error_def}), for a variety of different cases.  The
results, shown in figure \ref{fig:stat_importance}, indicate a weakly
negative correlation between the error and the scaled Reynolds stress
amplitude.  Perhaps a more reasoned approach to determining the
``importance'' of these statistics might reveal a stronger
correlation.


\section{Conclusion}
\label{sec:conclusion}

Two cases of developing ocean-surface boundary-layer turbulence were used to study the quasilinear
reduced-model representation of flow statistics.  In concrete terms,
the reduced quasilinear model represented certain second-order
turbulent transport and energy profiles in these cases qualitatively
well (less than 30\% error from the full-basis nonlinear profiles)
using $\mathcal{O}\left(10^{-3}\right)$ retained POD modes.  The
profiles that were represented well appear to be correlated with their role
driving the flow dynamics.  

The case of surface-forced thermal convection in a deepening boundary
layer exhibited nonuniform convergence of the statistical
profiles towards complete representation, causing substantial inaccuracies 
in flow statistics.  This was corrected for by restricting basis representation 
at large scales in a way that can be done dynamically at runtime.  The reduced-basis
quasilinear vertical statistics profiles in the Langmuir-turbulence
case, of which there are many more than in the thermal-convection case
that are non-trivial, exhibited intermediate convergence towards the
full-basis nonlinear profiles before converging on the full-basis
quasilinear profiles.  Both of these properties were observed more
thoroughly in an unrealistic miniature test case of the Langmuir
turbulence. 

What this means in broad terms is that a QL POD reduced model performs
better than would be expected based naively on the isolated
performances of the QL approximation and POD-based reduced-modeling:
not only does QL allow for a smaller problem size and better runtime
scalings than nonlinear POD reduced model; truncation of the QL
approximation can add multi-scale structure missing from the full-basis QL
solution, often producing flow features that resemble those of
full-basis NL more so than full-basis QL. By better understanding
these synergistic effects, it may be possible to craft QL POD reduced
models into a cost-effective SGS model. 

In addition to the performance of the cases studied, the potential for
generalizing this framework can be acknowledged.  Although only two
examples of turbulence were tested of the many that must be modeled in
an ESM, it should be noted that these were the only examples studied.
Because there were no assumptions made that are specific to these
cases in developing this QL POD reduced-model framework, it is
reasonable to expect the framework to work similarly well with other
types of boundary-layer turbulence in general, potentially addressing the drawbacks
of the current approach of modeling different turbulent phenomena
separately.

\section{Acknowledgements}
We thank Altan Allawala, Greg Chini, and Steve Tobias for helpful conversations.  The authors are grateful to the 
Kavli Institute for Theoretical Physics for supporting the Program ``Planetary Boundary Layers in Atmospheres, Oceans, and Ice on Earth and Moons'' in Spring 2018 where some of this work was carried out.  The research was supported in part by the National Science Foundation under Grant No. NSF PHY-1748958, NSF ONR N00014-17-1-2963, NSF OCE-1350795 as well as by the Institute at Brown for Environment and Society.

\appendix

\section{Proper Orthogonal Decomposition in Homogeneous Dimensions}
\label{app:pod_fourier}

Begin with a field, $\psi\left(\mathbf{x}\right)$, defined over $n$
continuous and statistically homogeneous periodic dimensions that span
$0$ to $L_i$ and define the volume $V = {\displaystyle \prod_i^n
  L_i}$.  For simplicity, inhomogeneous dimensions are not written
explicitly but all variables can be a defined over any number of
inhomogeneous dimensions without loss of generality. Because the
statistics are homogeneous in $\mathbf{x}$, the covariance, $c_2$,
indicating the second cumulant, must depend only on the field spacing:
\begin{equation}
c_2\left(\mathbf{x}_1 - \mathbf{x}_2\right) = \overline{\psi(\mathbf{x}_1)\psi(\mathbf{x}_2)}
\end{equation}
where the overline now specifically refers to a spatial average over
the $n$ homogeneous dimensions.  Introducing $\bar{\mathbf{x}} =
\frac{1}{2}\left(\mathbf{x}_1 + \mathbf{x}_2\right)$ and $\delta
\mathbf{x} = \mathbf{x}_1 - \mathbf{x}_2$, then this average can be
written
\begin{equation}
c_2\left(\mathbf{x}_1 - \mathbf{x}_2\right) = \frac{1}{V} \int_{\mathcal{V}_{\bar{x}}} d\bar{\mathbf{x}} \psi\left(\bar{\mathbf{x}} + \frac{\delta \mathbf{x}}{2}\right)\psi\left(\bar{\mathbf{x}} - \frac{\delta \mathbf{x}}{2}\right)
\end{equation}
Where $\mathcal{V}_{\bar{x}}$ spans the entire domain in position
space, $\left[0,L_i\right)$ for the $i$th homogeneous dimension.
  Then, writing the fields as a sum of their Fourier modes
\begin{align}
\psi(\mathbf{x}_1) &= \int_{\mathcal{V}_{k_1}}\frac{d\mathbf{k}_1}{\left(2 \pi\right)^n} e^{i \mathbf{k}_1 \cdot \mathbf{x}_1} \psi\left(\mathbf{k}_1\right) \\
\psi(\mathbf{x}_2) &= \int_{\mathcal{V}_{k_2}}\frac{d\mathbf{k}_2}{\left(2 \pi\right)^n} e^{i \mathbf{k}_2 \cdot \mathbf{x}_2} \psi\left(\mathbf{k}_2\right) 
\end{align}
where $\mathcal{V}_k$ spans the entire volume of wavenumber space,
$\left(-\infty,+\infty\right)$ in each homogeneous dimension.
Plugging this into the spatial average
\begin{align}
& c_2\left(\mathbf{x}_1 - \mathbf{x}_2\right) = \nonumber \\
&\qquad \frac{1}{V} \int_{\mathcal{V}_{k_1},\mathcal{V}_{k_2}} \frac{d\mathbf{k}_1}{\left(2 \pi\right)^n} \frac{d\mathbf{k}_2}{\left(2 \pi\right)^n} \int_{\mathcal{V}_{\bar{x}}} d\bar{\mathbf{x}} \nonumber \\
& \qquad \qquad e^{i \mathbf{k}_1 \cdot \left(\bar{\mathbf{x}}+\frac{\delta \mathbf{x}}{2}\right)} e^{i \mathbf{k}_2 \cdot \left(\bar{\mathbf{x}}-\frac{\delta \mathbf{x}}{2}\right)} \psi(\mathbf{k}_1)\psi(\mathbf{k}_2) \\
& c_2\left(\mathbf{x}_1 - \mathbf{x}_2\right) = \nonumber \\ 
& \qquad \int_{\mathcal{V}_k} \frac{d\mathbf{k}_1 d\mathbf{k}_2}{\left(2 \pi\right)^n} \delta\left(\mathbf{k}_1+\mathbf{k}_2\right) \nonumber \\
& \qquad \qquad e^{i \mathbf{k}_1 \cdot \frac{\delta \mathbf{x}}{2}} e^{- i \mathbf{k}_2 \cdot \frac{\delta \mathbf{x}}{2}} \psi(\mathbf{k}_1)\psi(\mathbf{k}_2) \\
& c_2\left(\mathbf{x}_1 - \mathbf{x}_2\right) = \nonumber \\
& \qquad \int_{\mathcal{V}_k} \frac{d\mathbf{k}_2}{\left(2 \pi\right)^n} e^{i \mathbf{k}_1 \cdot \delta\mathbf{x}} \psi(\mathbf{k}_1)\psi(\mathbf{k}_1)^*
\end{align}
so the covariance matrix is diagonal in Fourier space, and therefore
the POD modes are also Fourier modes in homogeneous dimensions.


\bibliographystyle{ametsoc2014}

\begin{thebibliography}{59}
\providecommand{\natexlab}[1]{#1}
\providecommand{\url}[1]{\texttt{#1}}
\renewcommand{\UrlFont}{\rmfamily}
\providecommand{\urlprefix}{URL }
\expandafter\ifx\csname urlstyle\endcsname\relax
  \providecommand{\doi}[1]{doi:\discretionary{}{}{}#1}\else
  \providecommand{\doi}{doi:\discretionary{}{}{}\begingroup
  \urlstyle{rm}\Url}\fi
\providecommand{\eprint}[2][]{\url{#2}}

\bibitem[{Adcroft et~al.(2018)}]{mitgcm}
Adcroft, A., and Coauthors, 2018: Mitgcm user manual. MIT department of EAPS,
  \urlprefix\url{http://mitgcm.org/download/manual/manual_20180114.pdf}.

\bibitem[{Allawala and Marston(2016)Allawala, and Marston}]{Allawala:2016hx}
Allawala, A., and J.~B. Marston, 2016: {Statistics of the stochastically-forced
  Lorenz attractor by the Fokker-Planck equation and cumulant expansions}.
  \textit{Physical review. E, Statistical physics, plasmas, fluids, and related
  interdisciplinary topics}, \textbf{94~(5)}, 052\,218.

\bibitem[{Allawala et~al.(2017)Allawala, Tobias,, and Marston}]{allawala17}
Allawala, A., S.~M. Tobias, and J.~B. Marston, 2017: Dimensional reduction of
  direct statistical simulation. \textit{arXiv preprint arXiv:1708.07805}.

\bibitem[{Bachman et~al.(2017)Bachman, Fox-Kemper, Taylor,, and
  Thomas}]{bachman2017parameterization}
Bachman, S., B.~Fox-Kemper, J.~Taylor, and L.~Thomas, 2017: Parameterization of
  frontal symmetric instabilities. i: Theory for resolved fronts. \textit{Ocean
  Modelling}, \textbf{109}, 72--95.

\bibitem[{Bartello and Holloway(1991)Bartello, and
  Holloway}]{bartelloholloway1991}
Bartello, P., and G.~Holloway, 1991: {Passive scalar transport in $\beta$-plane
  turbulence}. \textit{Journal of Fluid Mechanics}, \textbf{223}, 521--536,
  \doi{10.1017/S0022112091001532},
  \urlprefix\url{http://journals.cambridge.org/article_S0022112091001532}.

\bibitem[{Batchelor(1952)}]{batchelor1952diffusion}
Batchelor, G., 1952: Diffusion in a field of homogeneous turbulence: Ii. the
  relative motion of particles. \textit{Mathematical Proceedings of the
  Cambridge Philosophical Society}, Cambridge University Press, Vol.~48,
  345--362.

\bibitem[{Bauer et~al.(2015)Bauer, Thorpe,, and Brunet}]{Bauer_2015}
Bauer, P., A.~Thorpe, and G.~Brunet, 2015: {The quiet revolution of numerical
  weather prediction}. \textit{Nature}, \textbf{525~(7567)}, 47--55,
  \doi{10.1038/nature14956},
  \urlprefix\url{http://dx.doi.org/10.1038/nature14956}.

\bibitem[{Belcher et~al.(2012)}]{belcher2012global}
Belcher, S.~E., and Coauthors, 2012: A global perspective on langmuir
  turbulence in the ocean surface boundary layer. \textit{Geophysical Research
  Letters}, \textbf{39~(18)}.

\bibitem[{Bony et~al.(2015)}]{Bony:2015bd}
Bony, S., and Coauthors, 2015: {Clouds, circulation and climate sensitivity}.
  \textit{Nature Geoscience}, \textbf{8~(4)}, 261--268.

\bibitem[{Bouchet et~al.(2013)Bouchet, Nardini,, and
  Tangarife}]{bouchetnardinietal2013}
Bouchet, F., C.~Nardini, and T.~Tangarife, 2013: {Kinetic Theory of Jet
  Dynamics in the Stochastic Barotropic and 2D Navier-Stokes Equations}.
  \textit{Journal of Statistical Physics}, \textbf{153~(4)}, 572--625,
  \doi{10.1007/s10955-013-0828-3},
  \urlprefix\url{http://dx.doi.org/10.1007/s10955-013-0828-3}.

\bibitem[{{Canuto} and {Minotti}(2001){Canuto}, and
  {Minotti}}]{canutominotti2001}
{Canuto}, V.~M., and F.~{Minotti}, 2001: {Mixing and transport in stars - I.
  Formalism: momentum, heat and mean molecular weight}. \textit{Mon. Not. Roy.
  Ast. Soc.}, \textbf{328}, 829--838.

\bibitem[{Chaalal et~al.(2016)Chaalal, Schneider, Meyer,, and
  Marston}]{Chaalal:2016jx}
Chaalal, F.~A., T.~Schneider, B.~Meyer, and B.~Marston, 2016: {Cumulant
  expansions for atmospheric flows}. \textit{New Journal of Physics},
  \textbf{18~(2)}, 1--24.

\bibitem[{Craik(1977)}]{craik77}
Craik, A. D.~D., 1977: The generation of langmuir circulations by an
  instability mechanism. \textit{Journal of Fluid Mechanics}, \textbf{81~(2)},
  209--223.

\bibitem[{Doney et~al.(2004)}]{Doney_2004}
Doney, S.~C., and Coauthors, 2004: {Evaluating global ocean carbon models: The
  importance of realistic physics}. \textit{Global Biogeochem. Cycles},
  \textbf{18~(3)}, n/a--n/a, \doi{10.1029/2003gb002150},
  \urlprefix\url{http://dx.doi.org/10.1029/2003gb002150}.

\bibitem[{Farrell and Ioannou(2007)Farrell, and Ioannou}]{farrell07}
Farrell, B.~F., and P.~J. Ioannou, 2007: Structure and spacing of jets in
  barotropic turbulence. \textit{Journal of the Atmospheric Sciences},
  \textbf{64~(10)}, 3652--3665.

\bibitem[{Foken(2006)}]{foken200650}
Foken, T., 2006: 50 years of the monin--obukhov similarity theory.
  \textit{Boundary-Layer Meteorology}, \textbf{119~(3)}, 431--447.

\bibitem[{Fox-Kemper et~al.(2014)Fox-Kemper, Backman, Pearson,, and
  Reckinger}]{fox-kemper14}
Fox-Kemper, B., S.~D. Backman, B.~Pearson, and S.~Reckinger, 2014: Principles
  and advances in subgrid modelling for eddy-rich simulations. \textit{Clivar
  Exchanges}, \textbf{19~(2)}, 42--46.

\bibitem[{Fox-Kemper and Ferrari(2008)Fox-Kemper, and Ferrari}]{fox-kemper08}
Fox-Kemper, B., and R.~Ferrari, 2008: Parameterization of mixed layer eddies.
  part i: Theory and diagnosis. \textit{Journal of Physical Oceanography},
  \textbf{38~(6)}, 1145--1165.

\bibitem[{Hamlington et~al.(2014)Hamlington, Roekel, Fox-Kemper, Julien,, and
  Chini}]{Hamlington_2014}
Hamlington, P.~E., L.~P.~V. Roekel, B.~Fox-Kemper, K.~Julien, and G.~P. Chini,
  2014: {Langmuir{\textendash}Submesoscale Interactions: Descriptive Analysis
  of Multiscale Frontal Spindown Simulations}. \textit{J. Phys. Oceanogr.},
  \textbf{44~(9)}, 2249--2272, \doi{10.1175/jpo-d-13-0139.1},
  \urlprefix\url{http://dx.doi.org/10.1175/JPO-D-13-0139.1}.

\bibitem[{Harcourt(2013)}]{harcourt2013second}
Harcourt, R.~R., 2013: A second-moment closure model of langmuir turbulence.
  \textit{Journal of Physical Oceanography}, \textbf{43~(4)}, 673--697.

\bibitem[{Herring(1963)}]{herring63}
Herring, J.~R., 1963: Investigation of problems in thermal convection.
  \textit{Journal of Atmospheric Sciences}, \textbf{20}, 325--338.

\bibitem[{Holmes et~al.(2012)Holmes, Lumley, Berkooz,, and Rowley}]{holmes12}
Holmes, P., J.~L. Lumley, G.~Berkooz, and C.~W. Rowley, 2012:
  \textit{Turbulence, Coherent Structures, Dynamical Systems and Symmetry}. 2nd
  ed., Cambridge University Press, Cambridge.

\bibitem[{Kang et~al.(2009)Kang, Frierson,, and Held}]{kang2009tropical}
Kang, S.~M., D.~M. Frierson, and I.~M. Held, 2009: The tropical response to
  extratropical thermal forcing in an idealized gcm: The importance of
  radiative feedbacks and convective parameterization. \textit{Journal of the
  atmospheric sciences}, \textbf{66~(9)}, 2812--2827.

\bibitem[{Kolmogorov(1941)}]{kolmogorov1941dissipation}
Kolmogorov, A.~N., 1941: Dissipation of energy in locally isotropic turbulence.
  \textit{Dokl. Akad. Nauk SSSR}, Vol.~32, 16--18.

\bibitem[{Kraus and Turner(1967)Kraus, and Turner}]{KrausTurner67}
Kraus, E., and J.~Turner, 1967: A one-dimensional model of the seasonal
  thermocline. {II}: {The general theory and its consequences.}
  \textit{Tellus}, \textbf{19}, 98--106.

\bibitem[{{Krause} and {Raedler}(1980){Krause}, and
  {Raedler}}]{krauseraedler1980}
{Krause}, F., and K.~H. {Raedler}, 1980: \textit{{Mean-field
  magnetohydrodynamics and dynamo theory}}. Pergamon Press.

\bibitem[{Large et~al.(1997)Large, Danabasoglu, Doney,, and
  McWilliams}]{Large_1997}
Large, W.~G., G.~Danabasoglu, S.~C. Doney, and J.~C. McWilliams, 1997:
  {Sensitivity to Surface Forcing and Boundary Layer Mixing in a Global Ocean
  Model: Annual-Mean Climatology}. \textit{J. Phys. Oceanogr.},
  \textbf{27~(11)}, 2418--2447,
  \doi{10.1175/1520-0485(1997)027<2418:stsfab>2.0.co;2},
  \urlprefix\url{http://dx.doi.org/10.1175/1520-0485(1997)027<2418:stsfab>2.0.co;2}.

\bibitem[{Large et~al.(1994)Large, McWilliams,, and Doney}]{large94}
Large, W.~G., J.~C. McWilliams, and S.~C. Doney, 1994: Oceanic vertical mixing:
  A review and a model with a nonlocal boundary layer parameterization.
  \textit{Reviews of Geophysics}, \textbf{32~(4)}, 363 -- 403.

\bibitem[{Leibovich(1983)}]{Leibovich_1983}
Leibovich, S., 1983: {The form and Dynamics of Langmuir Circulations}.
  \textit{Annu. Rev. Fluid Mech.}, \textbf{15~(1)}, 391--427,
  \doi{10.1146/annurev.fl.15.010183.002135},
  \urlprefix\url{http://dx.doi.org/10.1146/annurev.fl.15.010183.002135}.

\bibitem[{Li et~al.(2019)}]{LiReichl19}
Li, Q., and Coauthors, 2019: Comparing ocean boundary vertical mixing schemes
  including {Langmuir} turbulence. \textit{Journal of Advances in Modeling
  Earth Systems (JAMES)}.

\bibitem[{Malkus(1954)}]{MALKUS:1954dh}
Malkus, W., 1954: {The Heat Transport and Spectrum of Thermal Turbulence}.
  \textit{Proceedings of the Royal Society of London. Series A, Mathematical
  and Physical Sciences}, \textbf{225~(1161)}, 196--212.

\bibitem[{Maltrud and Vallis(1991)Maltrud, and Vallis}]{maltrudvallis1991}
Maltrud, M.~E., and G.~K. Vallis, 1991: {Energy spectra and coherent structures
  in forced two-dimensional and beta-plane turbulence}. \textit{Journal of
  Fluid Mechanics}, \textbf{228}, 321--342.

\bibitem[{Marston et~al.(2016)Marston, Chini,, and Tobias}]{marston16}
Marston, J., G.~Chini, and S.~Tobias, 2016: Generalized quasilinear
  approximation: Application to zonal jets. \textit{Physical Review Letters},
  \textbf{116~(21)}, 214\,501.

\bibitem[{Marston(2010)}]{marston10}
Marston, J.~B., 2010: Statistics of the general circulation from cumulant
  expansions. \textit{Chaos}, \textbf{20~(4)}, 041\,107.

\bibitem[{{Marston} et~al.(2008){Marston}, {Conover},, and
  {Schneider}}]{marstonconoveretal2008}
{Marston}, J.~B., E.~{Conover}, and T.~{Schneider}, 2008: {Statistics of an
  Unstable Barotropic Jet from a Cumulant Expansion}. \textit{Journal of
  Atmospheric Sciences}, \textbf{65}, 1955, \doi{10.1175/2007JAS2510.1},
  \eprint{0705.0011}.

\bibitem[{Marston et~al.(2019)Marston, Qi,, and Tobias}]{marston14}
Marston, J.~B., W.~Qi, and S.~M. Tobias, 2019: \textit{Direct Statistical
  Simulation of a Jet {\rm in {\bf Zonal Jets: Phenomenology, Genesis, and
  Physics}} (arXiv:1412.0381)}. Cambridge University Press.

\bibitem[{McWilliams and Sullivan(2000)McWilliams, and
  Sullivan}]{McWilliams_2000}
McWilliams, J.~C., and P.~P. Sullivan, 2000: {Vertical Mixing by Langmuir
  Circulations}. \textit{Spill Science {\&} Technology Bulletin},
  \textbf{6~(3-4)}, 225--237, \doi{10.1016/s1353-2561(01)00041-x},
  \urlprefix\url{http://dx.doi.org/10.1016/s1353-2561(01)00041-x}.

\bibitem[{McWilliams et~al.(1997)McWilliams, Sullivan,, and
  Moeng}]{mcwilliams97}
McWilliams, J.~C., P.~P. Sullivan, and C.~Moeng, 1997: Langmuir turbulence in
  the ocean. \textit{Journal of Fluid Mechanics}, \textbf{334}, 1--30.

\bibitem[{Mellor and Yamada(1974)Mellor, and Yamada}]{mellor74}
Mellor, G.~L., and T.~Yamada, 1974: A hierarchy of turbulence closure models
  for planetary boundary layers. \textit{Journal of the Atmospheric Sciences},
  \textbf{31}, 1791--1806.

\bibitem[{O'Gorman and Schneider(2007)O'Gorman, and Schneider}]{ogorman07}
O'Gorman, P.~A., and T.~Schneider, 2007: Recovery of atmospheric flow
  statistics in a general circulation model. \textit{Geophysical Research
  Letters}, \textbf{34}, 524 -- 535.

\bibitem[{Pausch et~al.(2019)Pausch, Yang, Hwang,, and Eckhardt}]{pausch2019}
Pausch, M., Q.~Yang, Y.~Hwang, and B.~Eckhardt, 2019: Quasilinear approximation
  for exact coherent states in parallel shear flows. \textit{Fluid Dynamics
  Research}, \textbf{51}, 1--15.

\bibitem[{R{\"u}diger(1989)}]{ruediger1988}
R{\"u}diger, G., 1989: \textit{Differential rotation and stellar convection:
  Sun and solar-type stars}, Vol.~5. Gordon \& Breach, New York.

\bibitem[{Salmon(1998)}]{salmon98}
Salmon, R., 1998: \textit{Lectures on Geophysical Fluid Dynamics}. Oxford
  University Press, New York.

\bibitem[{Schmidt and Schumann(1989)Schmidt, and Schumann}]{schmidt89}
Schmidt, H., and U.~Schumann, 1989: Coherent structure of the convective
  boundary layer derived from large-eddy simulation. \textit{Journal of Fluid
  Mechanics}, \textbf{200}, 511--562.

\bibitem[{Sirovich(1989)}]{sirovich89}
Sirovich, L., 1989: Chaotic dynamics of coherent structures. \textit{Physica D:
  Nonlinear Phenomena}, \textbf{37~(1-3)}, 126--145.

\bibitem[{Skitka(2019)}]{skitka19_thesis}
Skitka, J., 2019: Quasilinear modeling of planetary boundary-layer turbulence
  [{P}h{D} {T}hesis]. \textit{Providence, RI: Brown University}.

\bibitem[{Spiegel(1962)}]{SPIEGEL:1962vf}
Spiegel, E.~A., 1962: {Thermal Turbulence at Very Small Prandtl Number}.
  \textit{Journal of Geophysical Research}, \textbf{67~(8)}, 3063--{\&}.

\bibitem[{Squire and Bhattacharjee(2015)Squire, and
  Bhattacharjee}]{Squire:2015fk}
Squire, J., and A.~Bhattacharjee, 2015: {Statistical Simulation of the
  Magnetorotational Dynamo}. \textit{Physical Review Letters},
  \textbf{114~(8)}, 085\,002--5.

\bibitem[{Srinivasan and Young(2012{\natexlab{a}})Srinivasan, and
  Young}]{srinivasan12}
Srinivasan, K., and W.~Young, 2012{\natexlab{a}}: Zonostrophic instability.
  \textit{Journal of the atmospheric sciences}, \textbf{69~(5)}, 1633--1656.

\bibitem[{Srinivasan and Young(2012{\natexlab{b}})Srinivasan, and
  Young}]{srinivasanyoung2012}
Srinivasan, K., and W.~R. Young, 2012{\natexlab{b}}: {Zonostrophic
  instability}. \textit{J. Atmos.\ Sci.}, \textbf{69}, 1633--1656.

\bibitem[{Stephens(2005)}]{2005JClimate}
Stephens, G.~L., 2005: {Cloud Feedbacks in the Climate System: A Critical
  Review.} \textit{Journal of Climate}, \textbf{18~(2)}, 237--273.

\bibitem[{Stocker et~al.(2013)}]{IPCC13}
Stocker, T.~F., and Coauthors, 2013: Climate change 2013: The physical science
  basis. Tech. rep. 1535 pp.

\bibitem[{Sullivan et~al.(1994)Sullivan, McWilliams,, and Moeng}]{sullivan94}
Sullivan, P.~P., J.~C. McWilliams, and C.-H. Moeng, 1994: A subgrid-scale model
  for large-eddy simulation of planetary boundary-layer flows.
  \textit{Boundary-Layer Meteorology}, \textbf{71~(3)}, 247--276.

\bibitem[{{Thomas} et~al.(2015){Thomas}, {Farrell}, {Ioannou},, and
  {Gayme}}]{thomasetal2015}
{Thomas}, V.~L., B.~F. {Farrell}, P.~J. {Ioannou}, and D.~F. {Gayme}, 2015: {A
  minimal model of self-sustaining turbulence}. \textit{Physics of Fluids},
  \textbf{27~(10)}, 105\,104, \doi{10.1063/1.4931776}.

\bibitem[{Thorpe(2004)}]{Thorpe_2004}
Thorpe, S., 2004: {{Langmuir} {Circulation}}. \textit{Annual Review of Fluid
  Mechanics}, \textbf{36~(1)}, 55--79,
  \doi{10.1146/annurev.fluid.36.052203.071431},
  \urlprefix\url{http://dx.doi.org/10.1146/annurev.fluid.36.052203.071431}.

\bibitem[{Tobias et~al.(2011)Tobias, Dagon,, and Marston}]{tobias11}
Tobias, S.~M., K.~Dagon, and J.~B. Marston, 2011: Astrophysical fluid dynamics
  via direct statistical simulation. \textit{The Astrophysical Journal},
  \textbf{727}, 127.

\bibitem[{Tobias and Marston(2013)Tobias, and Marston}]{tobias13}
Tobias, S.~M., and J.~B. Marston, 2013: {Direct Statistical Simulation of
  Out-of-Equilibrium Jets}. \textit{Physical Review Letters},
  \textbf{110~(10)}, 104\,502.

\bibitem[{Van~Roekel et~al.(2012)Van~Roekel, Fox-Kemper, Sullivan, Hamlington,,
  and Haney}]{van2012}
Van~Roekel, L., B.~Fox-Kemper, P.~Sullivan, P.~Hamlington, and S.~Haney, 2012:
  The form and orientation of langmuir cells for misaligned winds and waves.
  \textit{Journal of Geophysical Research: Oceans}, \textbf{117~(C5)}.

\bibitem[{Vedenov et~al.(1961)Vedenov, Velikhov,, and Sagdeev}]{Vedenov:1961us}
Vedenov, A.~A., E.~P. Velikhov, and R.~Z. Sagdeev, 1961: {Quasilinear theory of
  plasma oscillations}. \textit{Proceedings of IAEA Conference on Plasma
  Physics and Controlled Nuclear Fusion Research}, 465--475.

\bibitem[{Wang et~al.(2012)Wang, Akhtar, Borggaard,, and Iliescu}]{wang12}
Wang, Z., I.~Akhtar, J.~Borggaard, and T.~Iliescu, 2012: Proper orthogonal
  decomposition closure models for turbulent flows: a numerical comparison.
  \textit{Computer Methods in Applied Mechanics and Engineering}, \textbf{237},
  10--26.

\end{thebibliography}

\end{document}